\documentclass[]{aastex631}

\usepackage{graphicx}

\shorttitle{}
\shortauthors{Hoai et al.}
\graphicspath{{./}{figures}}

\begin{document}

\title{Stellar evolution along the AGB as revealed by the shape of Miras' visual light curves}

\author[0000-0002-3816-4735]{Do Thi Hoai}
\affiliation{Department of Astrophysics, Vietnam National Space Center (VNSC), Vietnam Academy of Science and Technology (VAST),\\
18 Hoang Quoc Viet, Cau Giay, Ha Noi, Vietnam}

\author[0000-0002-0311-0809]{Pham Tuyet Nhung}
\affiliation{Department of Astrophysics, Vietnam National Space Center (VNSC), Vietnam Academy of Science and Technology (VAST),\\
18 Hoang Quoc Viet, Cau Giay, Ha Noi, Vietnam}

\author[0000-0002-8979-6898]{Pierre Darriulat}
\affiliation{Department of Astrophysics, Vietnam National Space Center (VNSC), Vietnam Academy of Science and Technology (VAST),\\
18 Hoang Quoc Viet, Cau Giay, Ha Noi, Vietnam}

\author{Mai Nhat Tan}
\affiliation{Department of Astrophysics, Vietnam National Space Center (VNSC), Vietnam Academy of Science and Technology (VAST),\\
18 Hoang Quoc Viet, Cau Giay, Ha Noi, Vietnam}



\begin{abstract}
A new analysis of a sample of visual light curves of Mira variables is presented. The curves cover the past four decades and are selected from the AAVSO database as including a large number of high-density observations. The aim is to offer a more quantitative and more systematic picture than available from earlier studies. The results corroborate earlier descriptions and reveal new correlations between the shapes of the light curves and the evolution of the star along the Asymptotic Giant Branch (AGB). A family of nearly sinusoidal curves associated with M spectral types and displaying no sign of having experienced third-dredge-up events, is identified with confidence. A detailed study of its properties is presented and used to suggest possible interpretations. All other curves are clearly distinct from this family and usually start departing from it by displaying a broader luminosity minimum, progressively taking the form of a hump climbing the ascending branch as the star evolves along the AGB. This hump is studied in some detail and possible interpretations are considered. New correlations between parameters defining the shape of the light curve and the state of the star are revealed; while the average trend is established with confidence, deviations from it cause a significant scatter of the parameters of the curves. Comments aimed at shedding light on the underlying physics are presented together with speculative interpretations, in the hope that they could encourage and inspire new studies, in particular based on models of the inner star dynamics. 

\end{abstract}

\keywords{Asymptotic giant branch stars (2100) -- Mira variable stars (1066) -- Late stellar evolution (911) -- Light curves (918)}


\section{Introduction} \label{sec1}

For centuries, some stars have been known to display variable visual luminosity, some of which pulsating regularly. Such are, for example, the Cepheids and the Long-Period Variables (LPV) on the Asymptotic Giant Branch (AGB). The former were discovered at the end of the XVIIIth century \citep{Pigott1785, Goodricke1786} while the first pulsating AGB star, $o$ Ceti, was observed by Fabricius at the end of the XVIth century \citep{Hoffleit1997}. Over time, stars displaying similar variability, with periods at the scale of a few days for the former and of a year for the latter, have been discovered; they are now known as Cepheids and Miras respectively and the brighter of both families have been the targets of high-quality observations for many decades, often over a century. In the early years of the XXth century, Henrietta Leavitt discovered the period-luminosity relation obeyed by Cepheids in the Magellanic Clouds, for which the distance to Earth is well-defined \citep{Leavitt1907,Leavitt1912}, and Eddington proposed a mechanism governing their pulsation, which is now commonly accepted and is known under the name of kappa-mechanism \citep{Eddington1917}.

The kappa-mechanism implies a specific dependence of the opacity $\kappa$ (kappa) of the star interior as a function of density $\rho$ and temperature $T$. When energy transport is purely radiative, $\kappa$ obeys approximately the Kramers law, $\kappa \sim \rho T^{-7/2}$, which accounts for both bound-free and free-free photon absorption (ionisation and bremsstrahlung, respectively). A small contraction causes both $\rho$ and $T$ to increase but the effect of the latter on opacity dominates over that of the former, allowing for more photons to escape: the star is in a state of stable equilibrium, the effects of temperature and pressure balancing each other. However, as Eddington remarked, if the opacity increases with temperature, the star becomes unstable against pulsations; but it was not until the middle of the century that \citet{Zhevakin1953} identified ionised helium as the cause of the anomalous behaviour of the opacity. What happens is that the heat generated by compression is mostly used to ionise the medium and the temperature remains essentially constant as the density increases. Such a kappa-mechanism is at work in a region of the Hertzsprung-Russel (HR) diagram known as the instability strip and explains reasonably well the pulsations of Cepheids \citep[][ and references therein]{Guzik2023}.

In contrast, for AGB stars, the energy transport is dominantly convective rather than radiative, and therefore essentially independent of the opacity. A description of their pulsations cannot be as straightforward as in the case of Cepheids: to explain the pulsating instability in AGB stars one needs a dynamic theory of convection, for which a one-dimensional approximation, the so-called Mixing Length Theory (MLT), has been in use for over six decades \citep[][ and references therein]{Joyce2023}. In analogy with molecular heat transfer, it describes the bulk movement of the fluid in terms of the mean-free path of parcels in pressure equilibrium, but not in thermal equilibrium, with their surrounding, governed by the dependence of pressure on temperature. Recently, significant progress toward producing more realistic three-dimensional models of convection has been achieved \citep[][ and references therein]{Andrassy2022, Rizzuti2024} and the role played by convective cells in the evolution of AGB stars has become an important topic of study, both for observations \citep{RosalesGuzman2024, Paladini2018, Darriulat2024} and for modelling \citep[][and references therein]{Herwig2005, Marigo2013, Freytag2023}.

Another major complication makes the study of the light curves of LPV AGB stars particularly difficult: dust is abundantly produced in the cool atmospheres of the stars and, together with pulsations, is the main driver, by radiation pressure from the stellar surface, of the mass-losing wind \citep[][ and references therein]{McDonald2018,Liljegren2016}. As noted by \citet{Reid2002} for oxygen-rich stars and by \citet{Winters1994} for carbon-rich stars, such dust significantly hides the observation of the photosphere in the visible and dramatically affects the appearance of the visual light curve.

In spite of such difficult interpretation, much effort has been dedicated to the study of the visual light curves of LPV AGB stars, with the aim to relate the main features that they display to properties specific to the pulsating star. In particular, the very long intervals of time over which such curves have been observed, often over a century, have allowed for precise measurements of the pulsation period and for revealing its evolution with time. Fourier transforms and wavelet techniques are commonly used for such studies \citep{Templeton2005}. Period-luminosity relations have been identified \citep[][ and references therein]{Soszynski2007,Whitelock2008,McDonald2016}, showing evidence for Miras to be radial pulsators in the fundamental mode \citep{Wood1996, Trabucchi2017}. Particular attention has been given to the study of period changes over time. \citet{Zijlstra2002b} distinguish between three different types of period changes: continuous evolution, sudden changes and meandering periods. Indeed, strong period changes have been observed in a small fraction of Miras \citep{Templeton2005, Wood1981, Gal1995, Hawkins2001, Zijlstra2002a, MerchanBenitez2000, MerchanBenitez2002, Uttenthaler2011, Uttenthaler2016a, Uttenthaler2016b, Molnar2019} and are often, but not always, interpreted as evidence for recent thermal pulses in the helium burning shell. Correlations between the phase, traditionally measured from the maxima of the visual light curves and spanning one unit over a period, and temperature or diameter of the pulsating star have been measured in the near infrared in a few cases \citep{Woodruff2008, Lacour2009, VanBelle1996, VanBelle2002}; the temperature is found to be anti-correlated with the star diameter, as is also the case for Cepheids \citep{Andrievsky2005}, but important dependence of the luminosity on wavelength makes a reliable interpretation of these observations difficult.

More generally, the indicators of the evolution of LPVs along the AGB, in particular those that reveal the past occurrence of a third-dredge-up event, are found to be correlated not only with the period, but with several other features of the light curves. Such is the case of the $^{12}$C/$^{13}$C ratio \citep{Ohnaka1996, Abia1997, Greaves1997, Schoier2000, Ramstedt2014, Lebzelter2019}, of the presence of technetium in the spectrum \citep[][ and references therein]{Uttenthaler2019} and of the transition from oxygen-rich to carbon-rich \citep{Mennessier1997, Whitelock1999}. Similarly, in the case of more massive stars, the occurrence of hot bottom burning \citep[][and references therein]{GarciaHernandez2013} has been found to be correlated with peculiarities of the light curves. Other, weaker correlations, mostly related to the chemistry of the star atmosphere, have been reported: between OH emission and the steepness of the ascending branch of the visual light curve \citep{Bowers1975, Bowers1977}; between the appearance at 9.7 and 20 microns of silicate emission features \citep{Vardya1986, Onaka1989}, or also the dust mass loss rate evaluated using the K–[22] colour index \citep{MerchanBenitez2023} and the asymmetry of the light oscillation; between the detection of the H$_2$O vapour line at 1.35 cm \citep{Vardya1987}, or also of dust formation \citep{LeBertre1992, Winters1994}, and the shape of the oscillation.

Several authors, faced with the difficulty to devise a reliable and sensible interpretation of the visual light curves of Mira stars, have attempted to produce classifications in terms of the particular features that they display. It started early in the past century with \citet{Ludendorff1928} making a distinction between three main types of light curves: ($\alpha$), when the ascending branch is steeper than the descending branch and the minimum broader than the maximum; ($\beta$), when the curve is essentially symmetric with only minor differences between the steepness of the ascending and descending branches; and ($\gamma$), when the ascending branch displays steps or humps, or when the oscillations are double-peaked.  Indeed, the asymmetry between ascending and descending branches and the possible non-monotonic behaviour of the former, usually in the form of humps\footnote{These describe episodes of slowing down of the rate of luminosity increase on the ascending branch of an oscillation. They last a fraction of a period, typically 10 to 15\%. They are usually referred to as bumps or humps in the published literature and we use the latter in the present article.}, are, together with the amplitude of the oscillations, major features that have been the object of detailed consideration in all studies. Similar, but slightly different features have been observed in the near-infrared \citep[e.g.][]{Lockwood1971}, suggesting that they are indicators of intrinsic properties of the pulsating star. Subsequently, attempts at making the Ludendorff classification less subjective and at quantifying the relevant features by introducing adequate parameters have been made; in particular, \citet{Vardya1988} introduced the asymmetry parameter $f$, defined as the ratio of the rising time over the period, and found that 80\% of the Mira variables lie in the range 0.4$<$$f$$<$0.5 and that $f$ is correlated with the M, S or C spectral type of the pulsating star. \citet{Lebzelter2011} has parameterised the mean deviation of the oscillation profile from a sine wave on the ascending and descending branches separately, making a distinction between light curves displaying a broad minimum and a narrow maximum, and light curves displaying a narrow minimum and a broad maximum. More recently, \citet{MerchanBenitez2023} have separated Mira stars in two classes depending on both the symmetry and possibly non-monotonic shape of the oscillations and found a clear correlation with the dust mass-loss rate evaluated using the K–[22] colour index.

The study reported in the present article is in the wake of the work of these latter authors, motivated as we are by the feeling that the outstanding set of high-quality observations of visual light curves of LPV AGB stars has not yet been exploited as thoroughly as it deserves. Yet, we are conscious of the difficulty that such a study is facing and of the quantity and quality of previous investigations sharing the same ambition: we have defined accordingly which method to use and which sample of stars to study.

The article is organised as follows: in the next two sections we define the samples of curves on which the analysis is based and the parameterisation used to characterise each curve. Reaching reliable conclusions with confidence implies severe restrictions on the quality and density of the observations from which the curves are made. For this reason, we present analyses of two separate samples of light curves, one referred to as sample A is larger but the curve parameters are less rigorously defined than for the other, smaller sample, referred to as sample B. Sample A includes 71 stars but the density of observations is not sufficient to allow for the exclusive use of well-defined algorithms for a rigorous evaluation of the curve parameters; this is at variance with sample B, which includes 32 curves. For sample A, it is sometime necessary to exclude significant intervals of time where the density of observations is low. The curve parameters of sample A are simply evaluated from the plots accessible from the AAVSO website, using a number of well-tested graphic methods. Nevertheless, a comparison between the evaluations of the parameters of samples A and B, made on curves that belong to both samples, has shown that the less rigorous sample A results are indeed reliable. Section 4 discusses the main results obtained for oxygen-rich stars and Section 5 for carbon stars. Section 6 comments on the regularity of the light curves and Section 7 presents a discussion of the results attempting to correlate the properties of the light curves with the evolution of the star along the AGB. A brief summary is presented in Section 8.

\section{Selection of visual light curves}
\begin{figure*}
  \centering
  \includegraphics[width=0.85\linewidth]{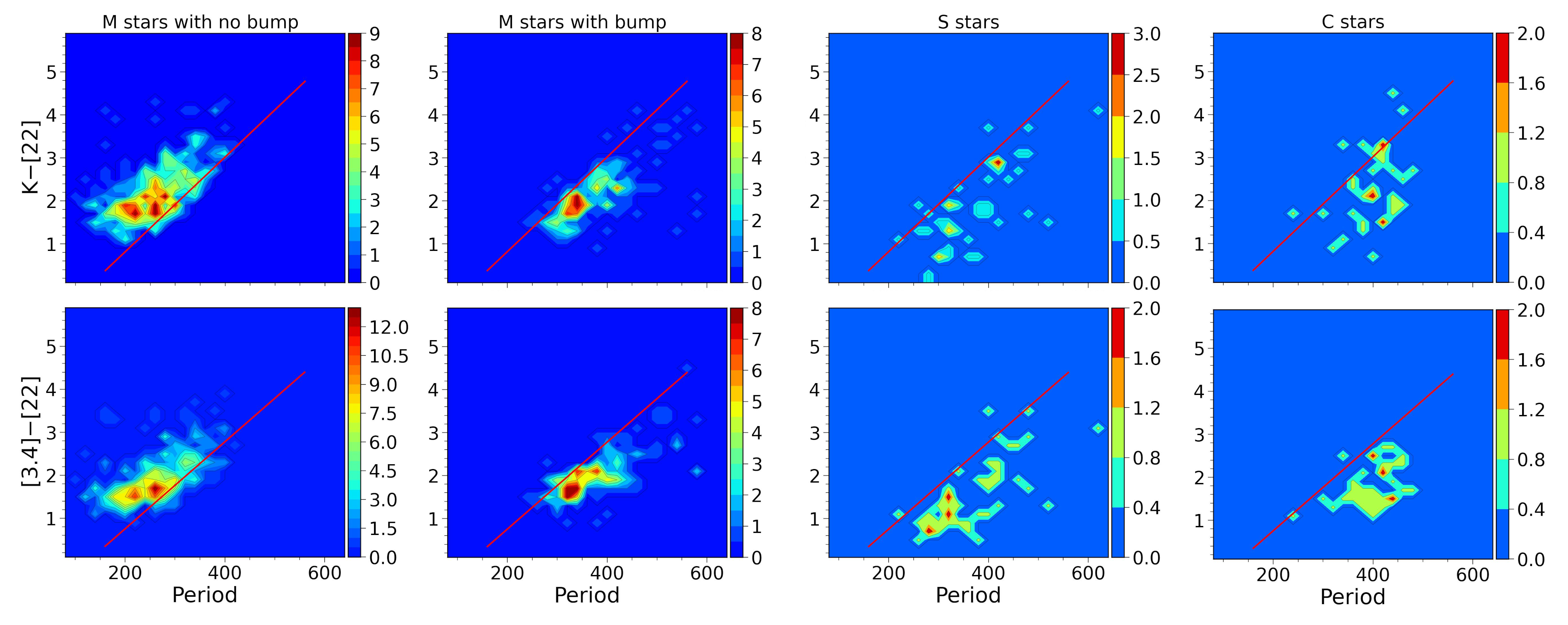}
  \caption{ Distribution of the light curves of the \citet{MerchanBenitez2023} study in the K$-$[22] (upper row) and [3.4]$-$[22] (lower row) vs period planes. The line shown in the upper row is defined by \citet{Uttenthaler2019} as K$-$[22]$\sim$$0.011(P-125)$. The same line, simply scaled down using the ratio (0.92) between the mean values of K$-$[22] and [3.4]$-$[22], is shown in the lower row.}
 \label{fig1}
\end{figure*}

{\label{sec2}}
Any progress in the understanding and interpretation of the visual light curves of Mira variables must build on, rather than repeat, results obtained by earlier studies \citep{Ludendorff1928, Vardya1988, Lebzelter2011, MerchanBenitez2023}. These studies identified several basic properties of such curves that help with their characterisation: the asymmetry between the ascending and descending branches, the irregularity between successive oscillations, both in time and magnitude, the width of the maxima and minima, the possible occurrence of non-linear features on the ascending branch, usually in the form of humps. They give evidence for correlations between the shape of the curve and the stage of evolution of the pulsating star on the AGB, possibly revealing the occurrence of recent helium flashes and third-dredge-up events, and making a distinction between stars of M, S and C spectral types. In the wake of such findings, the present study aims at providing a more precise and more detailed description of the shapes of the curves in the hope to better understand their relation to the intrinsic properties of the pulsating star. This requires restricting the study to light curves made from measurements of outstanding density and quality, implying that one must accept to work on relatively small samples, giving up the ambition to cover as much as possible of the available observations. Such an approach has guided the work reported in the present article. It has been arbitrarily restricted to observations covering the forty past years and to visual light curves from the AAVSO data base, with the understanding that curves from other sources and different wavelengths could be considered at a later stage if judged necessary. More precisely, we select stars from the sample used by \citet{MerchanBenitez2023}, covering 548 Mira stars with light curves of reasonable quality having a full V-band amplitude in excess of 2.5 mag. Such a sample provides a good coverage of the Mira family, however with the exception of binaries, which Merchán Benítez et al. considered judicious to exclude; such are the cases of $o$ Ceti and R Aqr. For each star of the sample, \citet{MerchanBenitez2023} provide the following information: the spectral type, the period and its variability, the possible presence of technetium in the spectrum, the $^{12}$C/$^{13}$C ratio when available, the values of the colour indices (in units of magnitude) K–[22] and [3.4]–[22], both of which are commonly used proxies of the dust mass-loss rate \citep{McDonald2016, McDonald2018}. In addition, they list the class, S (for symmetric) or A (for asymmetric), to which they assign the curve. Note that the meaning of the word symmetric by \citet{MerchanBenitez2023} differs from its use in the present work: here, like for \citet{Vardya1988}, it means equal times spent on the ascending and descending branches; for \citet{MerchanBenitez2023} it means monotonic behaviour of the ascending branch. Class A includes curves displaying a hump or a sharp change of slope on their ascending branch; class S includes all other curves. Figure \ref{fig1} displays the distribution of the curves in the colour vs period planes.

In a preliminary phase, in order to gain some familiarity with the features displayed by different light curves, we inspected visually a sample of 35 curves of nearby stars, selected for their large apparent brightness. It gave us the opportunity to design different forms of parameterisation and to test how they perform. We then extended such detailed inspection to four families: the 35 C-stars, the 26 S-stars, the 20 M-stars tagged has having technetium in their spectrum (referred to as Myes) and the 60 M-stars tagged as having no technetium in their spectrum (referred to as Mno) that the Merch\'{a}n Ben\'{i}tez sample contains. This preliminary phase, about which we do not report here, has nevertheless been essential in guiding us to optimally define how to select the samples on which the present article is based and to devise appropriate parameterisation of the light curves. After having rejected a few light curves made of observations of lesser density or quality, being particularly selective in the case of the large subsample of Mno curves, we were left with a final sample of 71 curves, referred to as sample A, containing 13 Mno stars, 14 Myes stars, 22 S stars and 22 C stars. Not included in this sample are the curves of three Mno stars, two of which display double-peak oscillations, R Cen and R Nor; the third one, X Oph, has so small oscillation amplitudes and so broad minima that it cannot be simply compared to the other curves. The clearly different nature of these three curves requires a separate analysis. Table \ref{tab5} of the appendix lists the sample A curves together with information of relevance.

Many curves of sample A do not allow for a rigorous parameterisation of their shape, failing to cover a sufficient density of observations over the whole phase range, in particular around minima of luminosity. For this reason, we selected a restricted sample of 24 light curves from sample A offering a high density of observations over most of the four past decades. It includes 6 Mno, 5 Myes, 7 S and 6 C spectral types. We added to it 8 light curves of spectral type M that had been discarded from sample A because technetium had not been searched for in their spectrum or, when it had been, could not be said to be present or absent with sufficient confidence. The resulting sample of 32 light curves is referred to as sample B. Table \ref{tab1} lists the values of the relevant parameters and Figure \ref{figA1} of the Appendix illustrates the main results.

\section{Defining the shapes of the selected light curves}\label{sec3}
\subsection{Parameterisation of the selected light curves}\label{sec3.1}

We start by measuring as accurately as possible the time and magnitude of each maximum and each minimum of the curve. For sample A, this can be done only approximately because of two difficulties that are met: one is the presence of flat maxima, with a very small enhancement usually located at its later edge; defining the time of the maximum as the time of the centre of the plateau or as the time of the enhancement implies differences reaching up to 8\% of the period. The same difficulty is met in the analysis of Cepheid light curves, as remarked and clearly illustrated in Figure \ref{fig2} of the study presented by \citet{Antonello1993}. This happens when a hump on the ascending branch reaches the maximum of the curve and eventually gets confused with it. The second difficulty relates to the evaluation of the time at minimum light and is mostly caused by the frequent lack of observations in this low apparent brightness region of the light curve.

For sample B, we collect the observations stored in the AAVSO database in time bins spanning typically 2\% of the period. Each bin contains the number of observations, the mean magnitude and its rms value with respect to the mean. To locate the maxima, we produce a parabolic fit to these data in a time interval of 7 bins centred on the date where the maximum is expected. In the rare cases where the fit gives a date deviating by more than a bin from the centre of the interval, we may increase the size of the interval and/or iterate with an interval shifted accordingly. This procedure defines maximum dates to better than 1\% of the period and maximum apparent luminosities to better than 0.1 units of magnitude. The main source of systematic uncertainty attached to the measurement of the dates is the occasional occurrence of very flat maxima.

Locating the minima proceeds similarly, except that we use broader intervals, typically of 17 bins, in order to cope with the much smaller number of observations, several bins being empty. We start with an interval centred at the phase corresponding to the mean asymmetry between ascending and descending branches and iterate the fit when necessary.  We estimate that this procedure defines minimum dates to better than 2\% of the period and minimum apparent brightness to better than 0.1 units of magnitude.

\begin{deluxetable*}{lcc ccc ccc ccc}
\tablenum{1}
\tablecaption{Parameterisation of the 32 light curves of sample B. Columns 2 to 7 list the values taken by the parameters defined in Section 3. Columns 8, 10 and 11 are taken from \citet{MerchanBenitez2023} and list the period variation in percent, the $^{12}$C/$^{13}$C ratio and the colour indices, respectively. Columns 9 and 12 list the profile type and the value taken by parameter $K_{\rm MB}$ as defined in Section 3.1. \label{tab1}}
\colnumbers
\tablehead{
\colhead{Name}&\colhead{$P/\Delta P$}&\colhead{$\varphi_{\rm min}/\Delta \varphi_{\rm min}$}
&\colhead{$M_{\rm max}/\Delta M_{\rm max}$}&\colhead{$M^\prime_{\rm max}$}&\colhead{$A/\Delta A$}
&\colhead{$\langle N_{\rm dis}/Rms_{\rm dis}\rangle$}&\colhead{$dP/P$}&\colhead{$Prof.$}
&\colhead{$^{12}$C/$^{13}$C}&\colhead{dust}&\colhead{$K_{\rm MB}$}}
\startdata
\multicolumn{12}{c}{Mno spectral type}\\
\cline{1-12}
R Aql	&	273.6/8.4	&	0.528/0.025	&	6.49/0.32	&	0.33	&	4.34/0.34	&	51/0.22	&	16.7	&	\edit2{a} 	&	8	&	2.3/2.2	&	$-$0.6	\\
R Boo	&	223.8/6.0	&	0.520/0.025	&	7.28/0.29	&	0.4	&	5.07/0.28	&	40/0.24	&	1.8	&	\edit2{a} 	&	-	&	1.6/1.8	&	$-$0.5	\\
R Cas	&	431.6/10.9	&	0.605/0.025	&	6.10/0.75	&	0.93	&	5.97/0.40	&	89/0.27	&	2.1	&	\edit2{b} 	&	12	&	2.4/1.8	&	1	\\
U Her	&	404.1/8.2	&	0.583/0.017	&	7.44/0.34	&	0.45	&	4.90/0.28	&	63/0.22	&	2.2	&	\edit2{b} 	&	19	&	2.4/2.5	&	0.7	\\
R Tri	&	266.0/6.0	&	0.522/0.022	&	6.24/0.32	&	0.46	&	5.35/0.29	&	44/0.20	&	1	&	\edit2{a} 	&	-	&	1.8/1.5	&	$-$0.2	\\
T UMa	&	256.1/10.2	&	0.583/0.030	&	7.77/0.53	&	0.7	&	5.28/0.33	&	62/0.23	&	2.3	&	\edit2{a} 	&	-	&	2.4/2.3	&	$-$1.0	\\
\cline{1-12}
\multicolumn{12}{c}{Myes spectral type}\\
\cline{1-12}
R Aur	&	457.1/8.4	&	0.458/0.020	&	7.57/0.45	&	0.55	&	5.92/0.43	&	42/0.26	&	4.4	&	\edit2{d}	&	33	&	2.4/1.8	&	1.3	\\
T Cep	&	338.4/13.0	&	0.467/0.038	&	6.10/0.29	&	0.28	&	4.12/0.48	&	117/0.23	&	7	&	\edit2{c} 	&	33	&	1.2/1.2	&	1.7	\\
RU Her	&	485.0/11.7	&	0.536/0.019	&	7.94/0.61	&	0.87	&	5.80/0.37	&	49/0.26	&	4.5	&	\edit2{b} 	&	25	&	2.2/2.4	&	1.8	\\
S Her	&	306.6/9.3	&	0.490/0.029	&	7.64/0.29	&	0.37	&	5.36/0.32	&	30/0.31	&	5.2	&	\edit2{d}	&	-	&	1.0/1.2	&	1	\\
R Ser	&	354.9/9.5	&	0.574/0.021	&	6.72/0.50	&	0.65	&	6.47/0.35	&	69/0.23	&	1.7	&	\edit2{a} 	&	14	&	1.9/2.0	&	0.6	\\
\cline{1-12}
\multicolumn{12}{c}{Other M spectral types}\\
\cline{1-12}
U UMi	&	324.3/14.9	&	0.502/0.043	&	8.29/0.28	&	0.2	&	3.26/0.37	&	23/0.24	&	3.6	&	\edit2{d}	&	-	&	1.5/1.7	&	0.7	\\
V Cas	&	228.7/5.7	&	0.513/0.025	&	7.66/0.27	&	0.28	&	4.80/0.43	&	34/0.24	&	3	&	\edit2{a} 	&	-	&	1.6/1.6	&	$-$0.5	\\
T Cas	&	441.4/25.2	&	0.441/0.054	&	8.30/0.31	&	0.17	&	3.12/0.55	&	50/0.33	&	2.5	&	\edit2{d}	&	33	&	2.1/1.7	&	1.4	\\
S CrB	&	360.3/11.3	&	0.636/0.022	&	7.20/0.47	&	0.56	&	5.46/0.37	&	65/0.22	&	1.9	&	\edit2{b} 	&	-	&	2.7/2.8	&	$-$0.1	\\
R UMa	&	300.8/6.8	&	0.609/0.022	&	7.40/0.38	&	0.47	&	5.49/0.25	&	78/0.23	&	1.7	&	\edit2{b} 	&	-	&	2.9/2.8	&	$-$1.0	\\
RT Cyg	&	190.3/11.1	&	0.536/0.027	&	7.38/0.45	&	0.62	&	4.65/0.32	&	33/0.25	&	1.1	&	\edit2{a} 	&	-	&	1.6/1.3	&	$-$0.9	\\
R Dra	&	246.8/5.5	&	0.546/0.022	&	7.60/0.42	&	0.49	&	5.13/0.27	&	46/0.21	&	2	&	\edit2{a} 	&	-	&	2.3/1.9	&	$-$1.0	\\
R CVn	&	328.9/6.9	&	0.509/0.024	&	7.56/0.37	&	0.45	&	4.44/0.34	&	34/0.20	&	3.3	&	\edit2{c} 	&	-	&	2.1/2.0	&	0.1	\\
\cline{1-12}
\multicolumn{12}{c}{S spectral type}\\
\cline{1-12}
R And	&	409.7/12.0	&	0.610/0.025	&	7.39/0.80	&	0.72	&	7.44/0.64	&	71/0.21	&	3.2	&	\edit2{a} 	&	40	&	2.9/2.3	&	0.2	\\
W And	&	397.5/8.8	&	0.575/0.021	&	7.92/0.68	&	0.9	&	6.29/0.31	&	36/0.21	&	1.8	&	\edit2{b} 	&	-	&	1.8/1.8	&	1.2	\\
R Cam	&	269.8/16.6	&	0.519/0.051	&	8.59/0.29	&	0.26	&	4.53/0.64	&	20/0.21	&	4.4	&	\edit2{d}	&	-	&	0.2/0.6	&	1.4	\\
T Cam	&	375.9/15.8	&	0.494/0.029	&	8.31/0.19	&	0.17	&	5.44/0.32	&	23/0.20	&	3.8	&	\edit2{d}	&	31	&	0.8/0.5	&	1.9	\\
Chi Cyg	&	406.9/9.2	&	0.549/0.018	&	5.14/0.68	&	0.91	&	8.31/0.37	&	227/0.23&	2	&       \edit2{c} 	&	36	&	2.4/1.7	&	0.7	\\
R Cyg	&	426.7/23.1	&	0.619/0.026	&	7.90/0.83	&	1.45	&	6.15/0.25	&	92/0.38	&	2.3	&	\edit2{a} 	&	29	&	2.9/2.9	&	0.4	\\
S UMa	&	226.3/6.8	&	0.498/0.031	&	8.01/0.09	&	0.08	&	3.69/0.38	&	55/0.25	&	3.5	&	\edit2{d}	&	-	&	1.2/1.1	&	$-$0.1	\\
\cline{1-12}
\multicolumn{12}{c}{C spectral type}\\
\cline{1-12}
S Cam	&	329.0/16.4	&	0.478/0.041	&	8.59/0.12	&	0.11	&	1.84/0.21	&	14/0.24	&	2.7	&	\edit2{d}	&	14	&	0.9/1.2	&	1.4	\\
W Cas	&	404.5/12.2	&	0.493/0.023	&	9.00/0.11	&	0.1	&	2.82/0.28	&	30/0.24	&	2.2	&	\edit2{d/c}	&	25	&	0.7/1.2	&	2.4	\\
S Cep	&	482.6/20.7	&	0.460/0.026	&	7.97/0.44	&	0.41	&	2.38/0.20	&	34/0.32	&	3.1	&	\edit2{c} 	&	-	&	2.7/1.7	&	1.2	\\
V CrB	&	357.3/14.9	&	0.583/0.025	&	8.26/0.57	&	0.47	&	3.13/0.23	&	39/0.32	&	4	&	\edit2{a} 	&	-	&	2.3/1.9	&	0.3	\\
U Cyg	&	466.3/12.3	&	0.512/0.026	&	7.61/0.47	&	0.43	&	3.18/0.40	&	51/0.34	&	4.5	&	\edit2{c} 	&	14	&	1.9/2.2	&	1.8	\\
T Dra	&	423.3/16.4	&	0.564/0.040	&	9.67/0.32	&	0.35	&	3.02/0.29	&	16/0.33	&	4	&	\edit2{e/a}	&	24	&	3.4/2.6	&	$-$0.1	\\
\enddata
\end{deluxetable*}

Having obtained for each maximum $i$ of the light curve a time $T_{\rm max,i}$ and a magnitude $M_{\rm max,i}$ and for its following minimum a time $T_{\rm min,i}$ and a magnitude $M_{\rm min,i}$, we define eleven parameters as follows:
\begin{itemize}
\item the period $P$, mean value of the time separating two successive maxima, $P$=$<$$T_{\rm max,i+1}-T_{\rm max,i}$$>$, and $\Delta P$, the rms value with respect to $P$, both measured in units of days.
\item the mean phase of the minimum, $\varphi_{\rm min}$=$<$$(T_{\rm min,i}-T_{\rm max,i})$/$(T_{\rm max,i+1}-T_{\rm max,i})$$>$. It is the complement to unity of the parameter $f$ introduced by \citet{Vardya1988} and is related to the asymmetry parameter $\alpha=1-2\varphi_{\rm min}$. $\Delta \varphi_{\rm min}$ is the rms value of this quantity with respect to $\varphi_{\rm min}$. In sample A, a direct evaluation of $\varphi_{\rm min}$ is often not possible, because of the insufficient phase coverage near luminosity minimum. We provide instead an approximate evaluation of the asymmetry parameter $\alpha$, obtained from a comparison of the slopes of the ascending and descending branches.
\item the mean brightness at maximum light, $M_{\rm max}$=$<$$M_{\rm max,i}$$>$, and its rms value with respect to the mean, $\Delta M_{\rm max}$. $M_{\rm max}$ depends on the distance of the star to the Sun, but $\Delta M_{\rm max}$ provides a first measure of the global irregularity of the curve. Another, independent measure is given by the mean of the absolute value of the difference between the magnitudes of two successive maxima, $\Delta'M_{\rm max}$=$|M_{\rm max,i+i}-M_{\rm max,i}|$. For random magnitude fluctuations having a Gaussian distribution, $\Delta M_{\rm max,i+i}$=1.08$\times$$\Delta M_{\rm max}$. A difference between the two parameters reveals the nature of the source of the observed magnitude fluctuations: when fluctuations between successive oscillations are small but when the magnitude of the maxima displays strong dependence on time, one expects $\Delta M_{\rm max}$$>$$\Delta'M_{\rm max}$; if instead the magnitude of the maxima stays constant on average, but fluctuates strongly from one oscillation to the next, $\Delta M_{\rm max}$$<$$\Delta'M_{\rm max}$. There are examples of both cases. In practice, rather than using $\Delta M_{\rm max}$ and $\Delta'M_{\rm max}$ separately, we may prefer to use their mean value, $\Delta M_0$=($\Delta M_{\rm max}$+$\Delta'M_{\rm max}$)/2 and their ratio, $R_{\rm \Delta M}$$=$$\Delta'M_{\rm max}/\Delta M_{\rm max}$.
\item The mean amplitude of the oscillations, measured in units of magnitude, $A$=$<$$M_{\rm min,i}-1/2(M_{\rm max,i}+M_{\rm max,i+1})$$>$; it is sufficient to average the amplitudes of the descending and ascending branches as their means must be equal when the number of oscillations is large enough. $\Delta A$ is the rms value of this quantity with respect to $A$; it provides a third measure of the global irregularity of the curve.  
\item A parameter $K_{\rm MB}$, referred to as the Merch\'{a}n Ben\'{i}tez index; it makes use of the finding of \citet{Uttenthaler2019} reported in Figure 13 of the \citet{MerchanBenitez2023} article, according to which a clear separation splits the curves in two distinct families in the K$-$[22] vs $P$ plane. Above a line of equation K–[22]$\sim$0.011($P$–125), the curves are mostly associated with spectral type M and are smooth, at strong variance with the population below the curve. We define accordingly $K_{\rm MB}$$=$$0.011(P-125)-(K-[22])$.
\item  Finally, we noted that some light curves display significantly larger dispersions than average between the magnitudes reported by different observations at a same time. At first sight, one would think that this simply reflects instrumental errors and is of no relevance to the present study. However, the effect is observed to depend on the nature of the star, being stronger for some carbon stars, which excludes an interpretation in pure terms of instrumental uncertainties. In order to study it, we select a time interval of $\sim$6\% of a period, centred on each maximum and record the number of observations that it contains, $N_{\rm dis,i}$, and the rms value of the measured magnitudes with respect to their mean, $\Delta M_{\rm dis,i}$, and retain their mean values as two additional global parameters, $N_{\rm dis}$=$<$$N_{\rm dis,i}$$>$ and $\Delta M_{\rm dis}$=$<$$\Delta M_{\rm dis,i}$$>$. 
\end{itemize}

\subsection{Normalised profiles}\label{sec3.2}
The visual inspection performed in the preliminary phase of the study had shown the difficulty to classify the light curves in terms of the lack of smoothness of the ascending branch. At first sight, many light curves appear to display regular featureless oscillations, while others display well-pronounced humps on most of the oscillations. However, a detailed inspection reveals continuity between these two groups: many curves display humps on only a small fraction of the oscillations, and the amplitude of such humps is often very small. We therefore produced for each curve of sample B a mean normalised oscillation profile, together with the distribution of the rms deviation of the individual oscillation profiles with respect to this mean profile. Normalised profiles cover from maximum $i$ to maximum $i+1$ and are defined as follows: the abscissa $x$ measures the phase in 50 bins of 0.02 width, defined as 0 at $T_{\rm max,i}$ and 1 at $T_{\rm max,i+1}$; on the descending branch, $x<\varphi_{\rm min,i}$, $y$=$(M-M_{\rm min,i})/(M_{\rm max,i}-M_{\rm min,i})$ and on the ascending branch, $x>\varphi_{\rm min,i}$, $y$=$(M-M_{\rm min,i})/(M_{\rm max,i+1}-M_{\rm min,i})$, $M$ being the mean magnitude in bin $x$. Like $x$, $y$ varies from 0 to 1, being equal to 0 at minimal apparent brightness and equal to 1 at maximal apparent brightness. In some cases, rather than normalising the oscillation profile over a whole period, we restrict it to a single branch, descending or ascending, with $x$ spanning from 0 to 1 for $\varphi$ spanning from 0 to $\varphi_{\rm min}$ or from $\varphi_{\rm min}$ to 1, respectively, each time using 25 bins.

\begin{figure*}
  \centering
  \includegraphics[width=15cm]{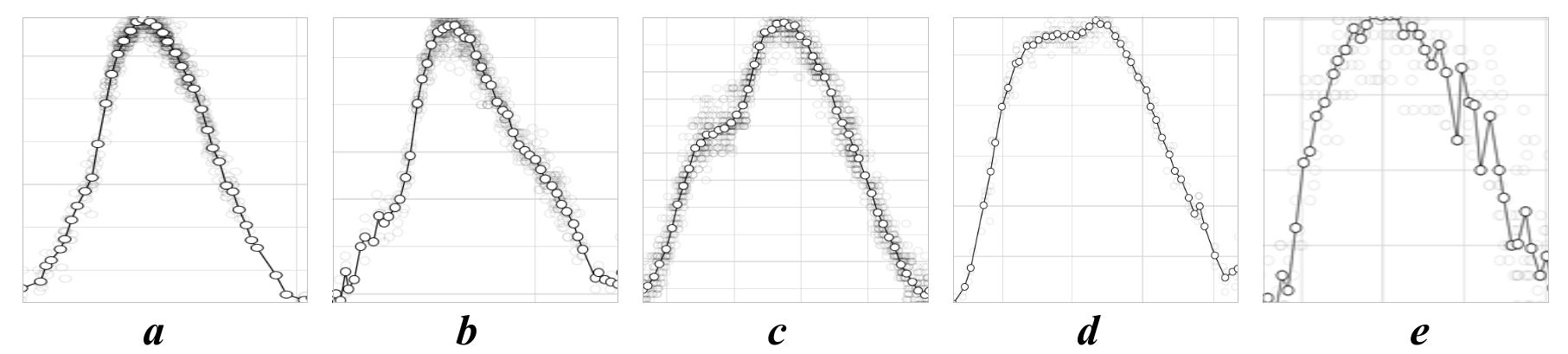}
  \caption{Typical oscillation profiles of types \edit2{a} to \edit2{e} (from left to right)}
 \label{fig2}
\end{figure*}

	 In the analysis of sample A curves, hoping to progress in the search for criteria efficient at discriminating between different light curves, we attempted at grouping them in classes associated with a specific shape of the oscillation profile. To this end, we defined five typical profile types illustrated in Figure \ref{fig2}.  Types \edit2{a} and \edit2{e} display no hump on the ascending branch, while types \edit2{b} to \edit2{d} do. The difference between types \edit2{a} and \edit2{e} is the width of the oscillation, larger in the latter case. The hump on the rising branch is clear for type \edit2{c}, typically between phases of 5/8 and 7/8, closer to minimum for type \edit2{b} and closer to maximum for type \edit2{d}. The exercise proved to be difficult in many cases, for which the choice between two successive profiles of a pair was not obvious. This gave evidence for continuity between the sequence of profiles depicted in Figure \ref{fig2}, a result that may suggest that stars on the AGB would typically evolve from profile \edit2{a} to profile \edit2{e} according to such a sequence. Part of the difficulty was that a same curve often displays a broader variety of oscillation profiles than the difference between the typical profiles of the pair of candidate types. In spite of such difficulty, evidence for significant correlations between the mean oscillation profile of \edit2{a} curve and its spectral type was obtained. In particular, the Mno spectral type was found to be present in profile types \edit2{a} and \edit2{b} exclusively, while profile type \edit2{e} is the exclusive domain of spectral type C. 

\subsection{Humpy and hump-free families}\label{sec3.3}
For the light curves of sample A that are also part of sample B, we performed a 6th degree polynomial fit of the normalised profile of each oscillation of the light curves; the result is always of very good quality and we used it to define humpy curves from their behaviour in the central phases of the ascending branch, excluding the minimum and maximum regions. The match with the assignment of profile types was in general excellent.

\begin{figure*}
  \centering
  \includegraphics[height=4.35cm,trim=0.cm 0.1cm 1.2cm 0cm,clip]{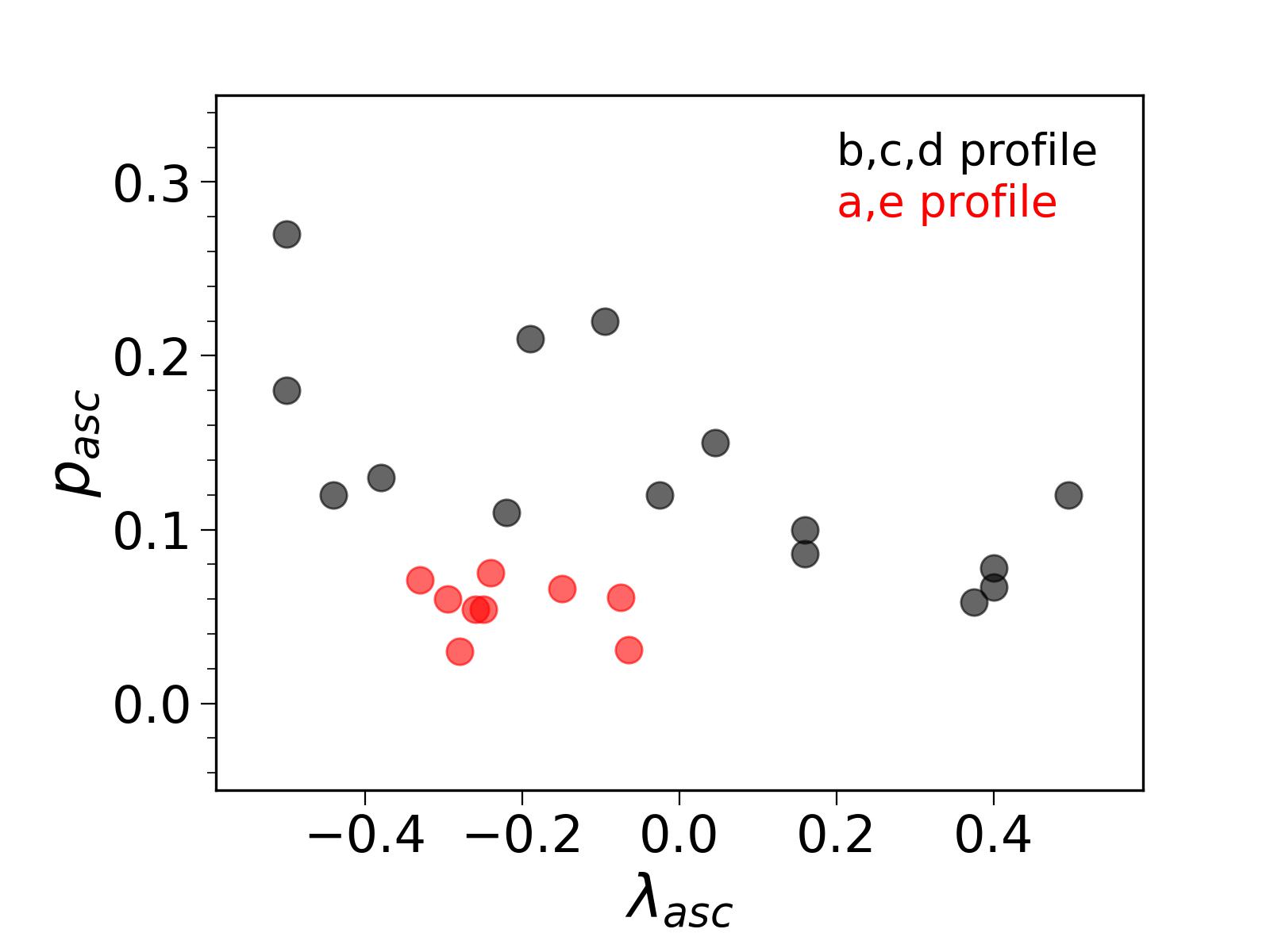}
  \includegraphics[height=4cm,trim=0.cm 0cm 0cm 0cm,clip]{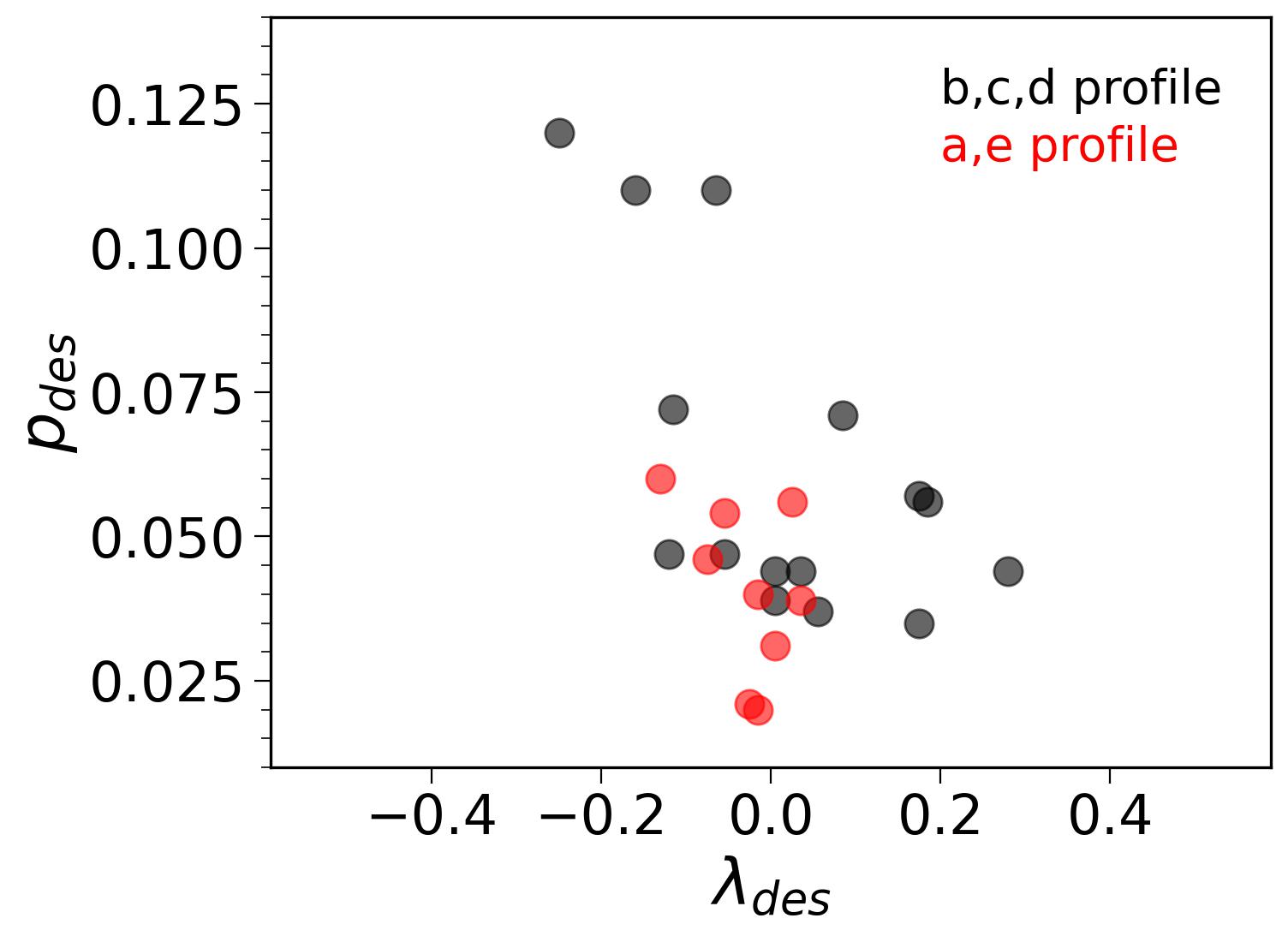}
  \includegraphics[height=4cm,trim=0.cm 0cm 0cm 0cm,clip]{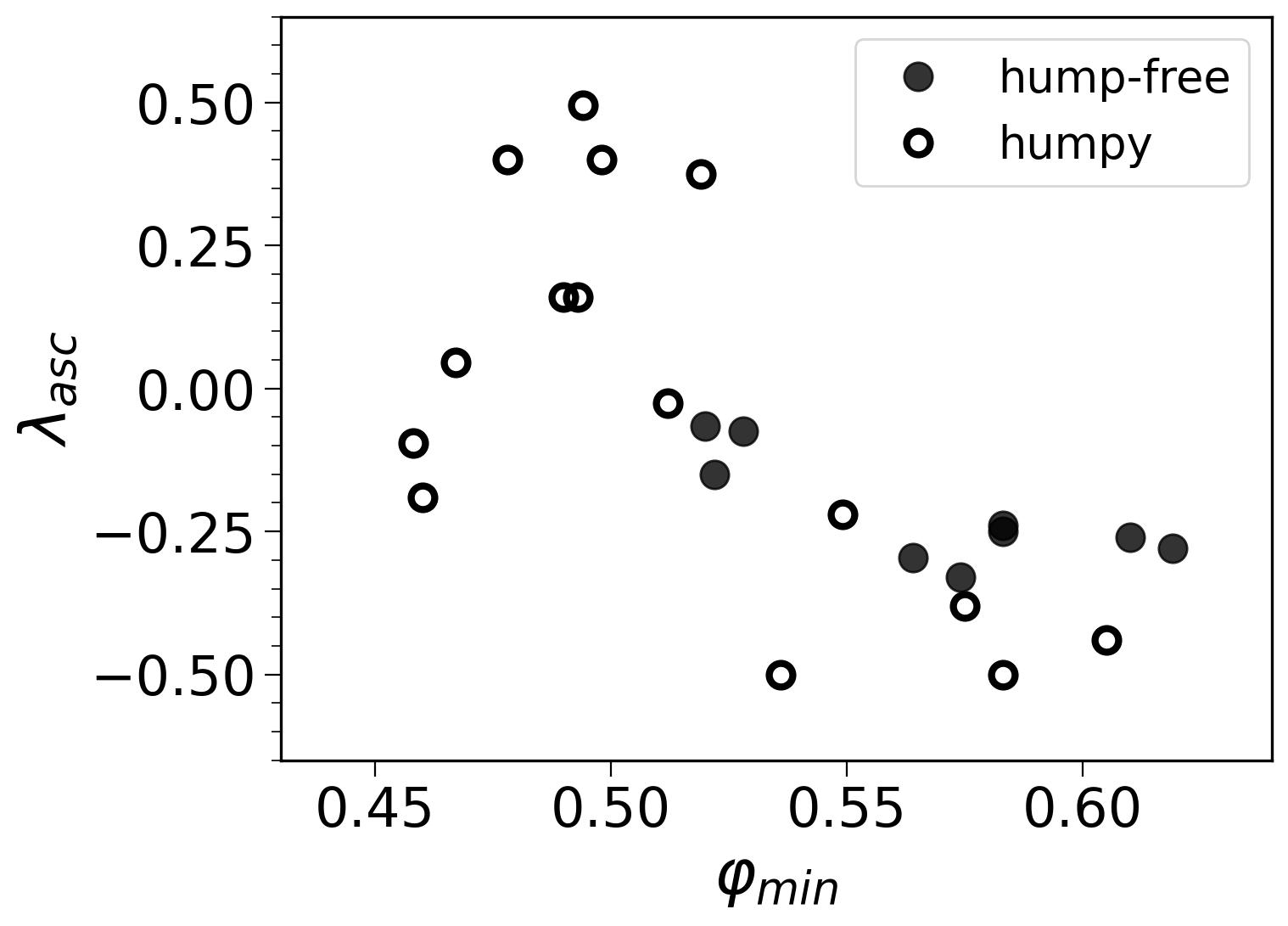}
  \caption{Dependence of $p$ on $\lambda$ for the ascending (left panel) and descending (central panel) branches. Black dots are for \edit2{b}, \edit2{c} and \edit2{d} profiles, red dots are for \edit2{a} and \edit2{e} profiles. The right panel displays the dependence of the value of $\lambda$ (ascending branch) on $\varphi_{\rm min}$. Open circles are for the humpy family, full circles for the hump-free family.}
  \label{fig3}
\end{figure*}

In the case of these 24 sample B curves, with the aim to obtain a simpler and more rigorous definition of humpy and hump-free curves than obtained from the assignment of profile types \edit2{a} to \edit2{e}, we tried to distinguish them on the only basis of the mean normalised profile of the ascending branch. To the normalised profile of each branch, we fit a curve depending on a single parameter $\lambda$: namely, writing $\xi=2x-1$, $\eta=2y-1$, $\eta$=Sgn$\times \frac{1}{2}\xi(\xi^2-3)+\lambda[1-\xi^2(2-\xi^2)]$ with Sgn=+1 on the descending branch and $-$1 on the ascending branch;  $\lambda$ is the value of $\eta$ at $\xi$=0  ($x=\frac{1}{2}$). They are the simplest possible polynomial forms having maxima and minima at the proper place and distorted by a single parameter, $\lambda$. Table \ref{tab2} lists the best fit values of $\lambda$ and $p$, the rms deviation between fit and data, for each branch separately and Figure \ref{fig3} displays the results in the $p$ vs $\lambda$ plane.

\begin{figure*}
  \centering
  \includegraphics[width=1\linewidth,trim=0.cm 0cm 0cm 0cm,clip]{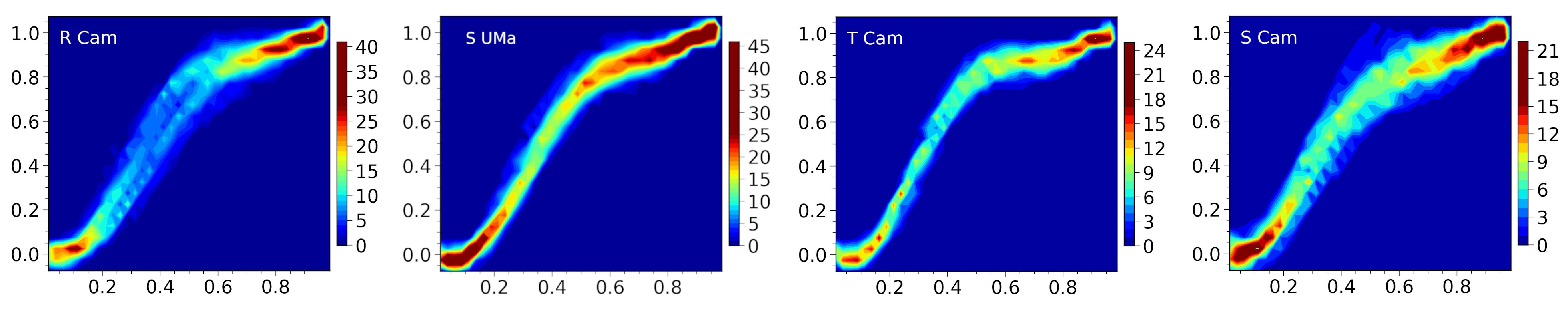}
  \caption{Superimposed normalised profiles of the ascending branch of the four curves having $\lambda>0.3$ on the ascending branch. From left to right: R Cam, S UMa, T Cam, S Cam.}
 \label{fig4}
\end{figure*}

They show a good match between the set of curves considered as hump-free from the assignment of profile types and the set having $p<0.1$ on the ascending branch, with the exception of three cases: S Her, R Cam and S UMa. These display clear humps near maximum light but their mean profiles are well described by a hump-free form having a large value of $\lambda$. This is better illustrated in Figure \ref{fig4}, which compares profiles of the four stars having $\lambda>0.3$ on the ascending branch (Figure \ref{fig3} left), R Cam, S UMa, T Cam and S Cam. All four have a \edit2{d} type profile, the first three are of spectral type S and S Cam is of spectral type C. This illustrates the difficulty to reveal the presence of a hump when it approaches maximal luminosity, as had been remarked earlier.

\begin{figure*}
  \centering
  \includegraphics[width=1\linewidth,trim=0.cm 0cm 0cm 0cm,clip]{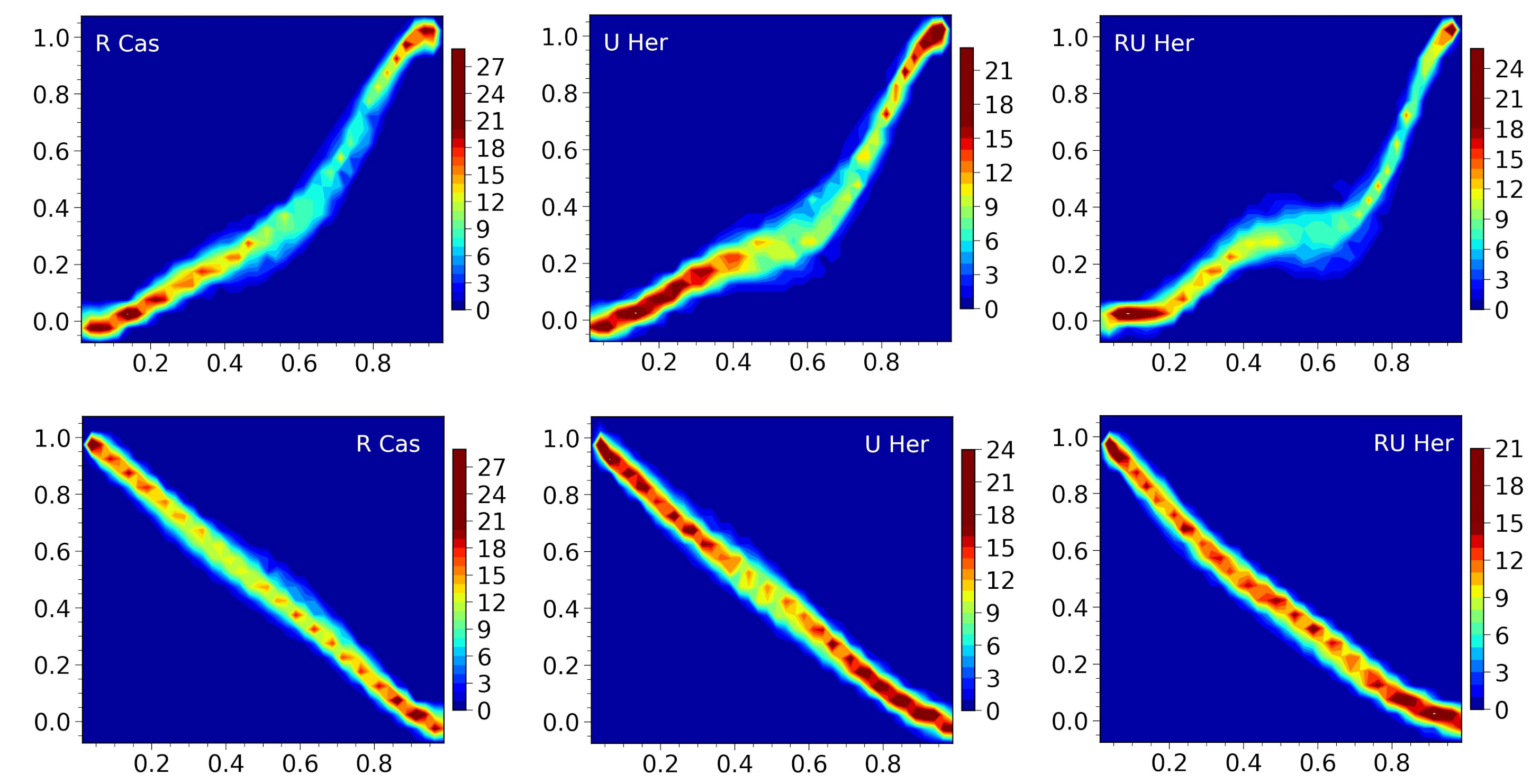}
  \caption{Superimposed normalised profiles of the ascending (upper panels) and descending (lower panels) branches of the three curves having $p>0.1$ on the descending branch (Figure \ref{fig3} centre); from left to right: R Cas (Mno/\edit2{b}), U Her (Mno/\edit2{b}) and RU Her (Myes/\edit2{b}).}
 \label{fig5}
\end{figure*}

A similar difficulty is met for humps near minimum luminosity. Figure \ref{fig5} displays the profiles of the three curves having $p>0.1$ on the descending branch (Figure \ref{fig3} centre), R Cas, U Her and RU Her, to which a \edit2{b} type profile had been assigned. They display clear humps on the lower part of the ascending branch but their descending branches are perfectly well-behaved. Indeed all 24 curves display a hump-free descending branch.

\begin{figure*}
  \centering
  \includegraphics[width=1\linewidth,trim=0.cm 0cm 0cm 0cm,clip]{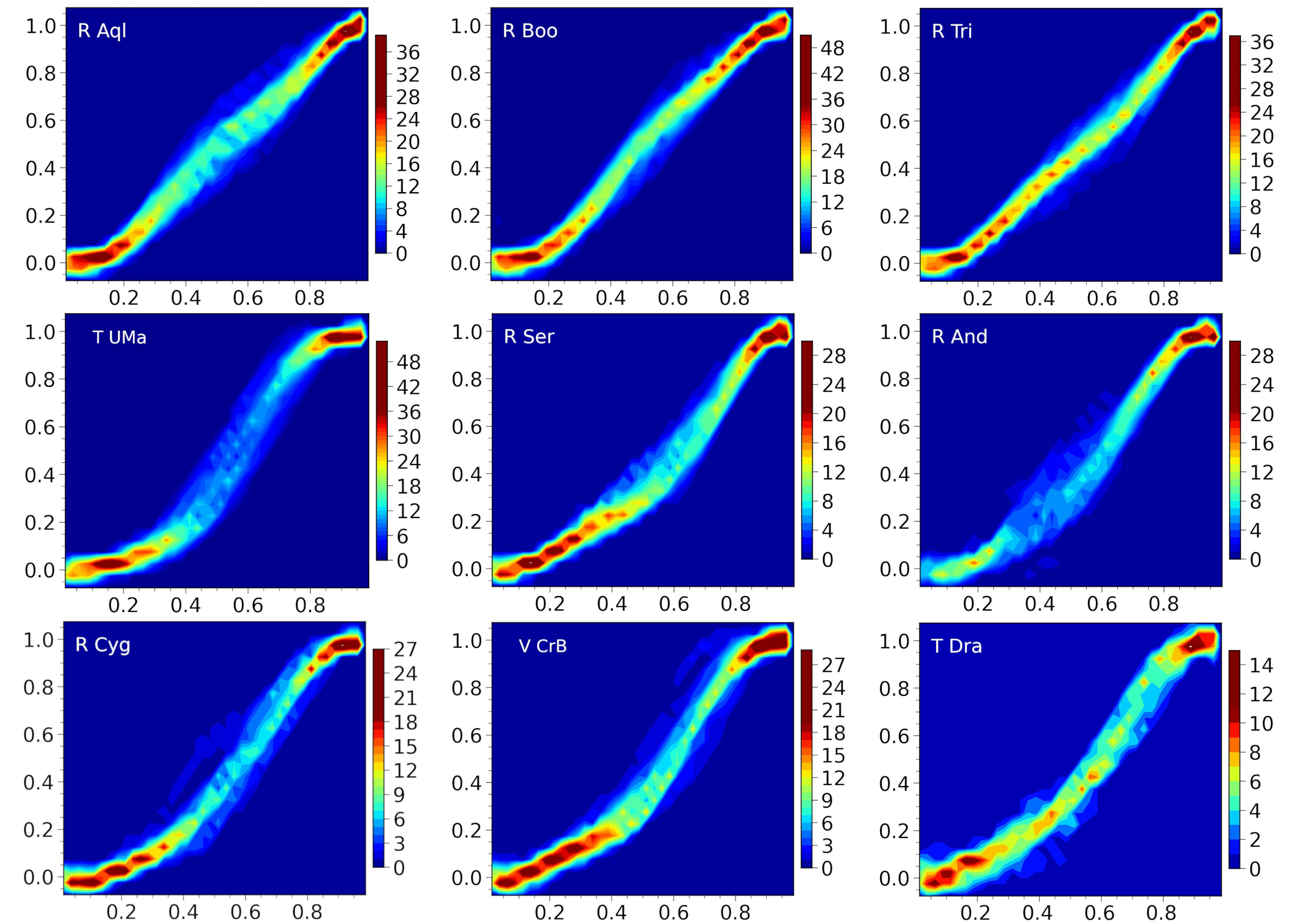}
  \caption{Superimposed normalised profiles of the ascending branches of the hump-free family. From left to right: R Aql, R Boo and R Tri (upper row), T UMa, R Ser and R And (middle row), R Cyg, V CrB and T Dra (lower row). }
 \label{fig6}
\end{figure*}

The risk of confusion of a hump with the minimum of luminosity is a warning that such a humpy curve may have been assigned an \edit2{a} profile type, in the same way as the S Cam curve might have been assigned an \edit2{e} profile type. In order to check on this, we display in Figure \ref{fig6} the normalised profiles of the ascending branches of the nine curves retained in the hump-free family. While none of these gives evidence for the presence of a clear hump, it is difficult to reliably ascertain their absence. The right panel of Figure \ref{fig3} displays the dependence of $\lambda$ (ascending branch) on the phase $\varphi_{\rm min}$ at minimal luminosity. In the hump-free family, for which the fit is good and $\lambda$ accordingly meaningful, three curves of spectral type Mno, R Aql, R Boo and R Tri, have $\lambda$ near zero and $\varphi_{\rm min}$ near 0.5, implying that they are close to a sine wave. But the other six curves are slightly asymmetric, with $\lambda$ between $-$0.24 and $-$0.33 and $\varphi_{\rm min}$ between 0.56 and 0.62. The three curves of the humpy family having $\varphi_{\rm min}$$>$0.57 are R Cas, U Her and W And, all having profiles showing a hump on the lower part of the ascending branch. While this confirms the validity of the assignment of the curves having $\varphi_{\rm min}$$>$0.56 to the humpy and hump-free families, respectively, it also illustrates how marginal is the distinction between these and invites prudence.    

\begin{deluxetable}{llc ccc c}
\tablenum{2}
\tablecaption{Best fit values of the single parameter fit to the two branches of the mean normalised profiles of each of the 24 sample B light curves selected from sample A (see text). \label{tab2}}
\tablehead{
\colhead{Name}&\colhead{Type/}&\multicolumn{2}{c}{Descending}&\multicolumn{2}{c}{Ascending}&\colhead{$\varphi_{\rm min}$}\\
\cline{3-4}
\cline{5-6}
\colhead{}&\colhead{profile}&\colhead{$\lambda$}&\colhead{$p$}&\colhead{$\lambda$}&\colhead{$p$}&\colhead{}}
\startdata
\multicolumn{7}{c}{Hump-free family}\\
\cline{1-7}
R Aql	&	Mno/\edit2{a}	&	$-$0.055	&	0.054	&	$-$0.075	&	0.061	&	0.53	\\
R Boo	&	Mno/\edit2{a}	&	$-$0.015	&	0.020	&	$-$0.065	&	0.031	&	0.52	\\
R Tri	&	Mno/\edit2{a}	&	$-$0.075	&	0.046	&	$-$0.15	&	0.066	&	0.52	\\
T UMa	&	Mno/\edit2{a}	&	$-$0.025	&	0.021	&	$-$0.24	&	0.075	&	0.58	\\
R Ser	&	Myes/\edit2{a} &	$-$0.015	&	0.040	&	$-$0.33	&	0.071	&	0.57	\\
R And	&	S/\edit2{a}	&	0.025	&	0.056	&	$-$0.26	&	0.054	&	0.61	\\
R Cyg	&	S/\edit2{a}	&	$-$0.13	&	0.060	&	$-$0.28	&	0.030	&	0.62	\\
V CrB	&	C/\edit2{a}	&	0.005	&	0.031	&	$-$0.25	&	0.054	&	0.58	\\
T Dra	&	C/\edit2{e-a}	&	0.035	&	0.039	&	$-$0.30	&	0.060	&	0.56	\\
\cline{1-7}
\multicolumn{7}{c}{Humpy family}\\
\cline{1-7}
R Cas	&	Mno/\edit2{b}	&	$-$0.065	&	0.11		&	$-$0.44	&	0.12	&	0.61	\\
U Her	&	Mno/\edit2{b}	&	$-$0.16	&	0.11		&	$-$0.50	&	0.18	&	0.58	\\
R Aur	&	Myes/\edit2{d}&	$-$0.12	&	0.047	&	$-$0.095	&	0.22	&	0.46	\\
T Cep	&	Myes/\edit2{c}&	$-$0.055	&	0.047	&	0.045	&	0.15	&	0.47	\\
RU Her	&	Myes/\edit2{b}&	$-$0.25	&	0.12		&	$-$0.50	&	0.27	&	0.54	\\
S Her	&	Myes/\edit2{d}&	0.005	&	0.044	&	0.16		&	0.086	&	0.49	\\
W And	&	Syes/\edit2{b}	&	0.085	&	0.071	&	$-$0.38	&	0.13	&	0.58	\\
R Cam	&	S?/\edit2{d}	 &	0.185	&	0.056	&	0.375	&	0.058	&	0.52	\\
T Cam	&	Syes/\edit2{d}	&	0.175	&	0.057	&	0.495	&	0.12	&	0.49	\\
Chi Cyg	&	Syes/\edit2{c}	&	$-$0.115	&	0.072	&	$-$0.22	&	0.11	&	0.55	\\
S UMa	&	Syes/\edit2{d}	&	0.175	&	0.035	&	0.40	&	0.067	&	0.50	\\
S Cam	&	C/\edit2{d}	        &	0.28	&	0.044	&	0.40	&	0.078	&	0.48	\\
W Cas	&	Cyes/\edit2{d-c}	&	0.035	&	0.044	&	0.16	&	0.10	&	0.49	\\
S Cep	&	C?/\edit2{c}	        &	0.055	&	0.037	&	$-$0.19	&	0.21	&	0.46	\\
U Cyg	&	Cyes/\edit2{c}	&	0.005	&	0.039	&	$-$0.025	&	0.12	&	0.51	\\
\enddata
\end{deluxetable}

\section{Oxygen-rich stars} \label{sec4}
\subsection{Mno spectral type} \label{sec4.1}

\begin{figure*}
  \centering
  \includegraphics[height=5cm,trim=0.cm 0cm 0cm 0cm,clip]{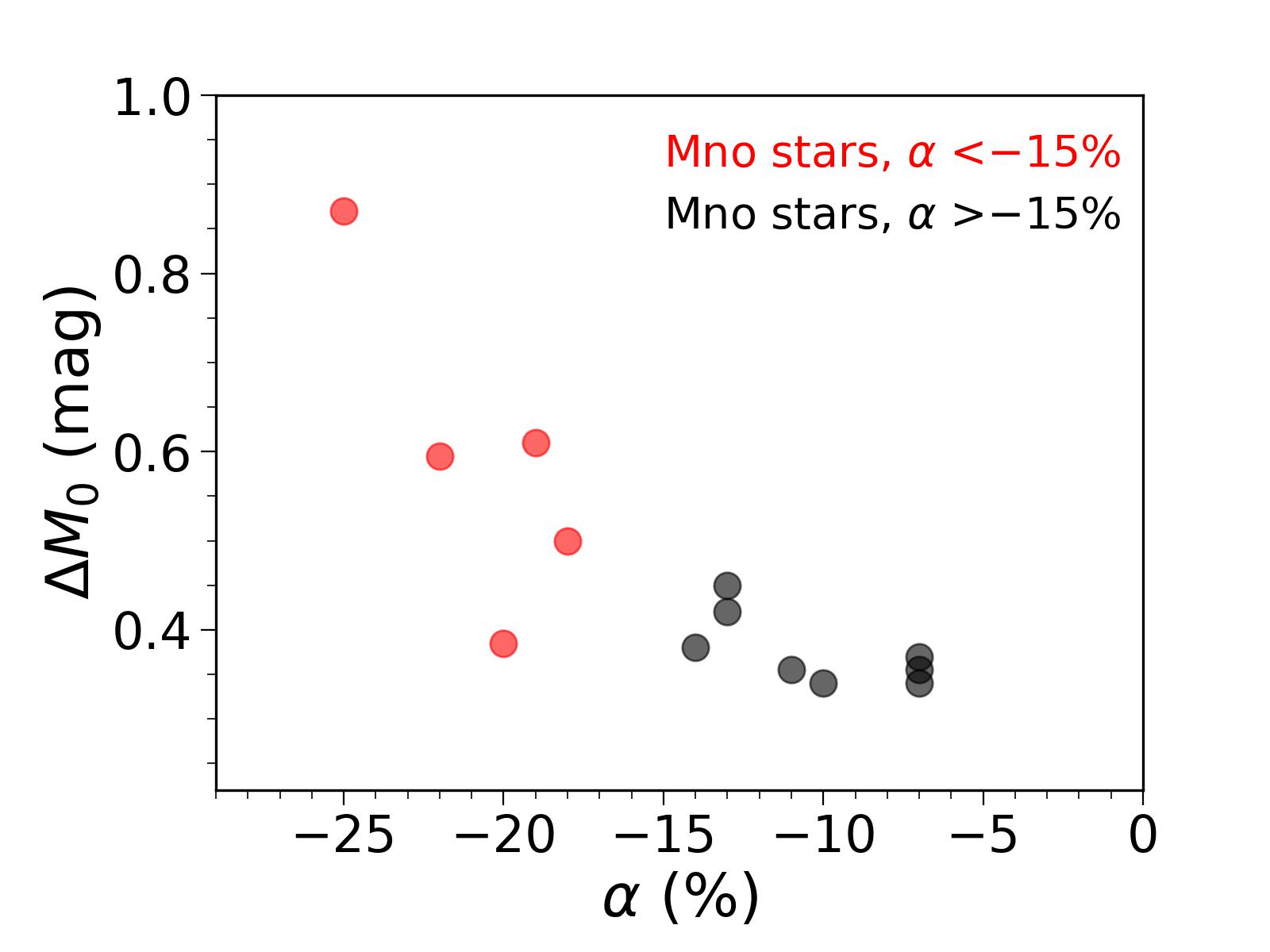}
  \includegraphics[height=5cm,trim=0.cm 0cm 0cm 0cm,clip]{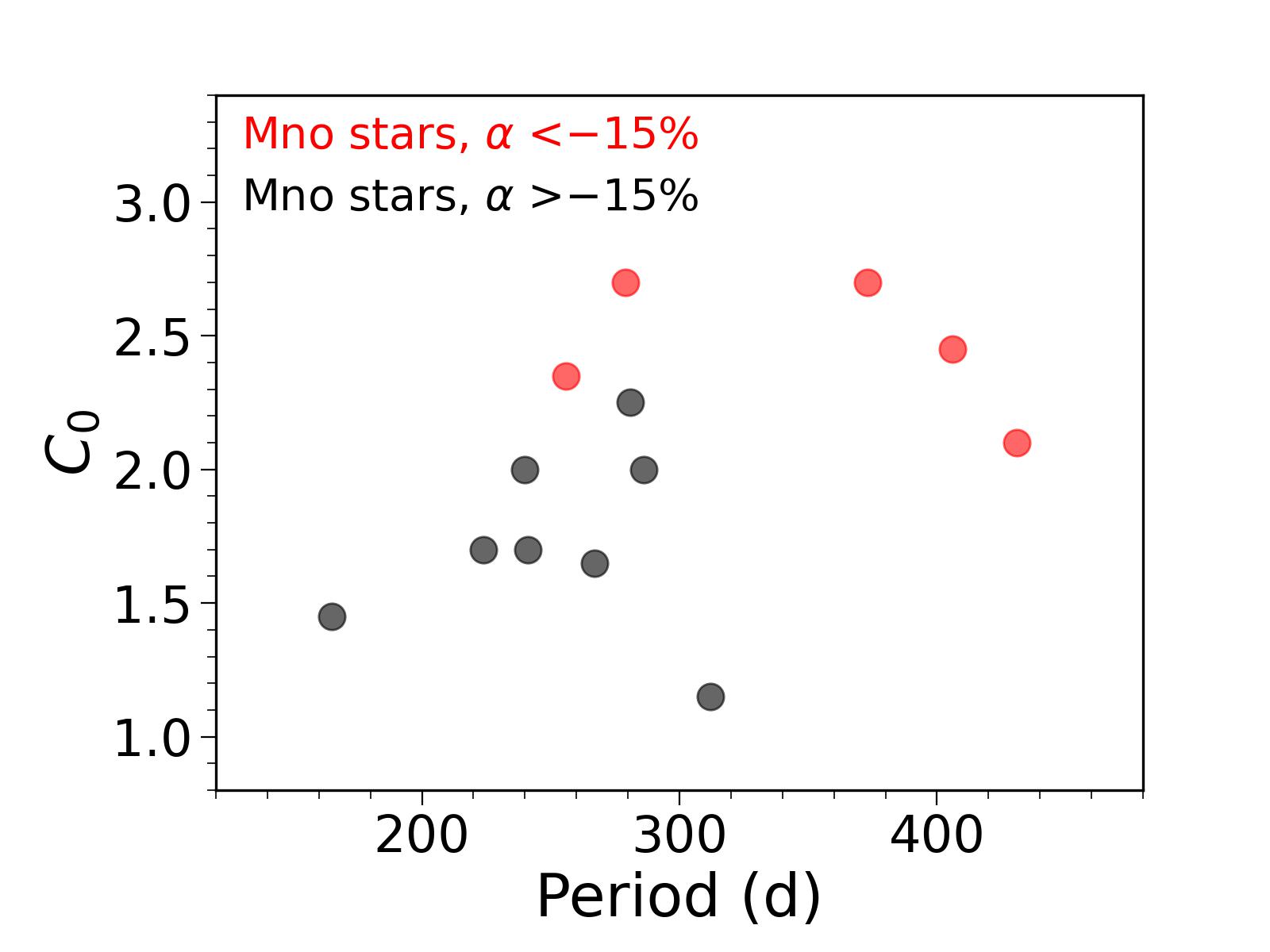}
  \caption{The Mno curves of sample A. Left: dependence of $\Delta M_0$ on $\alpha$. Right: dependence of $C_0$ on $P$. The red dots are for the second family. }
 \label{fig7}
\end{figure*}

In strong contrast with the other curves of sample A, which all display technetium in their spectrum when it has been searched for, the Mno spectral type family is expected to include stars that have recently entered the TP-AGB or, possibly, that are still in the E-AGB and will soon do so. As remarked by \citet{Uttenthaler2019}, it may include TP-AGB stars for which the absence of technetium in their spectrum would be caused by the thermal pulses being too weak to be followed by significant third-dredge up events: only the stronger thermal pulses on the upper AGB are expected to be followed by such third-dredge-up events. It is therefore interesting to learn whether an inspection of the shapes of the light curves may help with differentiating between stars on the E-AGB and stars on the TP-AGB. Also in strong contrast with the other curves of sample A, the Mno curves are exclusively made of \edit2{a} and \edit2{b} profile types. Yet, while having such strong identity, the Mno curves of sample A seem to host two different families of curves, as illustrated in Figure \ref{fig7}, which displays their distribution in the $\Delta M_0$ vs $\alpha$ and colour index, $C_0$=$\frac{1}{2}$(K+[3.4])$-$[22], vs period planes. 

A first nearly symmetric family, defined as having $\alpha$$>$$-$15\% ($<$$\alpha$$>$=$-10\pm3$\%), is relatively regular ($<$$\Delta M_{\rm max}$$>$=0.34$\pm$0.04, $<$$\Delta'M_{\rm max}$$>$=0.41$\pm$0.04), has small colour indices ($<$K$-$[22]$>$=1.8$\pm$0.4, $<$[3.4]$-$[22]$>$=1.7$\pm$0.3) and relatively short periods ($<$$P$$>$=252$\pm$42 days). The complementary family is of course asymmetric ($<$$\alpha$$>$=$-21\pm2$\%), less regular ($<$$\Delta M_{\rm max}$$>$=0.53$\pm$13, $<$$\Delta'M_{\rm max}$$>$=0.65$\pm$20) and has larger colour indices ($<$K$-$[22]$>$=2.6$\pm$0.3, $<$[3.4]$-$[22]$>$=2.3$\pm$0.2) and larger periods ($<$$P$$>$=349$\pm$55 days).  The first family contains eight curves, generally of profile type \edit2{a}. The only exception is R Leo, of profile type \edit2{b}, with the largest period of this first family (312 days) and accordingly a positive value of $K_{\rm MB}$, in contrast with the other curves of the first family, all having negative $K_{\rm MB}$ values.  The second family contains five curves, of which three are of profile type \edit2{b} and two, W Dra and T UMa, are of profile type \edit2{a}. While the cases of R Leo and W Dra may be considered as marginal, a higher density of observations being necessary for stating with strong confidence the validity of the curve parameters, the case of T UMa deserves special attention. Its curve displays high quality and density of observations, making it belong to sample B. It has a low value of $\alpha$ ($-$20\%), or equivalently a high value of $\varphi_{\rm min}$ (0.58), characteristic of the second family. But it has a low value of the period (256 days) characteristic of the first family. As all other parameters take values intermediate between those of the first and second family, it suggests that T UMa is in a state intermediate between the first and second family. If Mira variables would typically evolve from the first to second family, T UMa would then be currently doing so. Among the five other Mno-type curves selected in sample B, three are in the first family, R Aql, R Boo and R Tri, with $K_{\rm MB}$ values of $-$0.6, $-$0.5 and $-$0.2, respectively; the other two, R Cas and U Her, are in the second family with $K_{\rm MB}$ values of +1.0 and +0.7, respectively.

It is important at this point to remark that there is no reason to assume that stars evolve from one family to the other. At an early stage of the TP-AGB, or late stage of the E-AGB, we may expect the selected stars to be essentially characterised by their initial mass and their age on the AGB: being in general hosted in the solar neighbourhood, they are likely to have metallicity close to solar, $\sim$0.015. Such stars, if having a mass similar to or lower than the solar mass, are expected to spend a very long time in the Mno family and never reach the Myes, S and C spectral types \citep{Straniero2023, MillerBertolami2016}. Indeed, R Aql, a clear member of the first Mno family, is often quoted in the published literature as having recently experienced a thermal pulse \citep{Wood1981}. The Mno family must therefore host both such stars and stars that will later evolve to the Myes, S and C spectral types. A major question that needs therefore to be answered is whether, and if yes how, these two types of stars are related to the two types of Mno families that we have identified. Moreover, all stars are expected to transition from the E-AGB to the TP-AGB when they have exhausted the helium-burning layer around the inert core. Another major question that needs therefore to be answered is whether, and if yes how, the distinction between E-AGB and TP-AGB is related to the two types of Mno families that we have identified.  It is essential to keep both questions in mind when discussing the results obtained in the present work.

As clearly illustrated in the pair of leftmost panels of Figure \ref{fig1}, the main difference between ``M stars with no bump'' and ``M stars with bump'', as defined by \citet{MerchanBenitez2023}, is a difference of period. For stars evolving across the boundary, one needs to understand what causes the significant period increase associated with this transition. Our results, while broadly confirming those obtained by \citet{MerchanBenitez2023} offer a more detailed picture and imply that the line of equation K$-$[22]$\sim$0.011($P-125$) shown in Figure \ref{fig1} is in fact related to the distinction between the first family of Mno-type curves, having $K_{\rm MB}$ negative, and the other curves, the observed continuity between the two suggesting that stars may evolve from one family to the other. 

\subsection{Myes and S spectral types} \label{sec4.2}
\begin{figure*}
  \centering
  \includegraphics[height=5cm,trim=0.cm 0cm 0cm 0cm,clip]{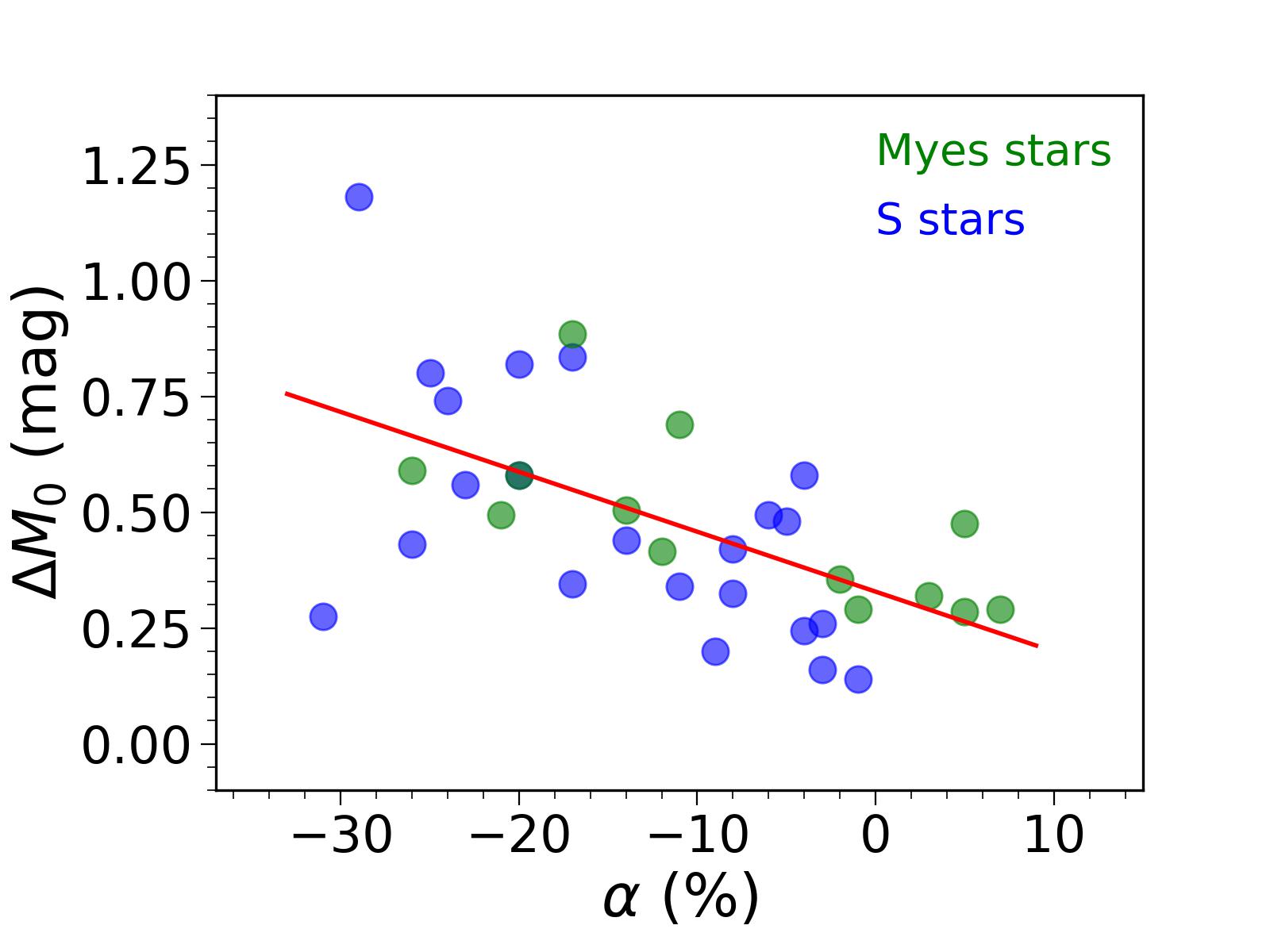}
  \includegraphics[height=5cm,trim=0.cm 0cm 0cm 0cm,clip]{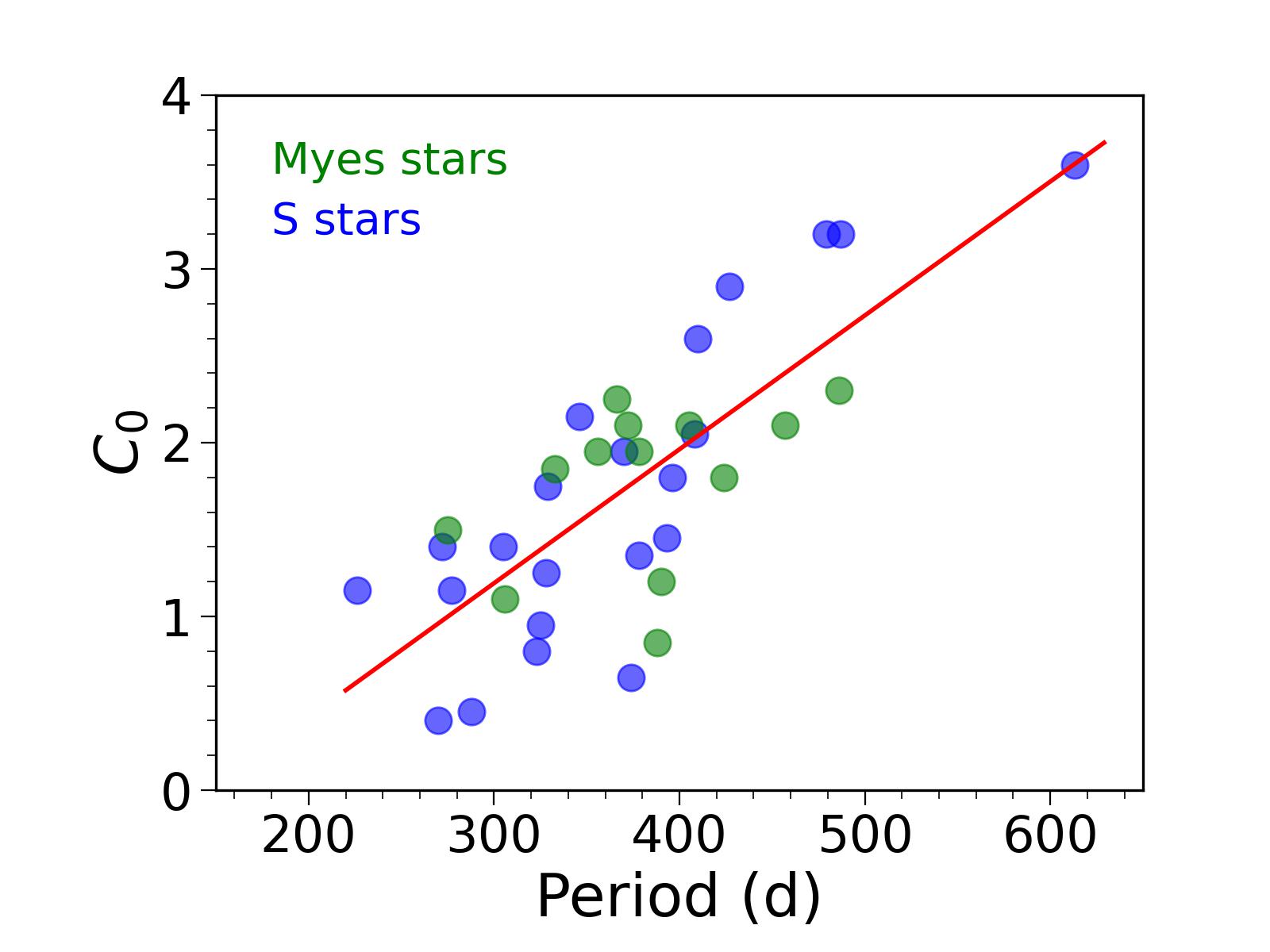}
  \caption{The Myes and S curves of sample A. Left: dependence of $\Delta M_0$ on $\alpha$. Right: dependence of $C_0$ on $P$. The green and blue dots are for Myes and S spectral types, respectively. The lines show linear best fits of equations $\Delta M_0$=$0.33-0.013\alpha$ and $C_0$=$-1.12+0.0077P$, respectively.}
 \label{fig8}
\end{figure*}

Figure \ref{fig8} displays the distribution in the irregularity $\Delta M_0$ vs asymmetry $\alpha$ and colour index $C_0$ vs period $P$ planes of the curves of sample A having an Myes or S spectral type. All have positive $K_{\rm MB}$ values, in agreement with the assignments made by \citet{MerchanBenitez2023}. Clear correlations are observed between the curve parameters. The more symmetric curves are also more regular and have smaller colour indices than the more asymmetric curves, which are irregular and have larger colour indices. Moreover, the more symmetric curves are exclusively of \edit2{c} and \edit2{d} profile types, while the more asymmetric are exclusively of the \edit2{a} and \edit2{b} profile types.

\begin{figure*}
  \centering
  \includegraphics[height=5cm,trim=0.cm 0cm 0cm 0cm,clip]{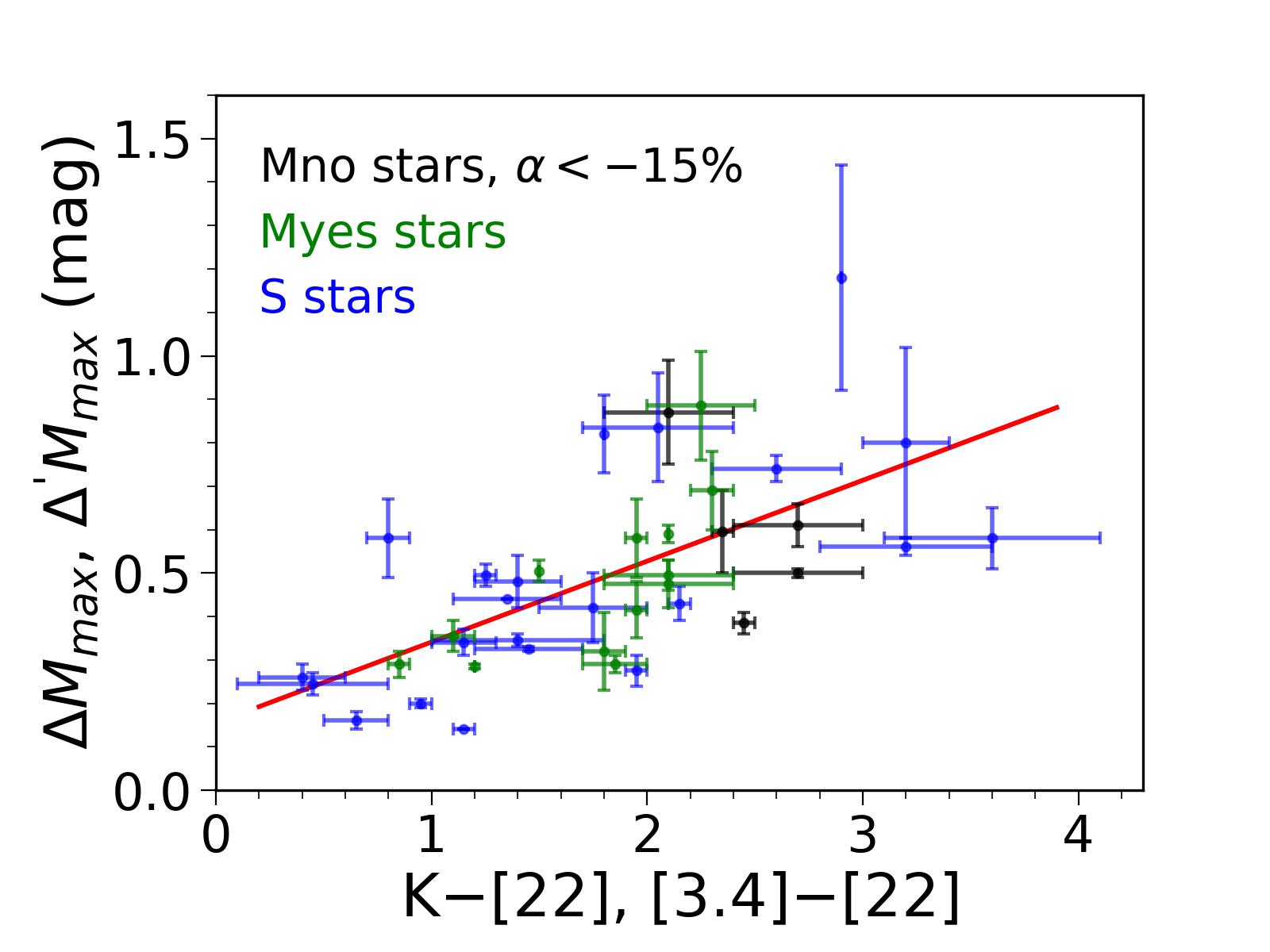}
  \includegraphics[height=5cm,trim=0.cm 0cm 0cm 0cm,clip]{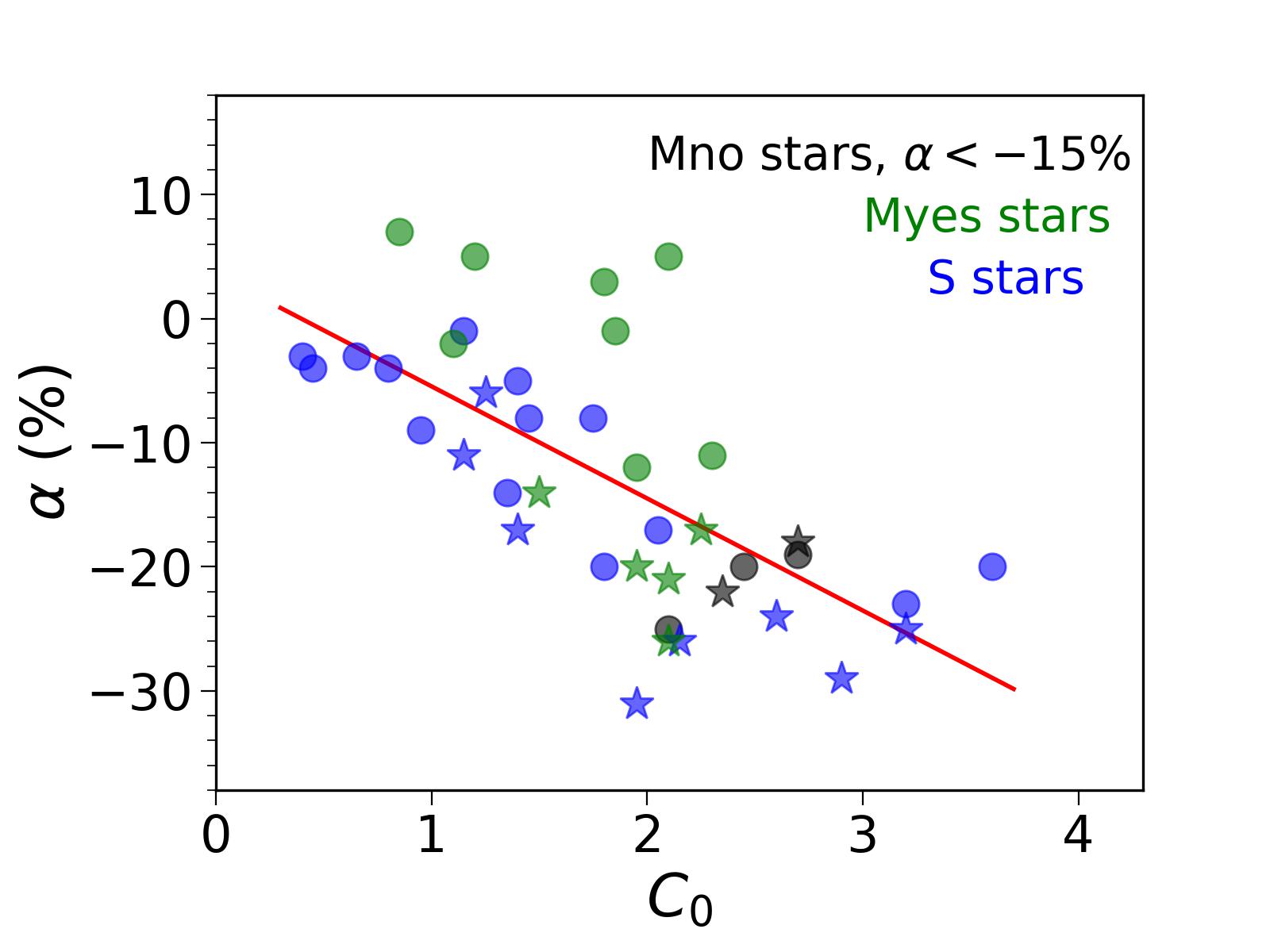}
  \caption{Curves of sample A excluding the first Mno family. Left: dependence of the regularity on colour. For each curve, we plot as ordinate a bar joining the values of $\Delta M_{\rm max}$ to $\Delta'M_{\rm max}$ and as abscissa a bar joining the values of K$-$[22] and [3.4]$-$[22]. Right: dependence of the asymmetry $\alpha$ (percent) on the mean colour parameter, $C_0$. Stars are for \edit2{a} profile types. On both panels, colours correspond to spectral types as indicated in the insert and the lines are the results of linear best fits with equation $\Delta M_0$=0.155+0.186$\times$$C_0$ on the left panel and $\alpha$(\%)=3.57–9.03$\times$$C_0$ on the right panel.}
 \label{fig9}
\end{figure*}

This encourages having a global look at all sample A curves of oxygen-rich star, including the second family of Mno spectral types but excluding the first family. This is done in Figure \ref{fig9}, the left panel of which displays the dependence of the regularity parameters, $\Delta M_{\rm max}$ and $\Delta'M_{\rm max}$, on the colour parameters, K$-$[22] and [3.4]$-$[22]; the right panel displays the dependence of the asymmetry parameter $\alpha$ on the mean colour parameter. They give clear evidence for significant correlations: more symmetric curves have smaller colour indices, probably meaning smaller dust mass-loss rate, and are more regular; more asymmetric curves, with an ascending branch faster than the descending branch, have larger colour indices and are more irregular. However, while Mno-type curves, associated with stars that just entered the TP-AGB, are clustered in the latter group, Myes- and S-type curves populate the whole range of parameters. Moreover, the Mno-type curves are of profile types \edit2{a} or \edit2{b}, while Myes- and S-type curves have profile types spanning from \edit2{a} to \edit2{d}. When having an \edit2{a} or \edit2{b} profile type, they tend to be on the asymmetric and irregular side, when having a \edit2{c} or \edit2{d} profile type, they tend to be on the symmetric and regular side. For stars evolving along the AGB from Mno to Myes and S types, this suggests that some of their light curves evolve all the way to a state of high symmetry and regularity with a low dust mass-loss rate while others span only a smaller path along such an evolution. In order to better understand this evolution, we study in the next sub-section how the hump, when present, evolves along the ascending branch.

\subsection{The hump} \label{sec4.3}
The evolution of type profiles for oxygen-rich stars of sample A is well illustrated by the mean values taken, for each of them separately, by the asymmetry parameter $\alpha$, the mean colour index $C_0$, and the mean regularity parameter, $\Delta M_0$. They are, respectively, for the \edit2{a} profile type, ($-$20\%, 2.1, 0.58), for the \edit2{b} profile type, ($-$19\%, 2.5, 0.59), for the \edit2{c} profile type, ($-$3\%, 1.4, 0.45) and for the \edit2{d} profile type, ($-$4\%, 1.2, 0.30): the \edit2{a} and \edit2{b} profiles are more asymmetric, more dusty and more irregular; the \edit2{c} and \edit2{d} profiles are more symmetric, less dusty and more regular. In order to obtain a more precise evaluation of this evolution, we use the curves of sample A that have been selected in sample B to study the evolution of the hump on the ascending branch.

\begin{figure*}
  \centering
  \includegraphics[width=2.59cm,trim=0.cm -9cm 0cm 0cm,clip]{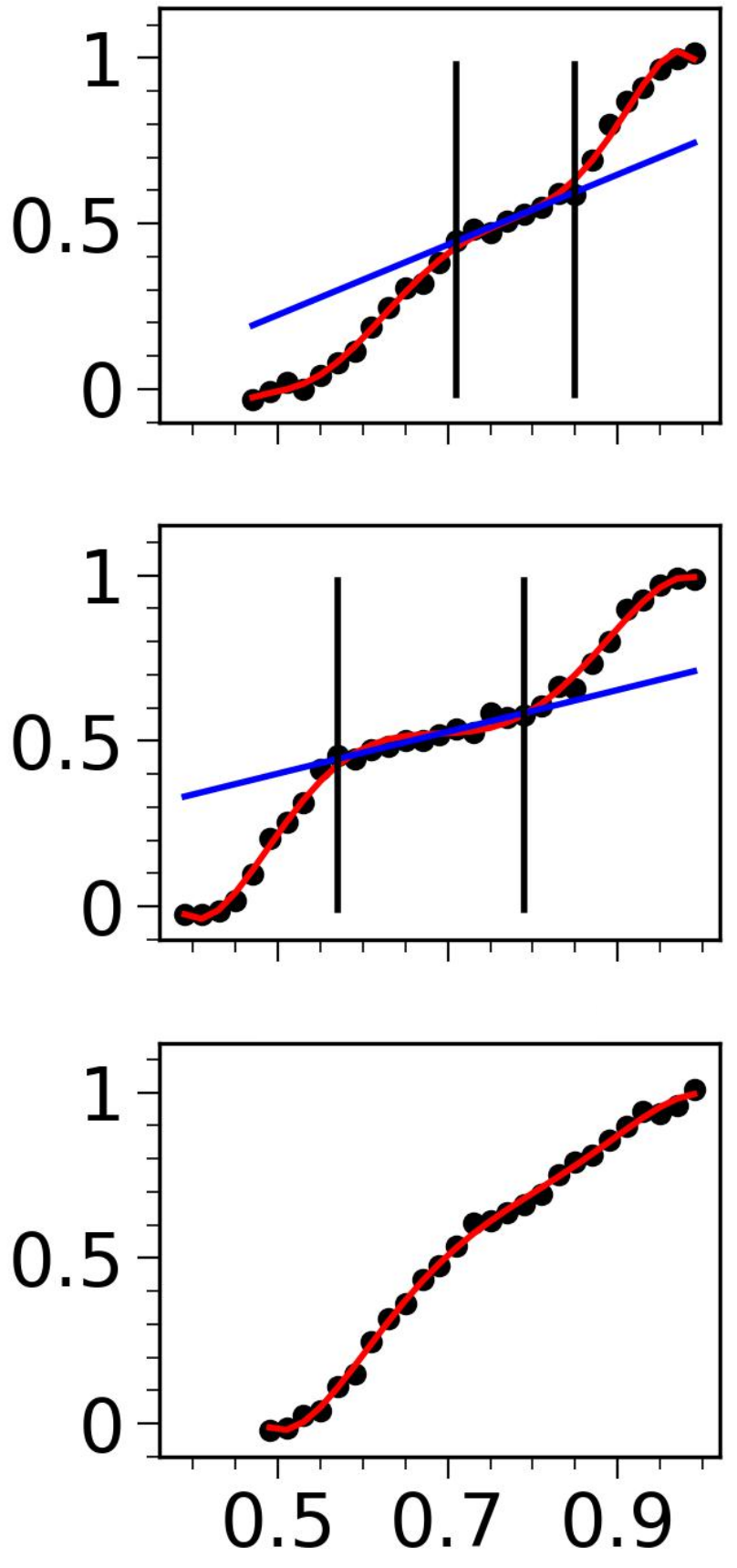}
  \includegraphics[height=7cm,trim=0.cm 0cm 1.6cm 0cm,clip]{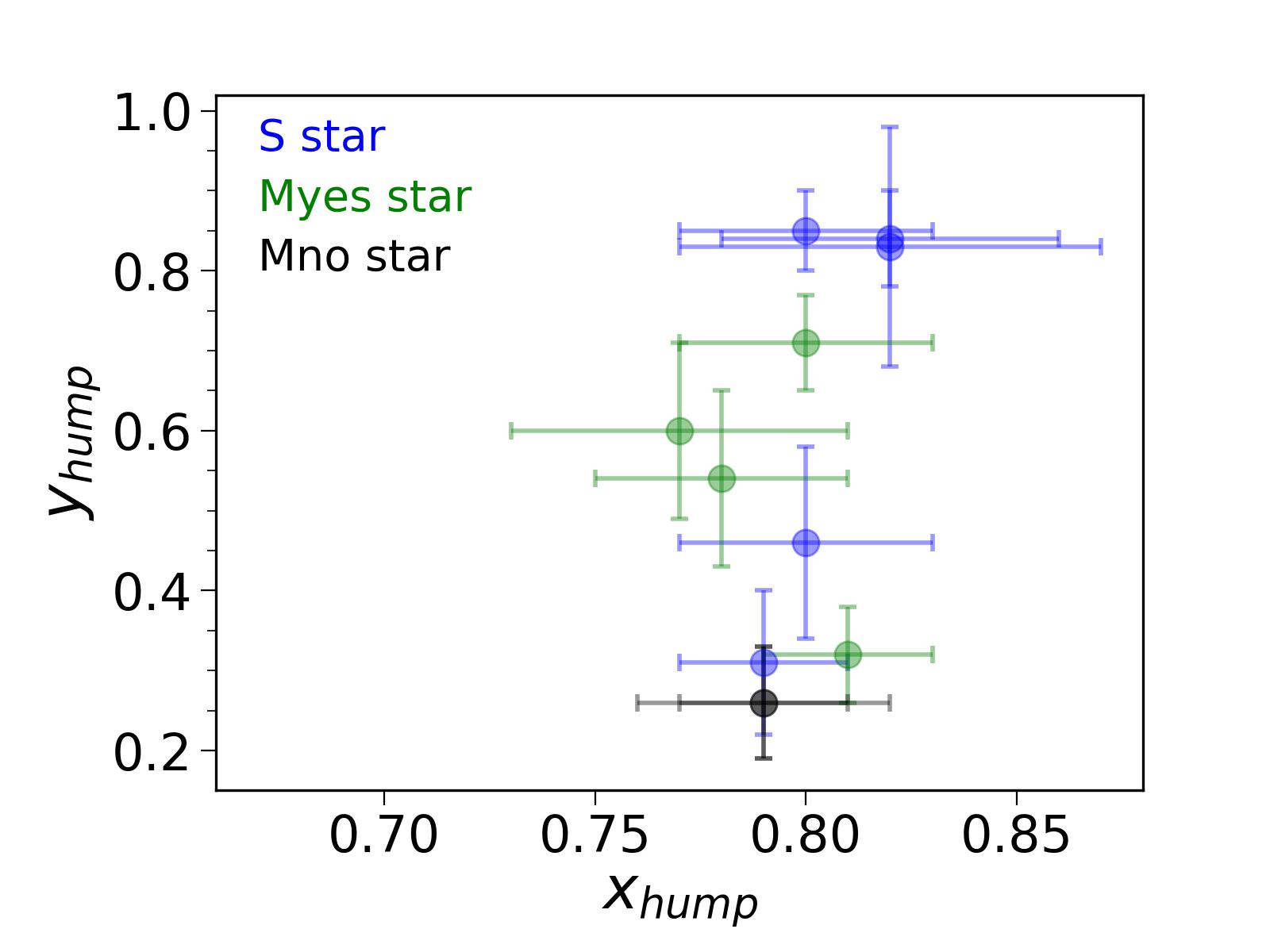}
  \includegraphics[height=7cm,trim=0.cm 0cm 0cm 0cm,clip]{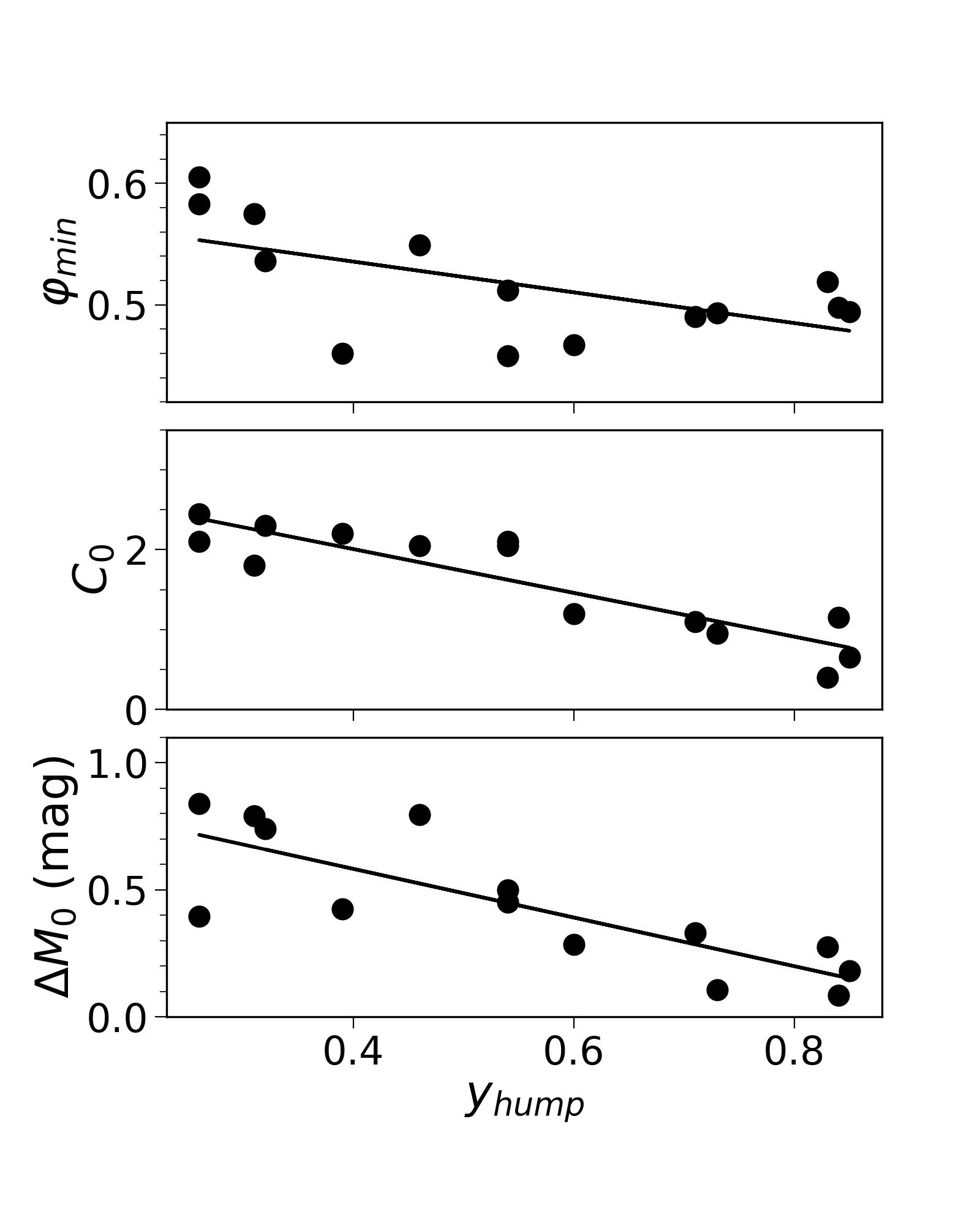}
  \caption{Left panel: illustration of the procedure used to define the hump parameters. The red curve shows the sixth-degree polynomial fit. The blue line shows the linear fit on the hump and the black lines show the hump interval. The lower panel was excluded from the sample of humpy oscillations as giving insufficient evidence for a hump. Central panel: distribution of the mean hump coordinates, $<$$y_{\rm hump}$$>$ vs $<$$x_{\rm hump}$$>$, for the selected curves of sample B listed in Table \ref{tab3}. The error bars show the rms deviations around the mean. Right panel: dependence on $y_{\rm hump}$ of the parameters $C_0$, $\varphi_{\rm min}$ and $\Delta M_0$ for the curves listed in Table \ref{tab3}. }
 \label{fig10}
\end{figure*}

We select a sample of light curves displaying a clear hump in many of their oscillations and, for each selected oscillation, fit a sixth-degree polynomial to the normalised profile of the ascending branch. The hump interval, [$x_{\rm hump1}$, $x_{\rm hump2}$], is then defined as giving the best linear fit in it, the normalised magnitude spanning accordingly from $y_{\rm hump1}$ to $y_{\rm hump2}$. The left panel of Figure \ref{fig10} illustrates the procedure. The result is summarised in Table \ref{tab3} and illustrated in the central panel of Figure \ref{fig10}. Defining $\Delta x_{\rm hump}$ as $|x_{\rm hump2}-x_{\rm hump1}|$ and $x_{\rm hump}$ as $\frac{1}{2}$($x_{\rm hump1}+x_{\rm hump2}$) and, similarly, $\Delta y_{\rm hump}$ as $|y_{\rm hump2}-y_{\rm hump1}|$ and $y_{\rm hump}$ as $\frac{1}{2}$($y_{\rm hump1}+y_{\rm hump2}$), we find that the mean values of $x_{\rm hump}$ are poor discriminants between different curves, the rms deviations of $x_{\rm hump}$ with respect to the mean taking values as large as the differences between the mean values of different curves. In contrast, the central panel of Figure \ref{fig10} shows that $y_{\rm hump}$ has rms dispersions around its mean significantly smaller than the differences between the mean values of different curves and its mean value is seen to span a broad range along the ascending branch. Typically, the rms dispersion of $y_{\rm hump}$ is of the order of a tenth, and $\Delta y_{\rm hump}$ is of the order of 14\% of the oscillation amplitude. The right panel of Figure \ref{fig10} displays the dependence on $y_{\rm hump}$ of three parameters: the mean colour index, $C_0$, the phase at minimal light, $\varphi_{\rm min}$ and the mean irregularity, $\Delta M_0$.

\begin{figure*}
  \centering
  \includegraphics[height=4cm,trim=0.cm 0cm 0cm 0cm,clip]{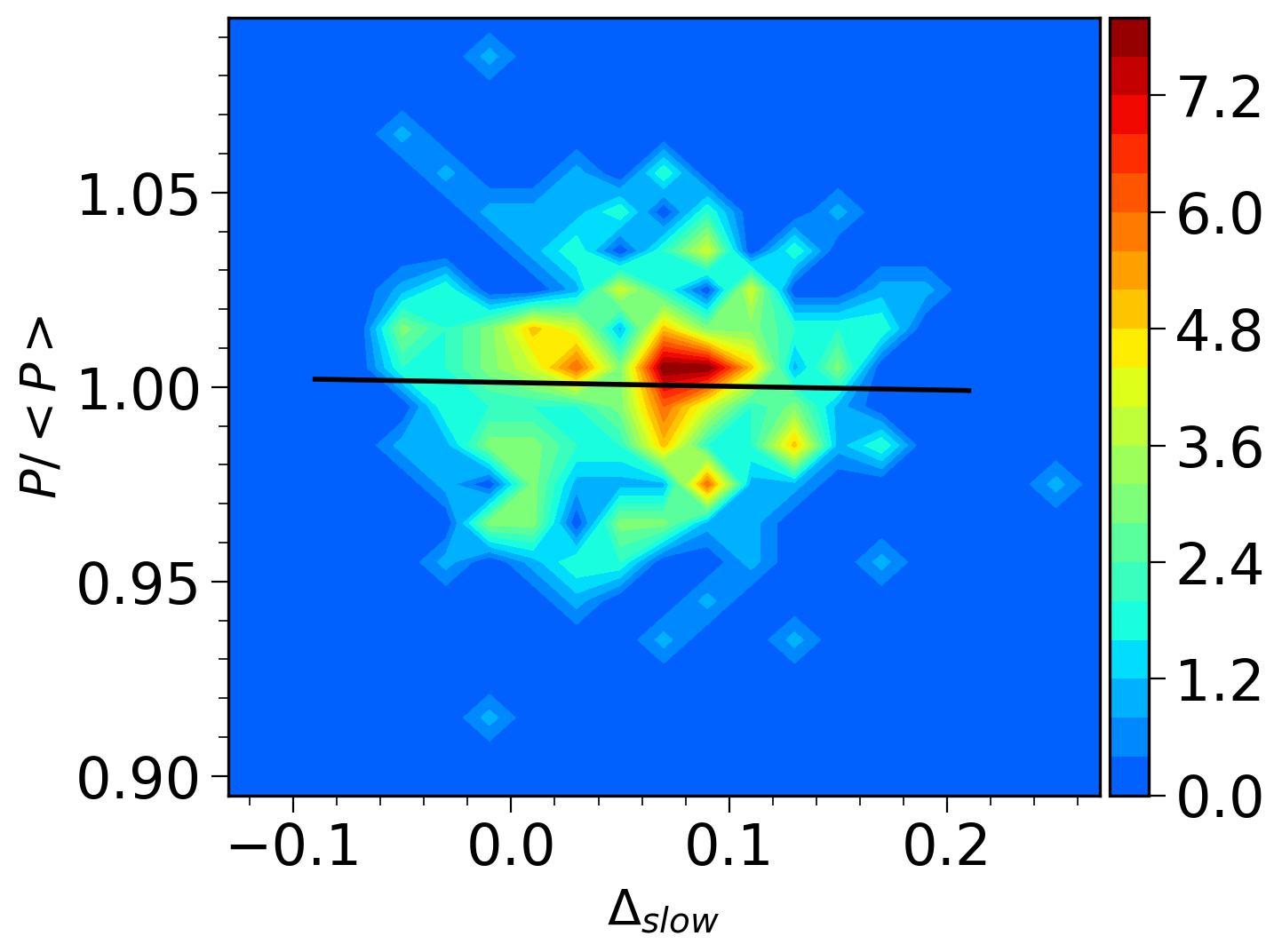}
  \includegraphics[height=4cm,trim=0.cm 0cm 0cm 0cm,clip]{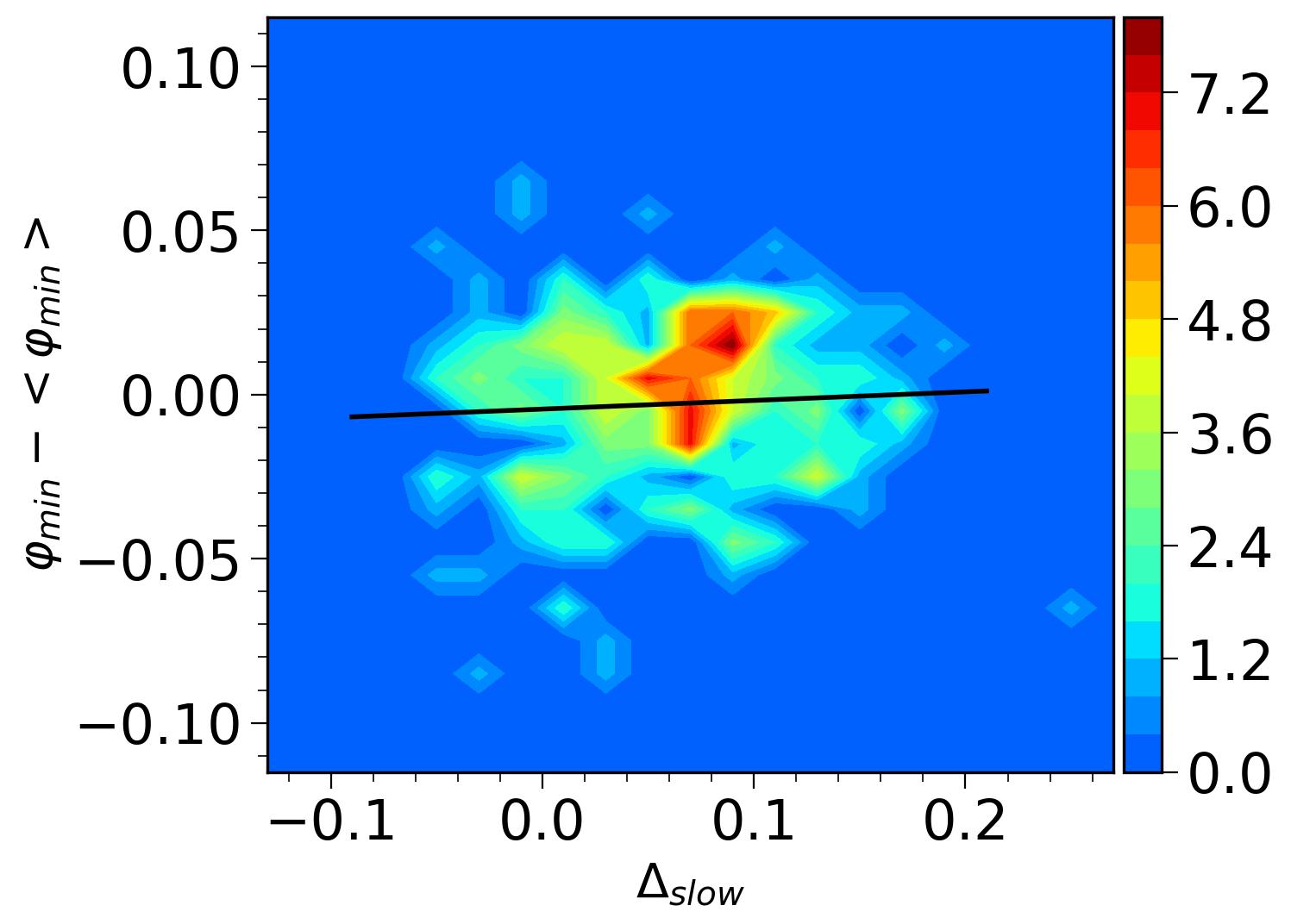}
  \includegraphics[height=4cm,trim=0.cm 0cm 0cm 0cm,clip]{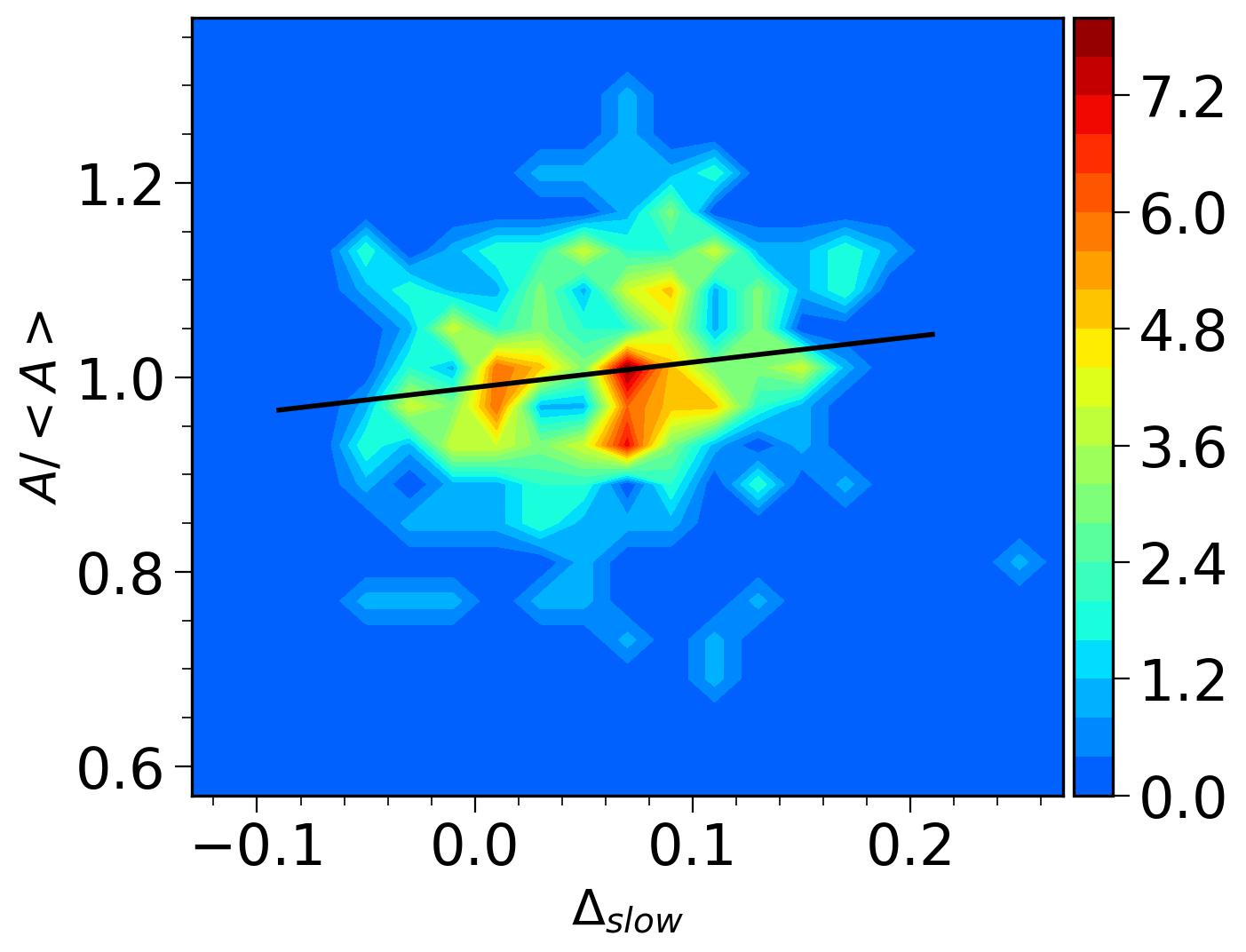}
  \caption{ Dependence on the delay caused by the hump, $\Delta_{\rm slow}$=$\Delta x_{\rm hump}-0.88(1-\varphi_{\rm min})\Delta y_{\rm hump}$, of the relative period (left), the phase shift at minimal light $\varphi_{\rm min}$$–$$<$$\varphi_{\rm min})$$>$ (centre) and the relative amplitude of the preceding descending branch (right) for eight light curves of sample B displaying a clear hump (0.3$<$$y_{\rm hump}$$<$0.7).}
 \label{fig11}
\end{figure*}

One may wonder whether the presence of a hump, during which the rate of increase of the luminosity is smaller than outside it, implies a lengthening of the period, the resulting slowing-down not being compensated by speeding-up either before or after it. To answer this question, we select 8 curves displaying a clear hump and plot, for each oscillation, the relative period, $P$/$<$$P$$>$, as a function of the delay $\Delta_{\rm slow}$ caused by the hump; this delay is the difference between $\Delta x_{\rm hump}$ and the time spent to climb $\Delta y_{\rm hump}$ in the absence of hump, namely $k(1–\varphi_{\rm min})\Delta y_{\rm hump}$; here, $k$ accounts for the faster rate of luminosity increase at the time of the hump compared with a linear increase ($k$=1). We evaluated $k$ separately for each of the 8 selected curves and obtained $k$=0.88$\pm$0.04. We checked carefully the validity of the procedure. The result is illustrated in the left panel of Figure \ref{fig11} and shows the absence of significant correlation: [d$P$/$<$$P$$>$]/d$\Delta_{\rm slow}$$\sim$0.0$\pm$0.2 while a value of 1 is expected in the absence of compensation. This implies that the hump delay is somehow compensated by speeding-up elsewhere. In order to understand whether this speeding-up occurs on the ascending or descending branch, we show in the central panel of the figure the dependence of $\varphi_{\rm min}$$-$$<$$\varphi_{\rm min}$$>$ on $\Delta_{\rm slow}$. The absence of correlation shows clearly that the compensation occurs on the ascending branch: the presence of a hump implies a change of shape, but not of duration of the ascending branch. 

One may also wonder whether the presence of a hump is related to the presence of a shock wave, in which case it might be enhanced by the strength of the preceding oscillation; we check on this by plotting the relative amplitude $A$/$<$$A$$>$ of the descending branch of the preceding oscillation as a function of $\Delta_{\rm slow}$. The result, displayed in the right panel of Figure \ref{fig11}, shows indeed a small, but barely significant correlation of the expected sign, [d$A$/$<$$A$$>$]/d$\Delta_{\rm slow}$$\sim$0.3$\pm$0.1.

\subsection{Recovering light curves that were ignored} \label{sec4.4}
\begin{figure*}
  \centering
  \includegraphics[height=5cm,trim=0.cm 0cm 0cm 0cm,clip]{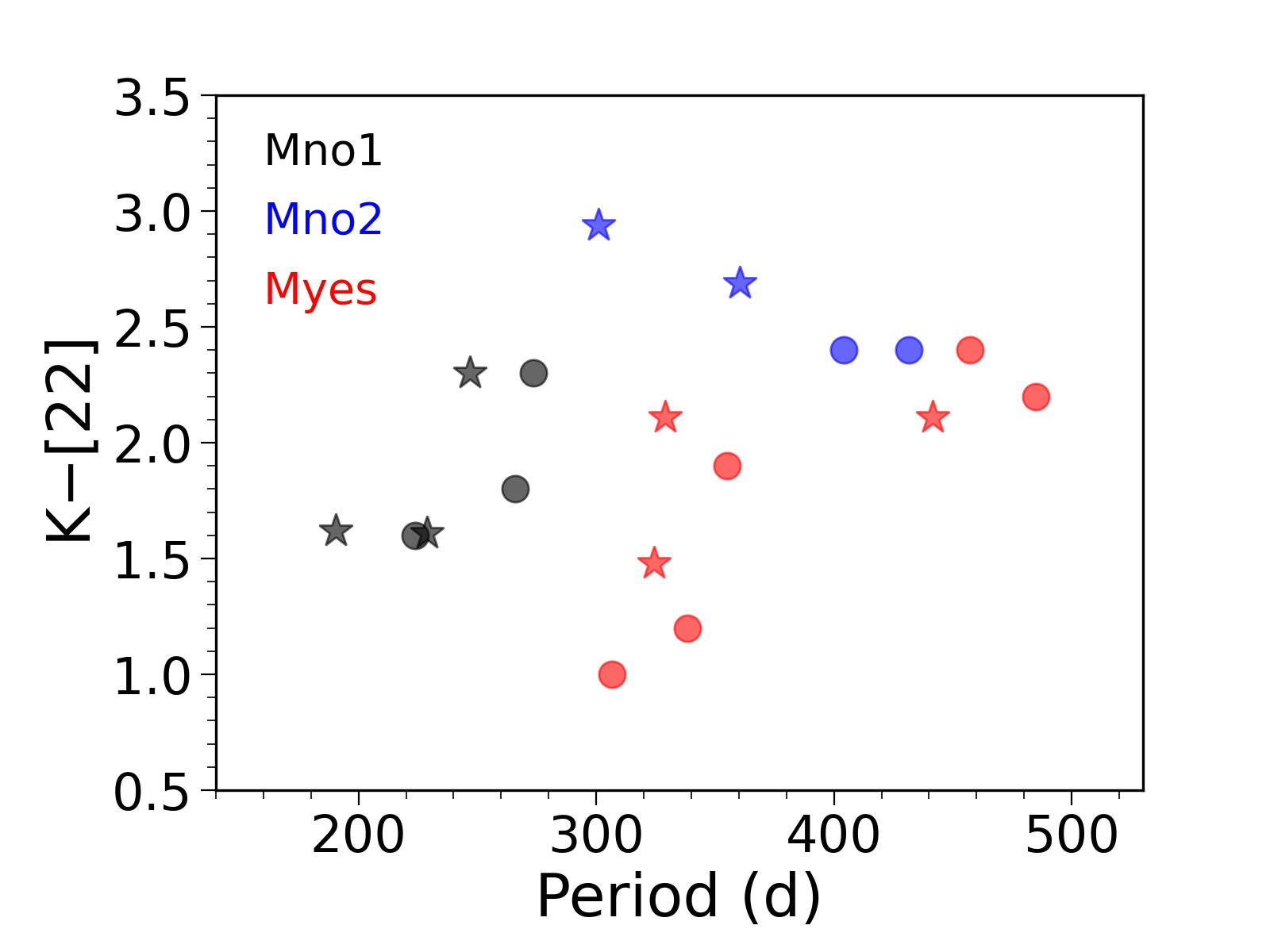}
  \includegraphics[height=5cm,trim=0.cm 0cm 0cm 0cm,clip]{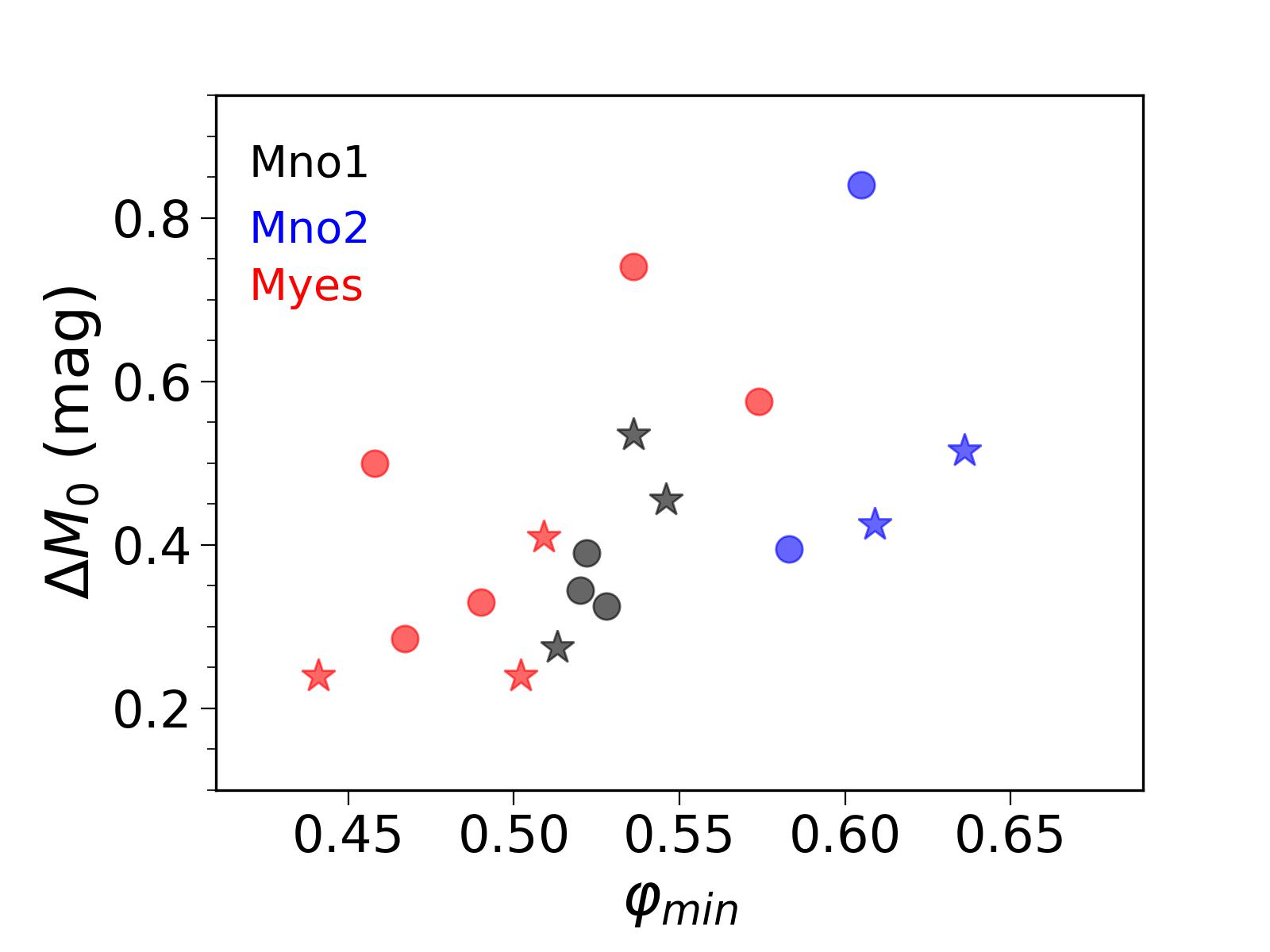}
  \caption{Comparing the M-type curves recovered in Section 4.4 (stars) to those of sample B selected from sample A (full circles). The left panel displays their distributions in the K$-$[22] vs $P$ plane, the right panel in the $\Delta M_0$ vs $\varphi_{\rm min}$ plane. Colours distinguish between different spectral types as indicated in the inserts. Spectral types of the eight curves recovered in Section 4.4 are as assigned in the text: V Cas, RT Cyg and R Dra to the first Mno family, S CrB and R UMa to the second Mno family and U UMi, T Cas and R CVn to Myes. }
 \label{fig12}
\end{figure*}

In the selection of sample A light curves, M-type curves for which technetium was either not searched for or, when searched for, was found neither clearly present nor clearly absent, were deliberately ignored. Having now obtained a clearer view of the general properties displayed by the analysed light curves, we can recover such curves from the Merchán Benitez sample of 548 Mira stars, requiring that they allow for the rigorous parameterisation of the sample B selection.  We select this way eight additional M-type curves: three for which technetium had not been searched for (R UMa, V Cas and U UMi), two for which the presence of technetium had been considered doubtful (R Dra and RT Cyg), two for which it had been considered possible (S CrB and R CVn) and one for which it had been considered probable (T Cas). The curve parameters are listed in Table \ref{tab1} and illustrated in Figure \ref{figA1}. Figure \ref{fig12} below compares the 8 new M-type curves to the M-type curves belonging to both sample A and sample B. It suggests the following assignments for the new curves: V Cas, RT Cyg and R Dra, with profiles of type \edit2{a}, to the first family of Mno spectral type; S CrB and R UMa, with profiles of type \edit2{b}, to the second family of Mno spectral type; U UMi, T Cas and R CVn, with profile types \edit2{c} and \edit2{d}, to the Myes spectral type.

\begin{deluxetable*}{lcc ccc ccc}
\tablenum{3}
\tablecaption{Hump parameters for curves of sample A that have been selected in sample B. The first column lists the star names, column 2 lists the spectral types, column 3 lists the profile types, column 4 lists the number of oscillations covered by the whole curve and column 5 the number of oscillations retained for the evaluation of the hump parameters, column 6 lists the mean values and rms deviations with respect to the mean of $y_{\rm hump}$ in percent, column 7 lists the phase at minimal light $\varphi_{\rm min}$, column 8 lists the mean irregularity parameter $\Delta M_0$ and column 9 the mean colour parameter $C_0$. \label{tab3}}
\colnumbers
\tablehead{
\colhead{Name}&\colhead{Type}&\colhead{Prof}&\colhead{$N_{\rm osc}$}&\colhead{$N_{\rm hump}$}&
\colhead{$y_{\rm hump}$}&\colhead{$\varphi_{\rm min}$}&\colhead{$\Delta M_0$}&\colhead{$C_0$} }
\startdata
U Her	&	Mno	&	\edit2{b}	&	34	&	33	&	26/7	&	0.58	&	0.4	&	2.5	\\
R Cas	&	Mno	&	\edit2{b}	&	33	&	32	&	26/7	&	0.61	&	0.84	&	2.1	\\
S Her	&	Myes	&	\edit2{d}	&	44	&	34	&	71/6	&	0.49	&	0.33	&	1.1	\\
RU Her	&	Myes	&	\edit2{b}	&	29	&	29	&	32/6	&	0.54	&	0.74	&	2.3	\\
R Aur	&	Myes	&	\edit2{d}	&	30	&	30	&	54/11	&	0.46	&	0.50	&	2.1	\\
T Cep	&	Myes	&	\edit2{c}	&	37	&	32	&	60/11	&	0.47	&	0.29	&	1.2	\\
S Uma	&	S	&	\edit2{d}	&	64	&	54	&	84/6	&	0.50	&	0.09	&	1.2	\\
Chi Cyg	&	S	&	\edit2{c}	&	35	&	28	&	46/12	&	0.55	&	0.80	&	2.1	\\
T Cam	&	S	&	\edit2{d}	&	37	&	27	&	85/5	&	0.49	&	0.18	&	0.7	\\
R Cam	&	S	&	\edit2{d}	&	52	&	42	&	83/15	&	0.52	&	0.28	&	0.4	\\
W And	&	S	&	\edit2{b}	&	30	&	23	&	31/9	&	0.58	&	0.79	&	1.8	\\
U Cyg	&	C	&	\edit2{c}	&	29	&	23	&	54/16	&	0.51	&	0.45	&	2.1	\\
S Cep	&	C	&	\edit2{c}	&	30	&	30	&	39/14	&	0.46	&	0.43	&	2.2	\\
W Cas	&	C	&	\edit2{d/c}	&	33	&	32	&	73/10	&	0.49	&	0.11	&	1.0	\\
\enddata
\end{deluxetable*}

\subsection{Other wavelengths} \label{sec4.5}
\begin{figure*}
  \centering
  \includegraphics[height=6.5cm,trim=0.cm 0cm 0cm 0cm,clip]{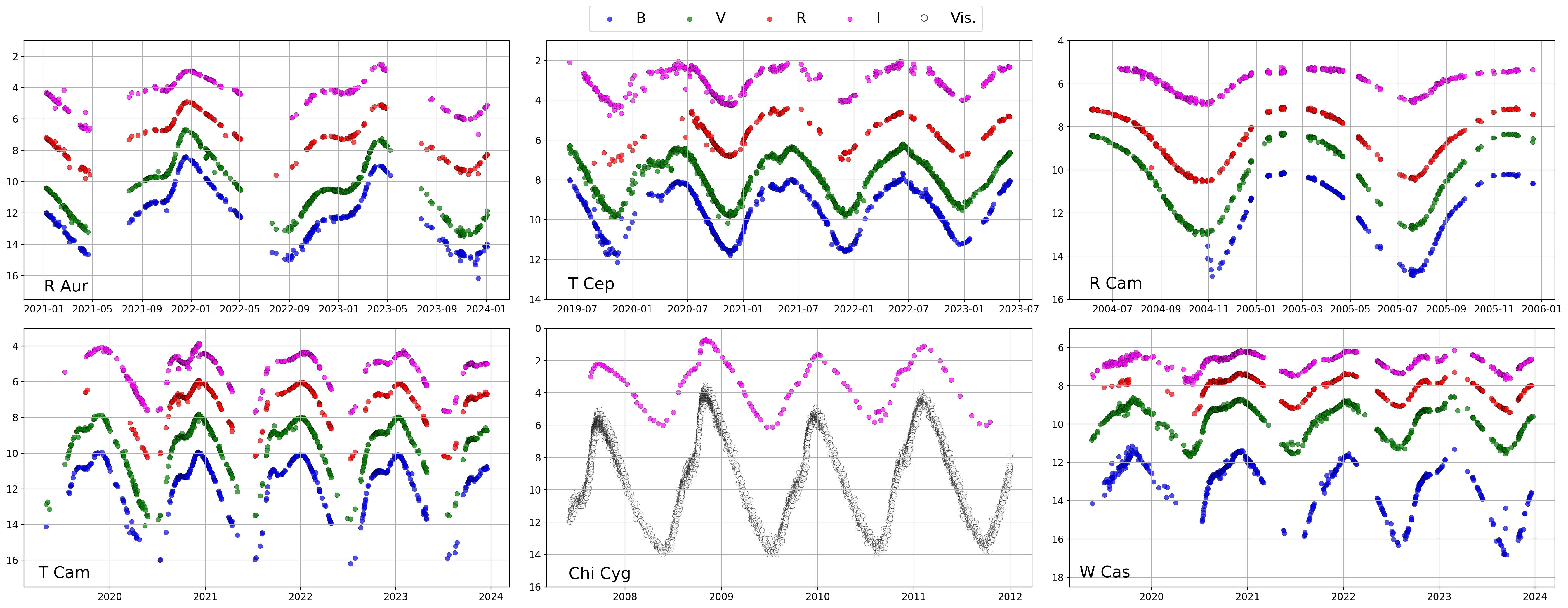}
  \caption{Intervals of the selected light curves (AAVSO data base) displaying significant observations in the B (blue), V (green), R (red) and I (magenta) band-passes. From left to right and up down, R Aur, T Cep, R Cam, T Cam, chi Cyg, W Cas.}
 \label{fig13}
\end{figure*}

It would, of course, be important to extend the studies presented here to bandpasses other than visible, but this is a major effort beyond the scope of the present article. Unfortunately, light curves at wavelengths outside the visible bandpass are scarce and include much smaller sets of observations. Figure \ref{fig13} displays the better examples of such observations for the curves of sample B. It shows, in addition to the visible, B, V and R, the Johnson infrared bandpass having a mean and full width at half maximum of 806 nm and 149 nm, respectively. The main observation is that when a hump is present (absent) on the visible curve, it is also present (absent) on the infrared curve but its detailed shape may be different. This observation is consistent with the generally accepted belief that the presence of a hump is related to global properties of the star pulsation but its detailed shape is affected by the nature of the outer atmosphere. Over 50 years ago, \citet{Lockwood1971} studied the near infrared ($\lambda$=1.04 $\mu$m) light curves of 24 Mira variables, 15 of which are part of our sample A, and measured the ratio $R_{\rm I/V}$ of their oscillation amplitude to that in the visible bandpass. Its mean value is 20\%. It is remarkable that the Mno curves of sample B are split in two families as clearly in terms of $R_{\rm I/V}$ as they are in terms of $\varphi_{\rm min}$ and $K_{\rm MB}$: R Cas and U Her have ($R_{\rm I/V}$, $\varphi_{\rm min}$, $K_{\rm MB}$)=(29\%, 0.61, 1.0) and (27\%, 0.58, 0.7), respectively, while R Aql and R Boo have ($R_{\rm I/V}$, $\varphi_{\rm min}$, $K_{\rm MB}$)= (17\%, 0.53, $-$0.6) and (15\%, 0.52, $-$0.5), respectively.

\section{Carbon stars} \label{sec5}
The 22 light curves of sample A carbon stars listed in Table \ref{tab5} display very different features from the curves of oxygen-rich stars studied in the preceding section, the most spectacular difference being their much smaller oscillation amplitude, 2.7 mag on average compared with 5.5 mag on average for the curves of other spectral types. For many of these, the assignment as a Mira star is marginal; they are close to semi-regular variables. What makes carbon stars different from oxygen-rich stars is not the mechanism at stake inside the star, both of them burning usually hydrogen in the hydrogen-rich layer with short interruptions during which they burn helium in the helium-rich layer; this happens when the phase of hydrogen burning has produced sufficient helium ashes; it starts with violent and brief thermal pulses. What makes the difference is the composition of the star atmosphere, with dust giving it a red and sooty appearance, causing a mass-loss rate typically one order of magnitude larger than for oxygen-rich stars, a few $10^{-6}$ rather than a few $10^{-7}$ solar masses per year. This shortens the time left for the star to survive by making the super-wind episode come sooner. A consequence of such differences is that the evaluation of shape parameters for carbon stars is more difficult and less reliable than it is for oxygen-rich stars. Recently, significant progress has been made in both observation and modelling of the relevant mixing mechanism responsible for the S to C spectral type transition, both of which are very challenging and still suffer of several unanswered questions \citep{Abia2022, Straniero2023}.

\begin{figure*}
  \centering
  \includegraphics[height=4.7cm,trim=0.cm 0cm 1.5cm 0cm,clip]{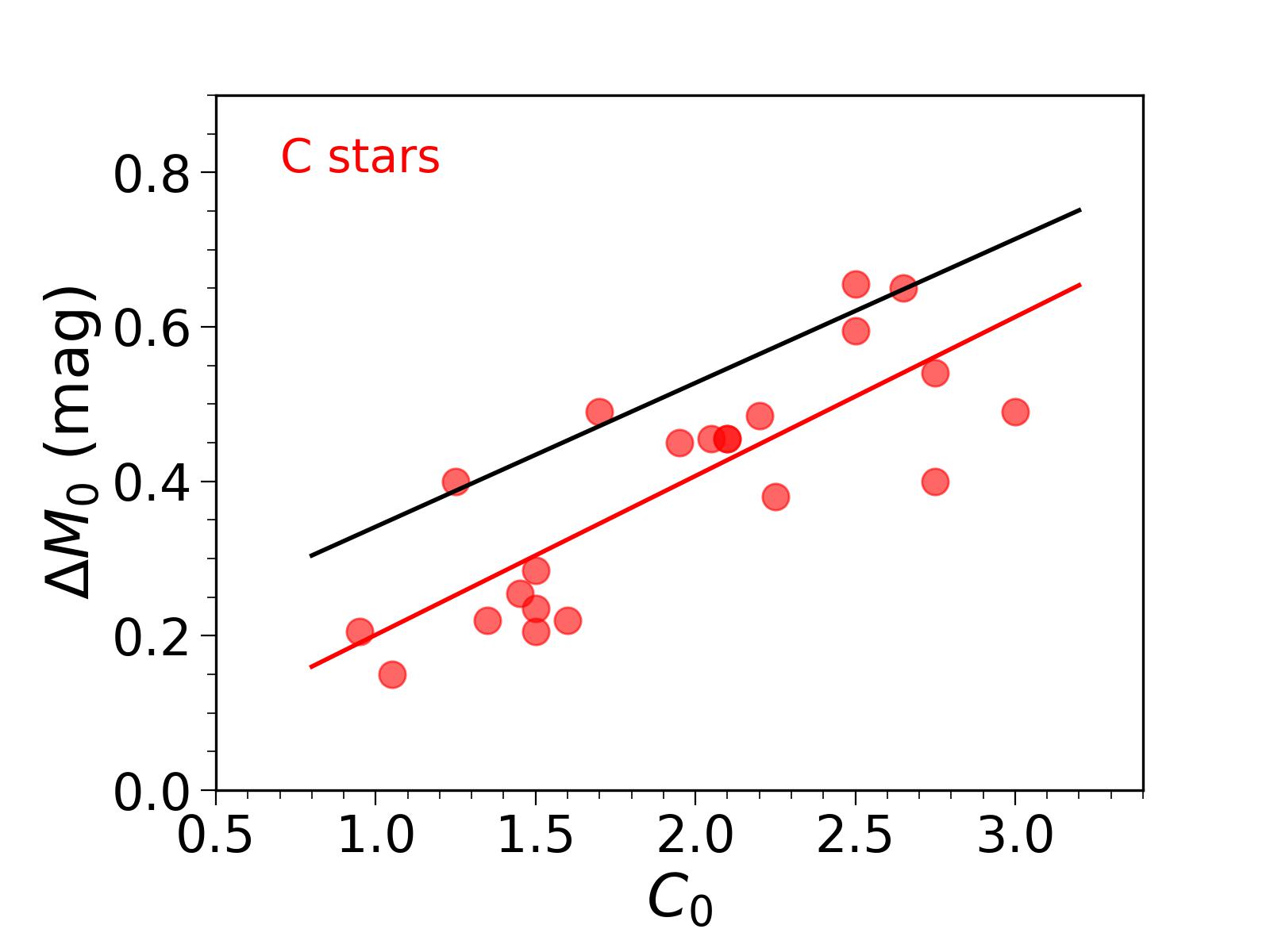}
  \includegraphics[height=4.7cm,trim=0.cm 0cm 1.5cm 0cm,clip]{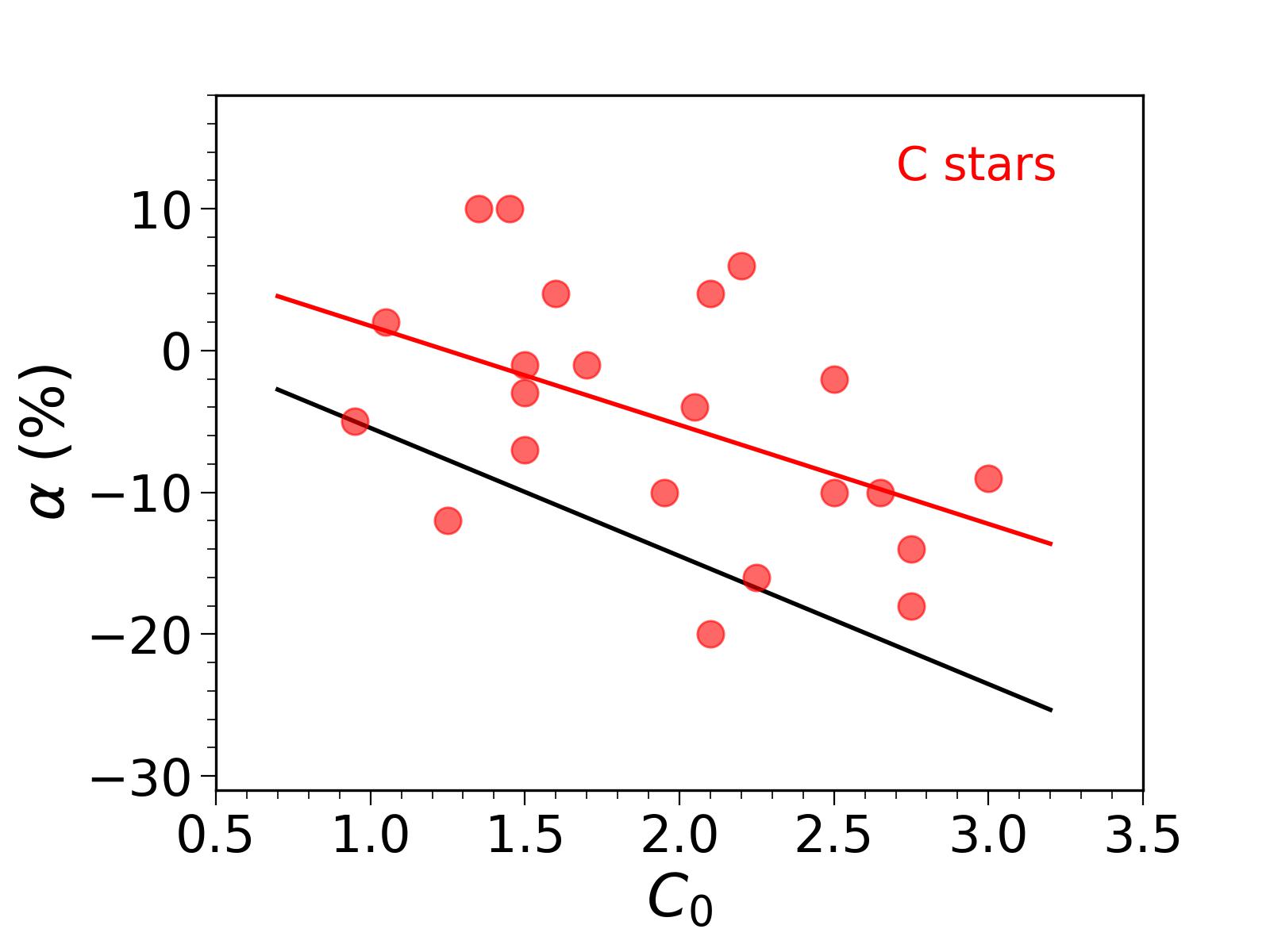}
  \includegraphics[height=4.7cm,trim=0.cm 0cm 1.5cm 0cm,clip]{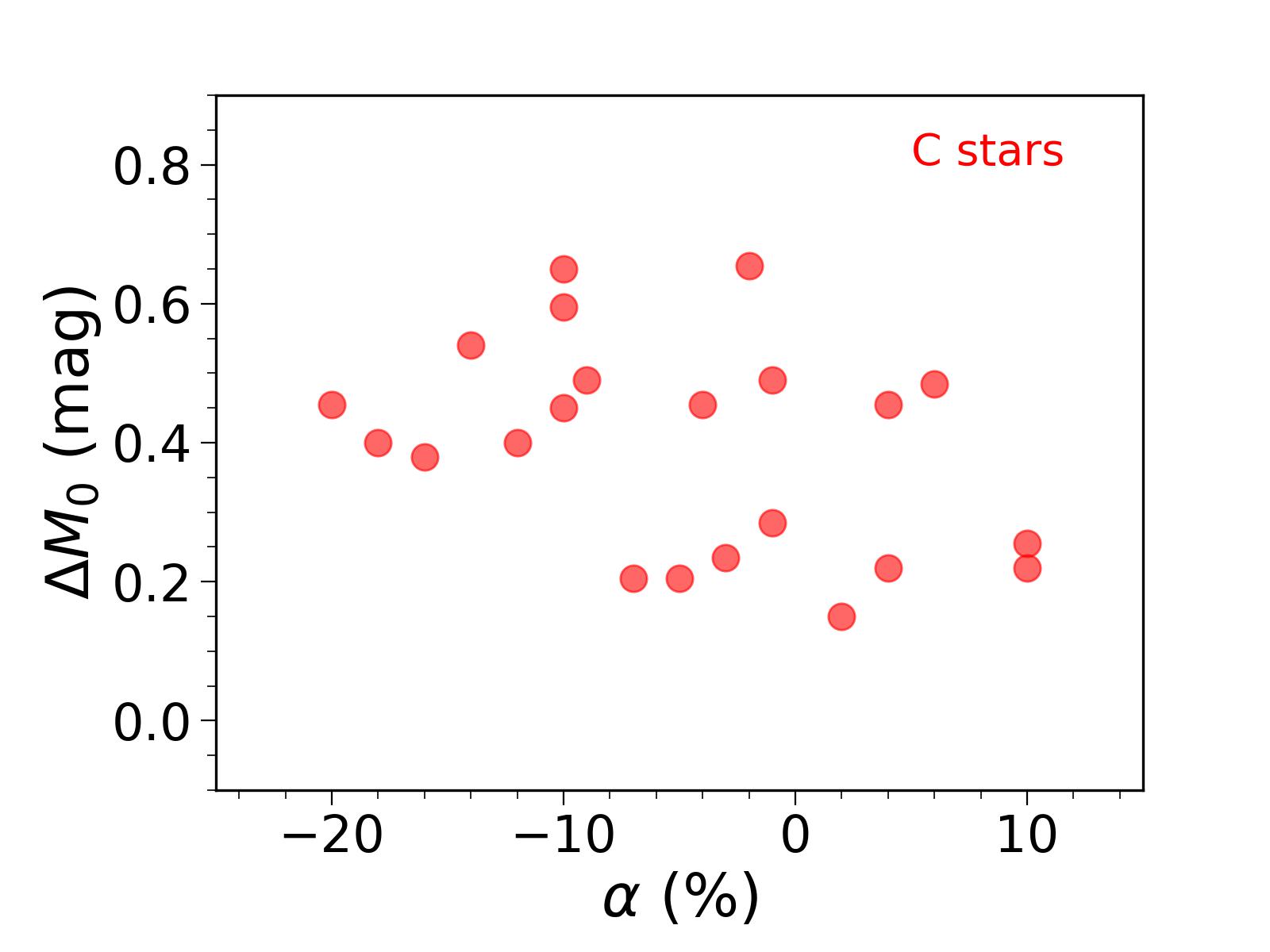}  
  \caption{Sample A curves of carbon stars. Distributions of $\Delta M_0$ vs $C_0$ (left), $\alpha$ vs $C_0$ (centre) and $\Delta M_0$ vs $\alpha$ (right). The black lines are the best fit results to the curves of oxygen-rich stars (Figure \ref{fig9}), the red lines are the best fit results to the C star curves illustrated in the figure.}
 \label{fig14}
\end{figure*}

Figure \ref{fig14} displays the dependence of the irregularity $\Delta M_0$ and asymmetry $\alpha$ parameters on colour index $C_0$ for the 22 curves of sample A having a C spectral type. In spite of the lesser reliability of the evaluation of the curve parameters, implying more scatter among them, the curves are seen to obey similar correlations as for oxygen-rich stars: more regular curves are also more symmetric and have smaller colour indices. However, for a same colour index, they are more regular, by typically 0.12 mag, and more symmetric, by a difference in $\alpha$ of typically 10\%, than oxygen-rich stars, possibly suggesting that once having crossed the S- to C-type boundary, the curves keep evolving a bit in such a direction.

\begin{figure*}
  \centering
  \includegraphics[width=1\linewidth,trim=0.cm 0cm 0cm 0cm,clip]{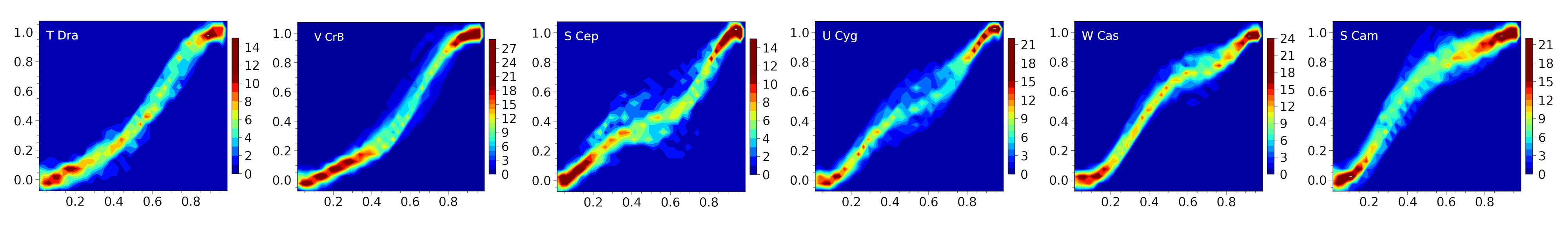}
  \caption{The ascending branches of the curves of the six carbon stars selected in sample B span the whole range of hump phases: from left to right, T Dra, V CrB, S Cep, U Cyg, W Cas and S Cam. }
 \label{fig15}
\end{figure*}

The assignment of profile types to curves of carbon stars cannot be done reliably in many cases and the profiles listed in Table \ref{tab5} of the Appendix are often nothing more than best guesses. It is therefore preferable to limit the study of profile types to the curves selected in sample B.  These include T Dra, V CrB, S Cep, U Cyg, W Cas and S Cam and the profiles of their ascending branches are displayed in Figure \ref{fig15}. Remarkably, their humps cover the whole range of phases. An interpretation that may come to mind is that such stars are extrinsic carbon stars which acquired their carbon excess from a white dwarf companion long ago and are currently evolving along the AGB. In practice, however, only T Dra \citep{Ramstedt2012} has been observed to emit X rays, a distinctive feature of such binarity. It displays, in addition, a UV excess, as does also RU Her \citep{AlonsoHernandez2024}. But such an interpretation would not apply to the other C-stars and seems quite unlikely. Listing the values of ($\varphi_{\rm min}$, $C_0$, $\Delta M_0$) for the six stars of sample B, ranked in order of increasing hump phase as shown in Figure \ref{fig15}, we obtain: (0.56, 3.0, 0.31), (0.58, 2.1, 0.28), (0.43, 2.2, 0.28), (0.51, 2.0, 0.37), (0.49, 1.0, 0.26) and (0.48, 1.1, 0.23), respectively. Although not perfectly, they follow the general trend that was displayed in Figure \ref{fig9} for oxygen-rich stars, giving support to the suggestion that the state of such carbon stars, the end point of their evolution on the TP-AGB, is also the end point of the evolution of their light-curves along a common path on the ascending branch, sometimes very close to its start, sometimes reaching up to its end.

\section{Regularity}  \label{sec6}

The above analyses, in both Sections 4 and 5, have occasionally revealed relations between the regularity of the observed light curves, defined as the level of similarity between successive cycles, and other parameters. We have been using two parameters to measure the regularity of the curves, $\Delta M_{\rm max}$ and $\Delta'M_{\rm max}$. Figure 16 illustrates how the curves of sample A are distributed in the $\Delta M_{\rm max}$ vs $\Delta'M_{\rm max}$ plane. It shows that the ratio $R_{\rm \Delta M}$=$\Delta'M_{\rm max}$/$\Delta M_{\rm max}$, which has value 1.08 in the case of random Gaussian fluctuations, is larger for the Mno spectral type and smaller for the C spectral type. This figure gives evidence for the presence of very irregular light curves, which have been accordingly excluded from sample B, but the existence of which must not be forgotten. The three worst cases, W Aql, AX Cep and R Lep have respective values of ($\Delta M_{\rm max}$, $\Delta'M_{\rm max}$) equal to (1.0, 0.6), (0.8, 0.5) and (0.8, 0.5) and are illustrated in Figure \ref{fig17}. The distribution of the sample B curves in the $\Delta M_{\rm max}$ vs $\Delta'M_{\rm max}$ plane is displayed in the left panel of Figure \ref{fig18}.  It shows the outstanding behaviour of R Cyg, with alternatively larger and smaller maxima of luminosity \citep{Wallerstein1985, Kiss2002, Roberts2020}. The right panel of the figure provides another illustration of this feature by introducing a new parameter, $\Delta''M_{\rm max}$=$<$$|M_{\rm max,i+2}-M_{\rm max,i}|$$>$, the mean of the absolute value of the difference between two maxima separated by one maximum in between, which would be equal to $\Delta'M_{\rm max}$ for random fluctuations. It shows that like R Cyg, but at a much lower level, RU Her and chi Cyg display some significant correlation between the maximal luminosities of successive cycles.

As a further illustration of the regularity of the light curves, we display in Figure \ref{fig19} the dependence on the mean colour index $C_0$, of the dispersion at maximal luminosity, $\Delta_{\rm dis}$, and of the amplitude of oscillation, $A$. Both show the strong homogeneity of the triplet of curves belonging to the first family of spectral type Mno. Most curves have $\Delta_{\rm dis}$$<$0.3 mag, with a mean$\pm$rms value of 0.23$\pm$0.02 mag; with $\Delta_{\rm dis}$$>$0.30 are an Myes star, S Her, and four C stars, S Cep, V CrB, U Cyg and T Dra. Together with the lower oscillation amplitude of C-type curves, the larger value of $\Delta_{\rm dis}$ gives them often the appearance of lower quality observations. Most stars have oscillation amplitudes smaller than 7 mag, with a clear split between C-type curves, $A$$<$3.5 mag, and other spectral types, $A$$>$3.5 mag; the two curves having $A$$>$7 mag are of S-type stars R And and chi Cyg. 

\begin{figure}
  \centering
  \includegraphics[width=0.5\linewidth,trim=0.cm 0cm 0cm 1cm,clip]{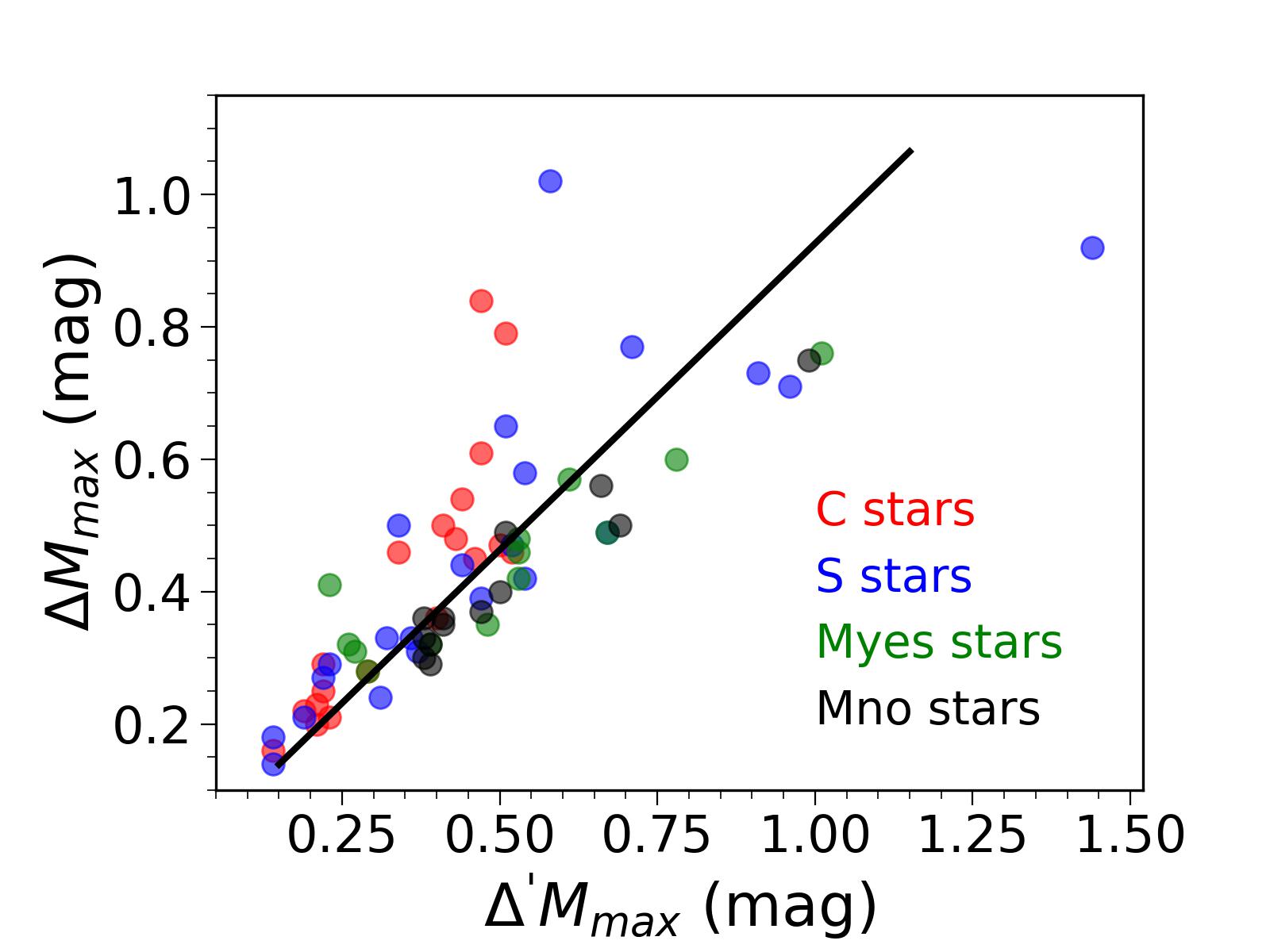}
  \caption{ Distribution of the curves of sample A in the plane of irregularity parameters $\Delta M_{\rm max}$ vs $\Delta'M_{\rm max}$. The line corresponds to the relation expected for random Gaussian fluctuations. Colours distinguish between different spectral types as indicated in the insert.}
 \label{fig16}
\end{figure}

\begin{figure*}
  \centering
  \includegraphics[width=1\linewidth,trim=0.cm 0cm 0cm 0cm,clip]{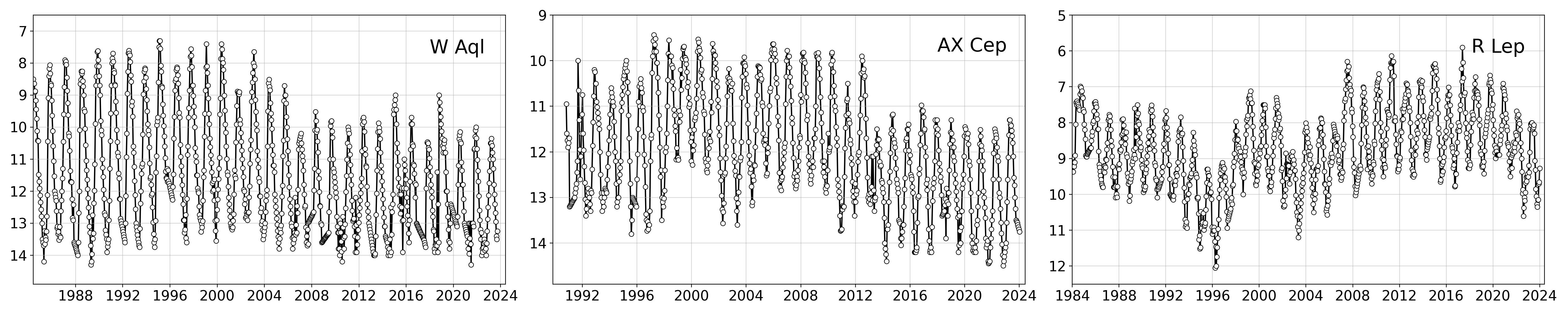}
  \caption{Irregular light curves of W Aql (left), AX Cep (centre) and R Lep (right) excluded from sample B.}
 \label{fig17}
\end{figure*}

\begin{figure*}
  \centering
  \includegraphics[height=5cm,trim=0.cm 0cm 0cm 1cm,clip]{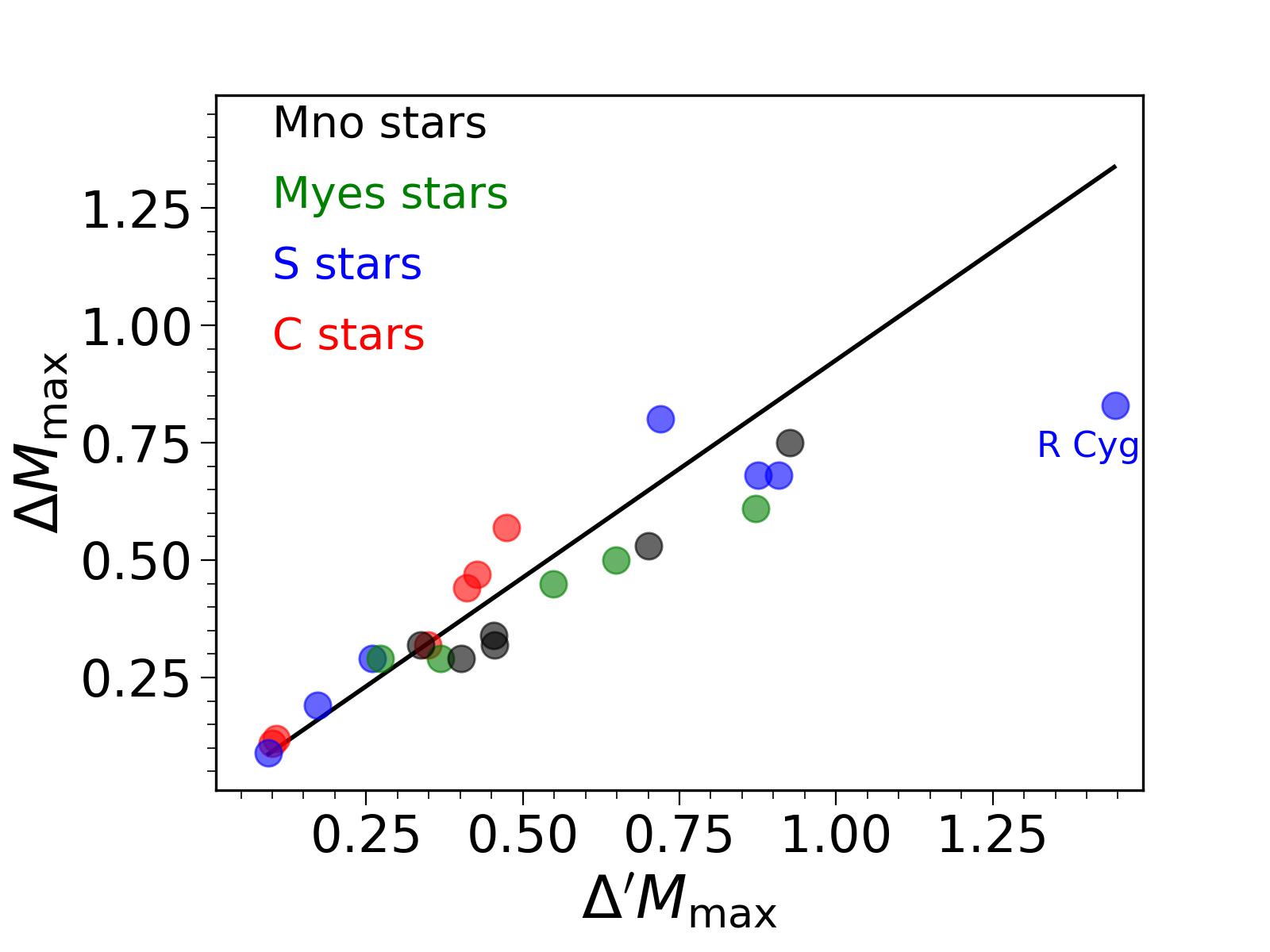}
  \includegraphics[height=5cm,trim=0.cm 0cm 0cm 1cm,clip]{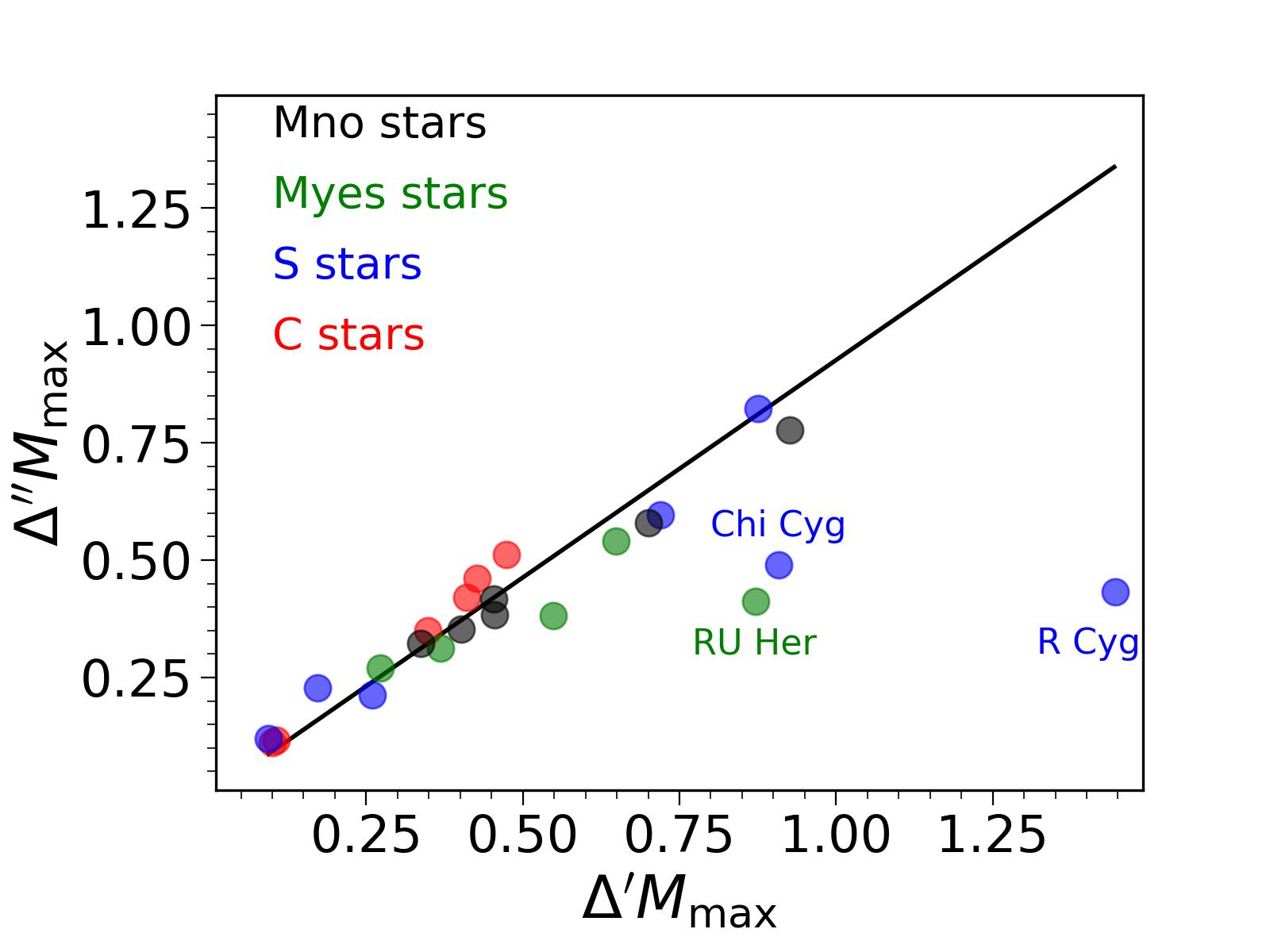}
  \caption{Distributions of the sample B curves in the $\Delta M_{\rm max}$ vs $\Delta'M_{\rm max}$ plane (left) and in the $\Delta''M_{\rm max}$ vs $\Delta'M_{\rm max}$ plane (right).}
 \label{fig18}
\end{figure*}

\begin{figure*}
  \centering
  \includegraphics[height=5cm,trim=0.cm 0cm 0cm 1cm,clip]{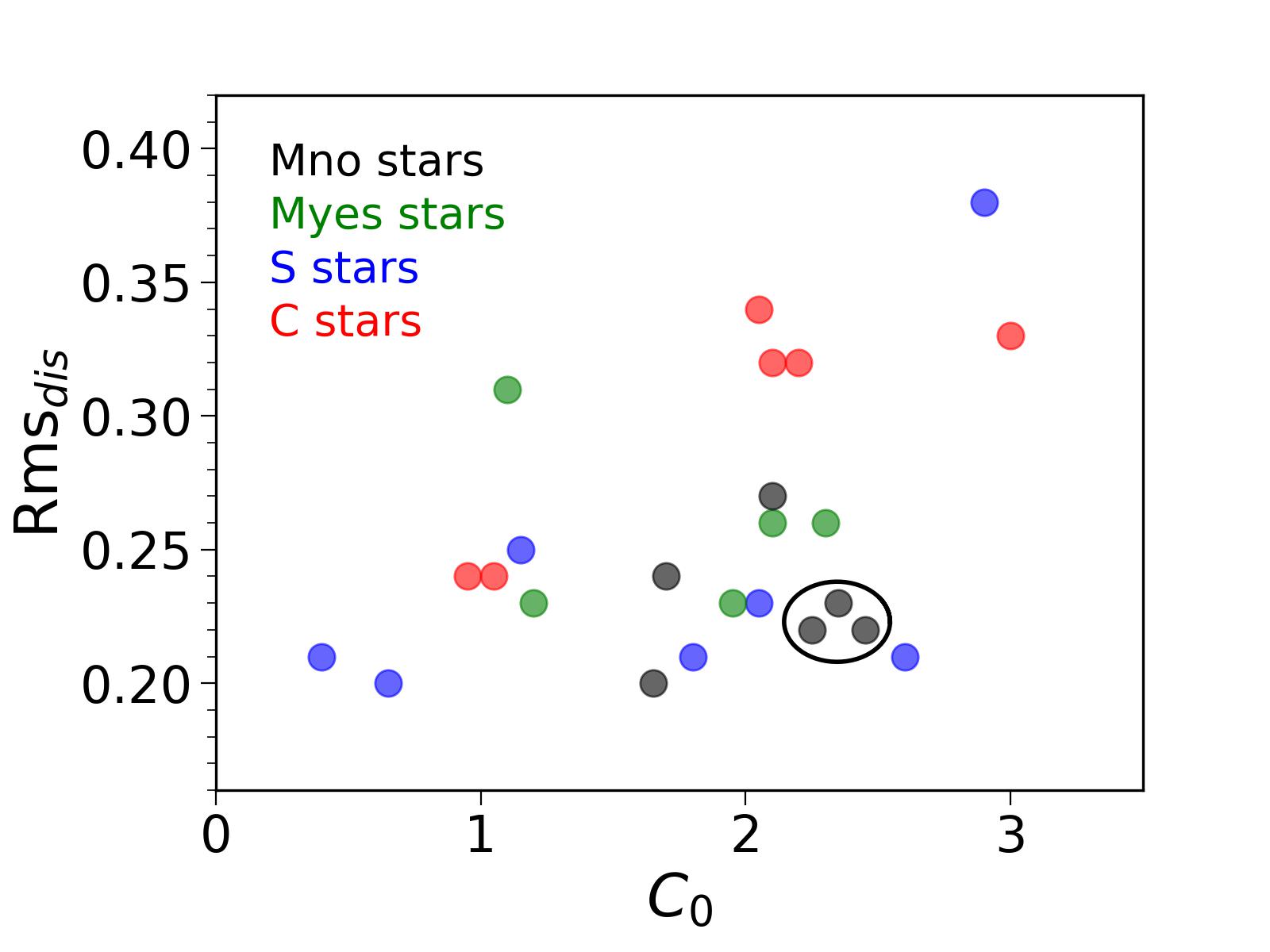}
  \includegraphics[height=5cm,trim=0.cm 0cm 0cm 1cm,clip]{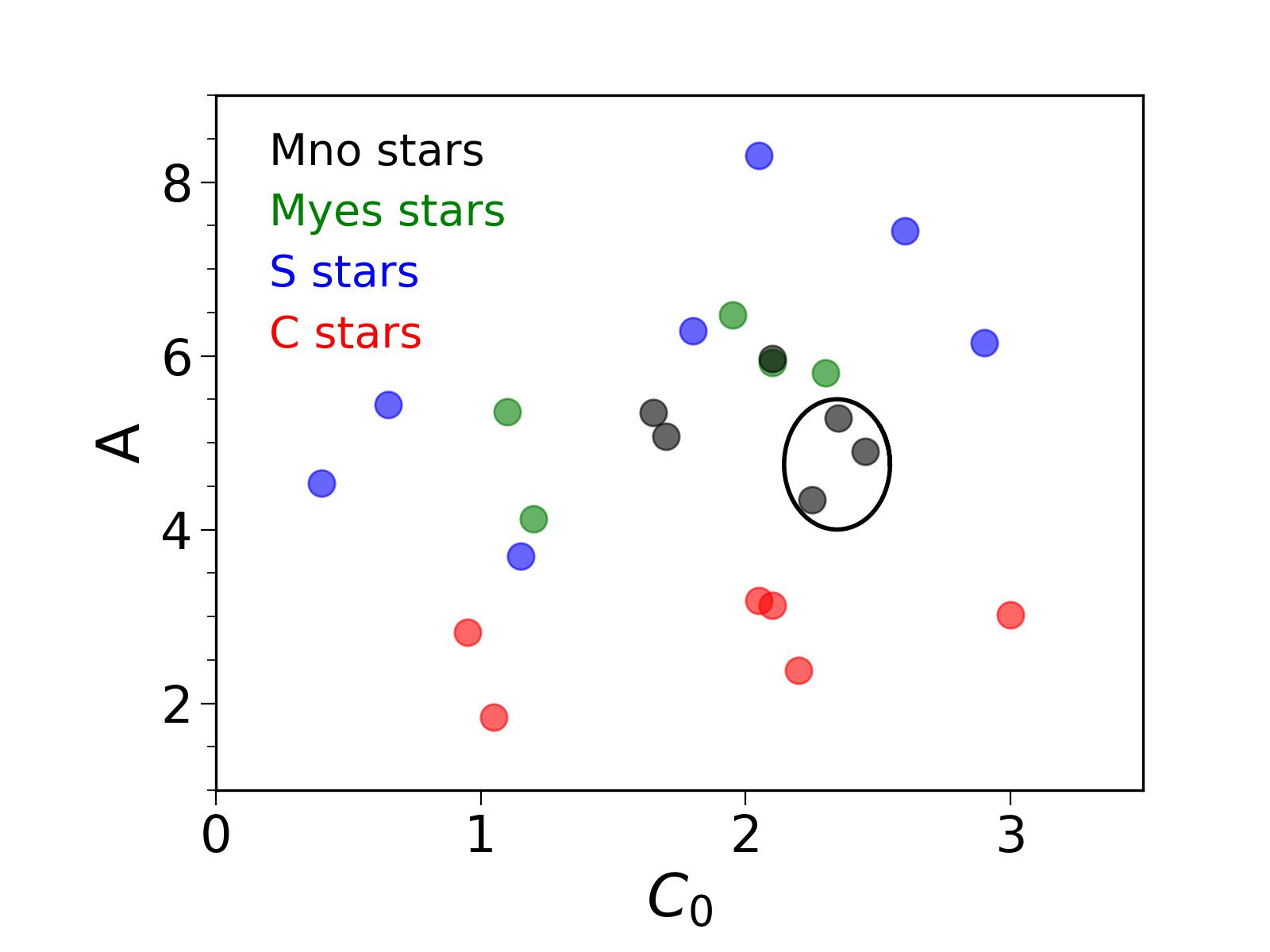}
  \caption{Distributions of the 24 sample A light curves selected in sample B in the $\Delta_{\rm dis}$ (left) and $A$ (right) vs $C_0$ planes. The triplet of curves of the first Mno family is circled. }
 \label{fig19}
\end{figure*}

Period changes have been abundantly studied in earlier work \citep[][ and references therein]{MerchanBenitez2023}. In particular \citet{Zijlstra2002b} have given evidence for three different types of period changes: continuous evolution, sudden changes and meandering periods. As mentioned in the introduction, strong period changes have been observed in a small fraction of Miras. The left panel of Figure 20 displays the distribution of the light curves of sample B in the $\Delta P$ vs $P$ plane, showing that most stars have $\Delta P$$<$18 days. In order to estimate how much of this is due to uncertainties in the evaluation of the time of maximal light, we show in the right panel the values taken in each curve by the mean value of the ratio ($P_{\rm i+1}$/$<$$P$$>$$-$1)/($P_{\rm i}$/$<$$P$$>$$-$1), the idea being that a measurement error in the time of maximal brightness of an oscillation will increase the value of one of the adjacent periods and decrease the other. Precisely we define $a$=$<$$(P_{\rm i+1}$/$<$$P$$>$$-$1)$\times$($P_{\rm i}$/$<$$P$$>$$-$1)$>$/$<$($P_{\rm i}$$>$/$<$$P$$>$$-$1)$^2$$>$, which takes the value of $-$0.46 for random measurement errors. The majority of curves have $a$ negative as expected from the uncertainty on $P_{\rm i}$ (if $P_{\rm i}$ is measured larger than real, $P_{\rm i+1}$ will be measured smaller than real on average). Four curves have instead $a$ positive, giving evidence for a systematic variability of the period: R Aur, T Cep, RU Her, all three of S spectral type, and R Aql, from the first family of Mno curves.

\begin{figure*}
  \centering
  \includegraphics[width=7.2cm,trim=0.cm 0cm 0cm 1cm,clip]{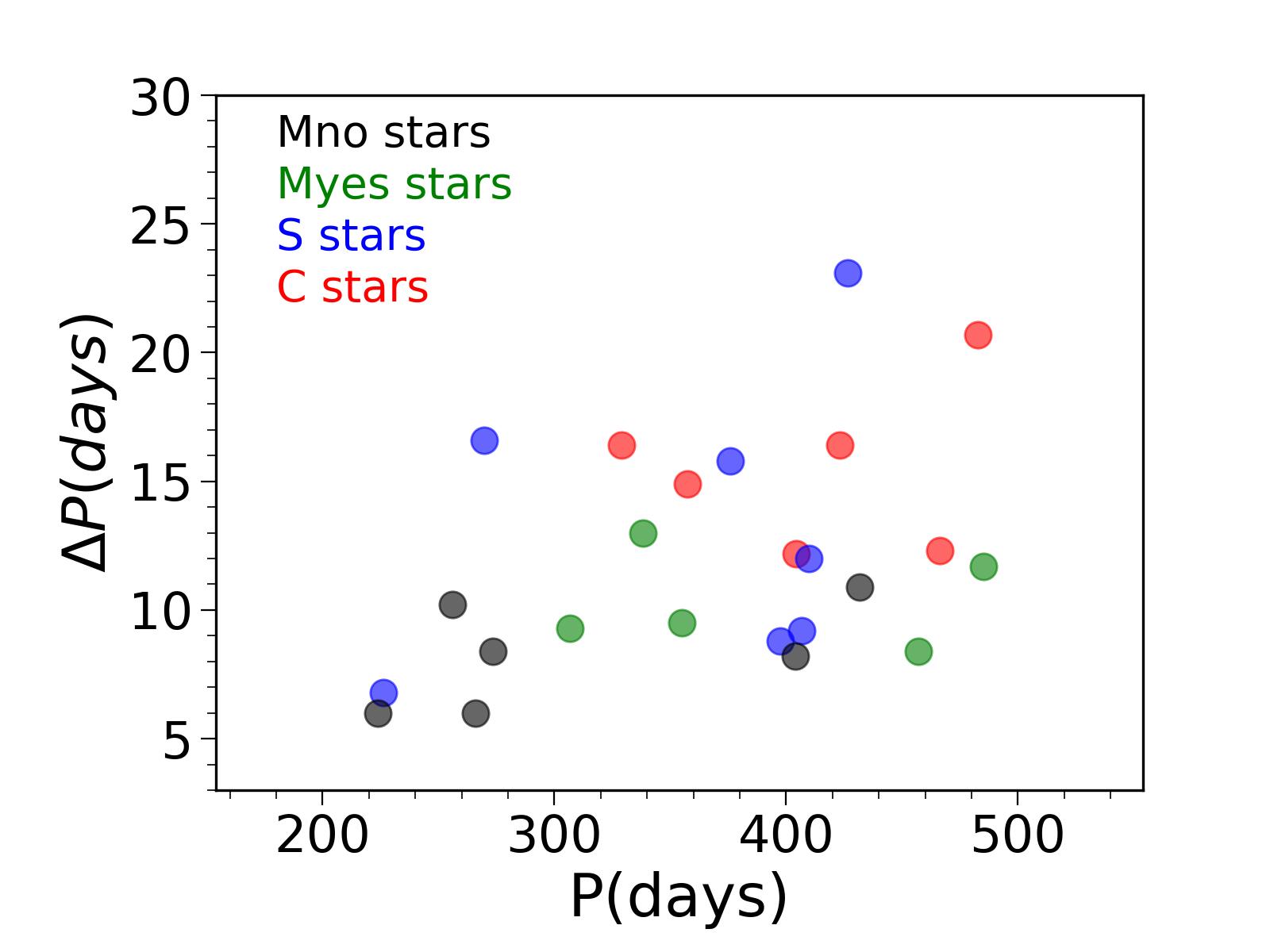}
  \includegraphics[width=6.66cm,trim=0.cm -1.5cm 0cm 0cm,clip]{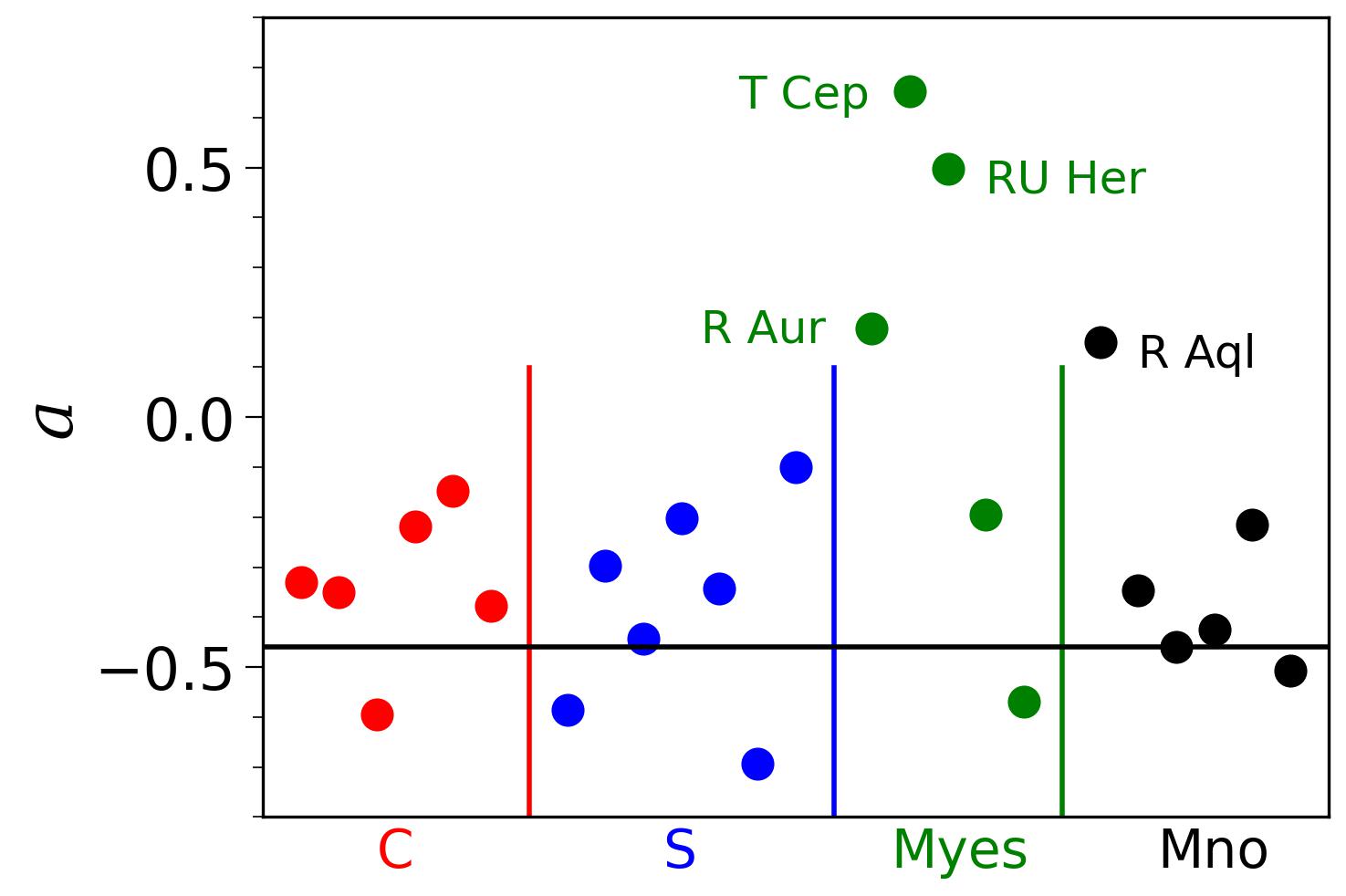}
  \caption{Regularity of the periods of the 24 sample A curves selected in sample B. Left: distribution in the $\Delta P$ vs $P$ plane. Right: distribution of parameter $a$ (see text) for each spectral type separately. The horizontal line is the value expected for random measurement errors. }
 \label{fig20}
\end{figure*}

\section{Discussion of the main results} \label{sec7}
The analyses presented in the preceding sections, of light curves of Mira variables selected in samples A and B, have confirmed and corroborated most of the results of earlier studies \citep{Ludendorff1928, Vardya1988, Lebzelter2011, Uttenthaler2019, MerchanBenitez2023}. At the same time, they have contributed new results, the main ones being: 1) the evidence for two different families of curves of spectral type M before the occurrence of third-dredge-up events; 2) the presence of a hump on the ascending branch of many curves, its properties and its evolution; 3) the presence of correlations between the regularity, the colour index and the symmetry of the curves; 4) the uniformity of the descending branch. In the present section, we briefly summarize these results and discuss possible interpretations. 

\subsection{Evidence for two distinct families of Mno type curves} \label{sec7.1}
\begin{figure*}
  \centering
  \includegraphics[height=4cm,trim=0.cm 0cm 0cm 1cm,clip]{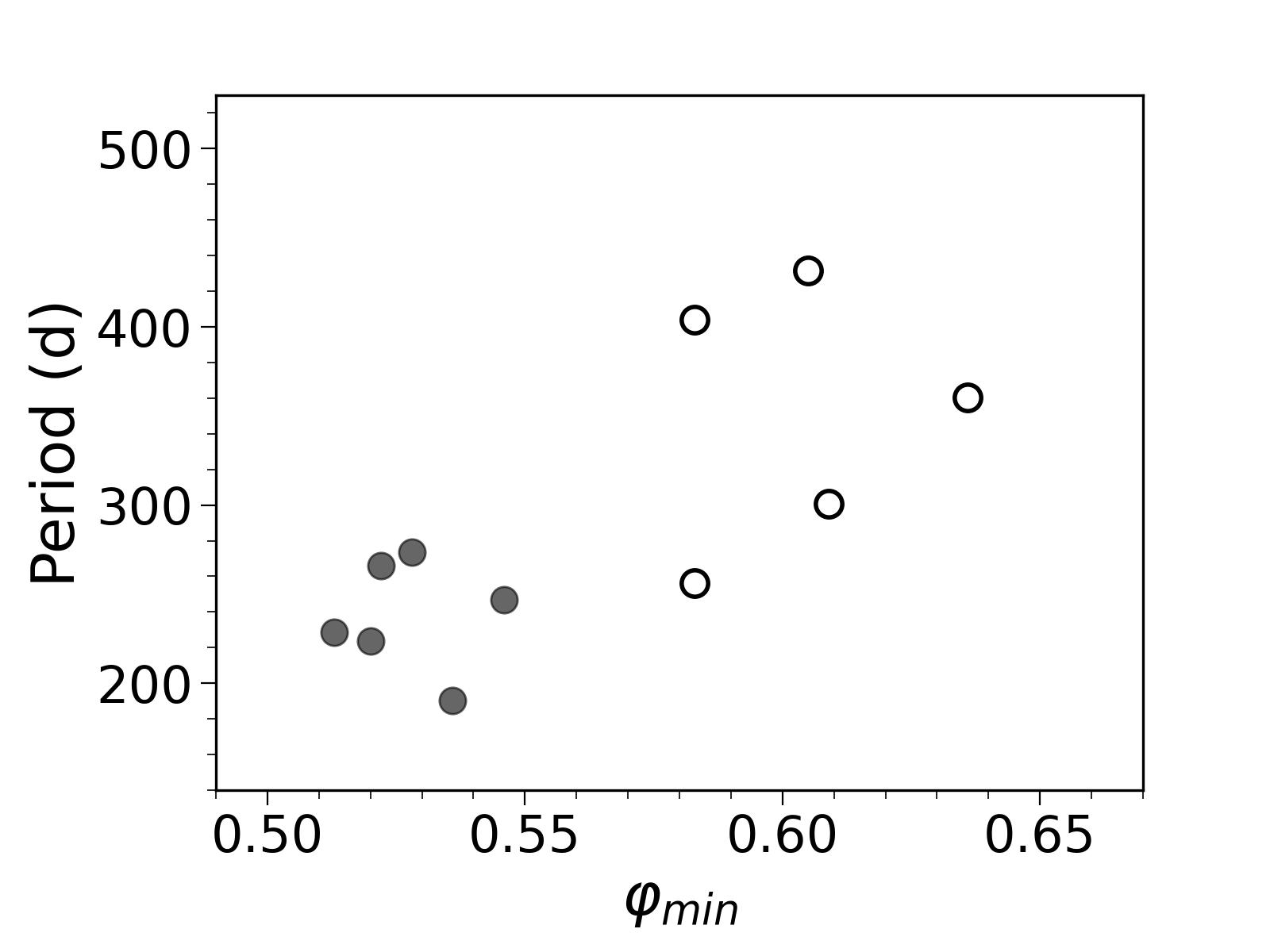}
  \includegraphics[height=4cm,trim=0.cm 0cm 0cm 1cm,clip]{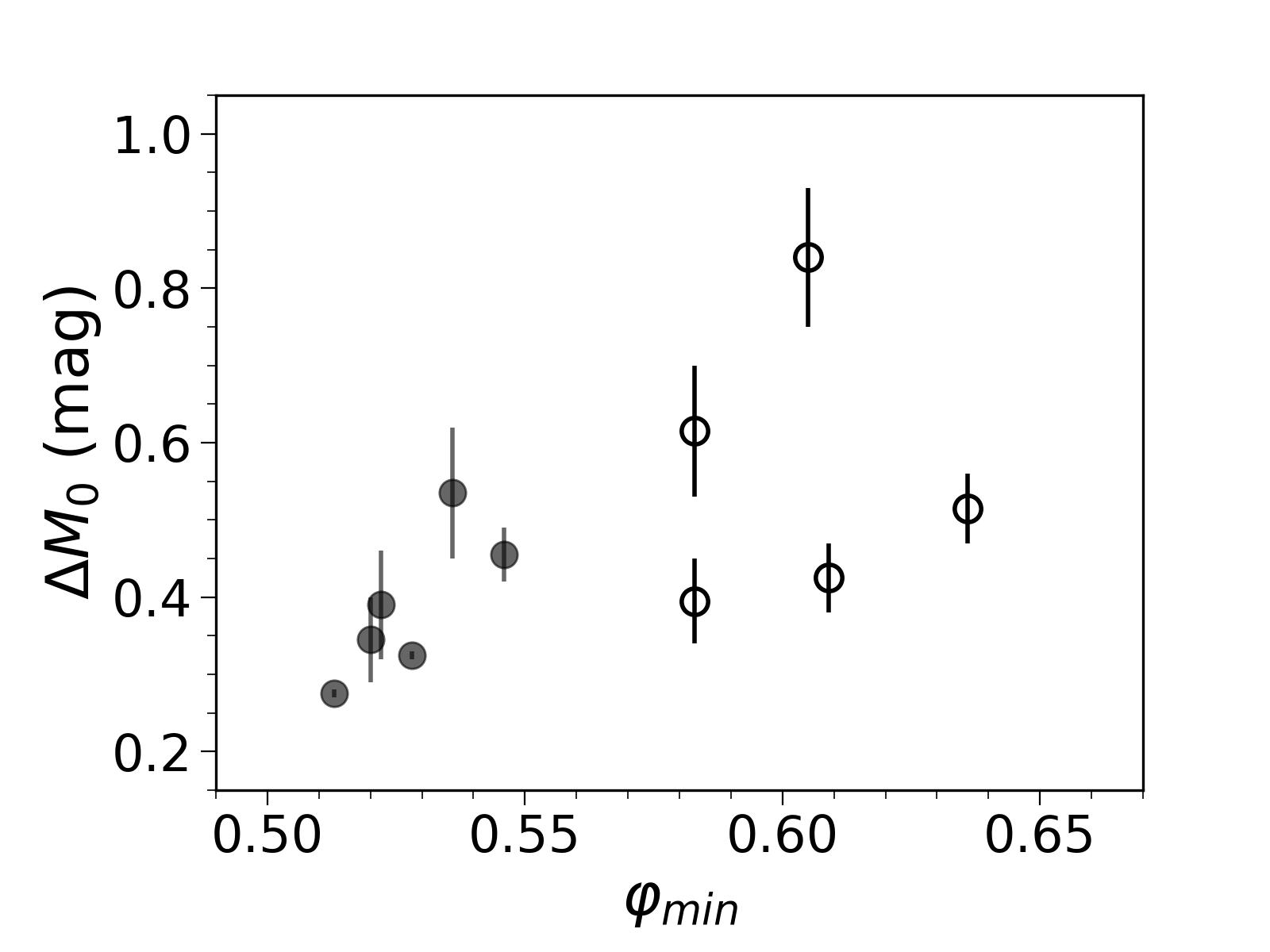}
  \includegraphics[height=4cm,trim=0.cm 0cm 0cm 1cm,clip]{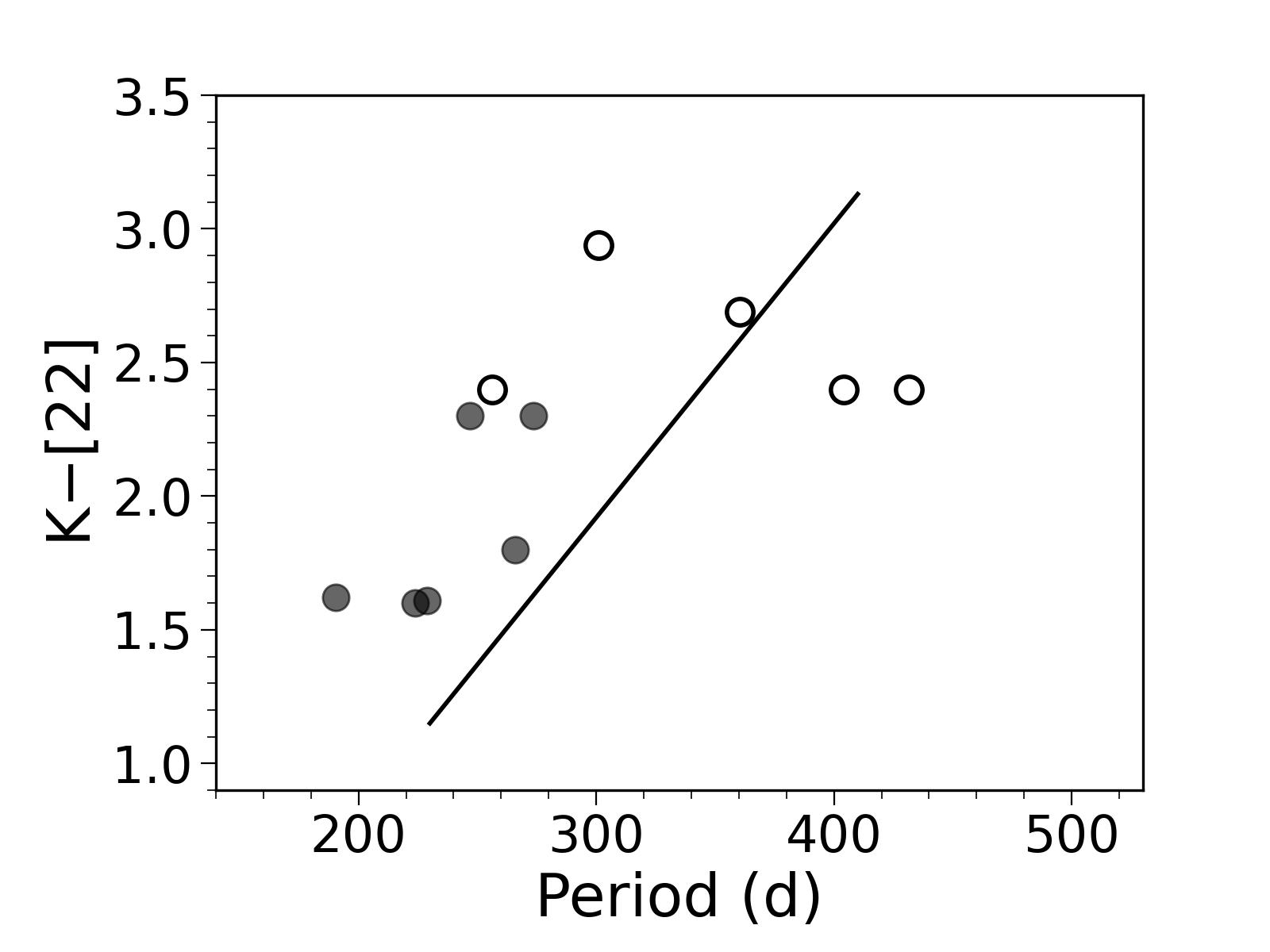}
  \caption{Sample B curves of spectral-type M identified as preceding the occurrence of third-dredge-up events. Left: period $P$ vs phase at minimal light, $\varphi_{\rm min}$. Centre: irregularity parameter $\Delta M_0$ vs $\varphi_{\rm min}$. Right: colour index K$-$[22] vs $P$; the line is the separator defined by \citet{Uttenthaler2019}. Full circles are for the first family, open circles are for the second family.}
 \label{fig21}
\end{figure*}

The analysis of sample A curves of spectral type Mno presented in Section 4.1 has given clear evidence for the presence in this group of two distinct families. Figure \ref{fig21} illustrates this result for sample B. The first family is more symmetric with $\varphi_{\rm min}$ between 0.50 and 0.55, compared with between 0.57 and 0.65 for the second family. It has smaller periods, below 300 days, is more regular and has smaller colour indices than the second family. It is exclusively made of \edit2{a} type profiles, while the second family contains also \edit2{b} type profiles. It sits on the left-hand side of the colour vs period diagrams first produced by \citet{Uttenthaler2019}, with $K_{\rm MB}$ negative, while the second family occupies also the right-hand side. When observed in infrared (Section 4.5), the ratio $R_{\rm I/V}$ of the infrared oscillation amplitude to that in the visible bandpass is smaller than 20\% while it exceeds 25\% for the second family. It includes the curves of R Aql, R Boo, R Tri, V Cas, RT Cyg, and R Dra. The second family includes the curves of R Cas, U Her, T UMa, S CrB and R UMa. 

All these curves have large oscillation amplitudes, on average 4.9 mag for the first family and 5.4 mag for the second family: they are clearly identified as Mira variables. When analysed in terms of Fourier transforms, a very powerful tool abundantly used in the published literature, they are seen to pulsate in the fundamental mode. However, for what concerns us here, Fourier transforms are not helpful; AGB stars are not organ pipes and what we are after is to understand the shape of a single oscillation: expressing the presence of a hump in terms of overtone contributions would not help with revealing the underlying physics. Yet, it is important to understand at which stage of their evolution along the AGB, stars can be expected to pulsate as Mira variables. For a star of solar metallicity having an initial mass between 1 and 3 solar masses, the time spent on the E-AGB is an order of magnitude larger than the time spent on the TP-AGB as oxygen-rich. Precisely, the ratio takes values of 17, 9, 7 and 10 for initial masses of 1, 1.5, 2 and 3 solar masses, respectively \citep{MillerBertolami2016}. While, on most of the E-AGB, stars are not expected to pulsate as Mira variables but rather as low-amplitude semi-regulars \citep{Yu2020}, it is conceivable that they do at the end of the E-AGB, say in the last 10\% or so. By then, their radius and luminosity have considerably increased.

We recall that at the end of the E-AGB, the star has a small degenerate core of carbon and oxygen in its centre, surrounded by a thin helium-rich shell, itself surrounded by a thin hydrogen-rich shell, both shells being separated by a thin inter-shell layer \citep{Herwig2005}. During the whole TP-AGB, the star is in a state fully governed by what happens in this pair of shells. When running out of helium fuel at the end of the E-AGB, the star starts burning hydrogen around the depleted helium shell: when this will have produced enough helium ashes in the inter-shell layer, the star will experience its first thermal pulse, which will be followed by several others at intervals typically two orders of magnitude shorter than the time spent on the E-AGB. What happens in the convective envelope surrounding them and in the stellar atmosphere further out is, of course, essential to decide how the star looks like when observed from far away and to understand the mass loss mechanism that defines its lifetime on the AGB; but it obscures the basic dynamics causing the evolution of the star and considerably complicates the interpretation of what is observed.

In order to suggest possible interpretations of the presence of two different families in the curves of Mno spectral type, we need to identify a physical mechanism that distinguishes between two different groups among Mira variables that have not yet experienced a significant third-dredge-up event. Two examples of such a possible distinction, suggesting two different scenarios, come immediately to mind: between stars at the end of the E-AGB and stars on the TP-AGB, and between low mass stars that will not evolve to the Myes spectral type and higher mass stars that will.

In the first scenario, one of the Mno families would include E-AGB stars, the other TP-AGB stars. All E-AGB stars are expected to transition to the TP-AGB phase when they run out of helium fuel around the inert core. The E-AGB group includes therefore stars of all possible masses; in contrast, the TP-AGB group includes all low mass stars that do not experience strong enough third-dredge-up events to allow their evolution to the Myes spectral type but excludes the higher mass stars that have already evolved to the Myes spectral type. The E-AGB group is therefore more homogeneous than the TP-AGB group and must be identified with the first Mno family. In favour of such a scenario, is its good qualitative match with what we observe: the E-AGB stars are expected to be characterised by only a few parameters, essentially their mass and their age on the AGB, and can therefore be expected to experience relatively regular and simple oscillations; for the same reason, their transition to the TP-AGB can be expected to take place at a well-defined stage of their evolution, suggesting a clear distinction between the two families in a two-parameter space such as the K$-$[22] vs $P$ plane. In contrast, the second family would be made of stars that experience thermal pulses, which implies contributions from complex convection mechanisms, and would include low mass stars that are experiencing mass-loss, another complex mechanism, up to the end of their AGB life. The second Mno family would therefore be expected to cover larger domains than the first Mno family in any parameter space that characterizes the shape of the light curve and/or the evolutionary state of the star.  Such an interpretation would then imply that when transiting from E-AGB to TP-AGB, namely from helium- to hydrogen-burning around the inert core, the star experiences significant changes in the dynamics that govern its interior, resulting, in particular, in an increase of the pulsation period. The transition would also cause a significant change in the shape of the light curve, taking the form of a slowing-down episode near minimal light, that would progressively evolve to higher phases and become a hump in the cases of the stars evolving to the Myes spectral type.  The simultaneity between the appearance of such a slowing-down episode and the increase of the period might have suggested that the two events are related, the slowing-down not being compensated by speeding-up during the remaining part of the period, which would then increase accordingly. However, the study of the delay caused by the hump that was presented in Section 4.3 invalidates such a suggestion.

In contrast, the second scenario would imply that the transition from E-AGB to TP-AGB does not result in major changes of the pulsation properties of the star or, if it does, that the transition occurs before the star having evolved to a state that enables its pulsating as a Mira variable; namely, in this latter case, the whole sample of stars studied in the present article would be made of TP-AGB stars. Then, the second Mno family would probably be made of stars that will soon experience sufficient third-dredge-up events to enter the Myes spectral type and the first Mno family of low mass stars that will never enter the Myes spectral type but spend their entire lifetime on the AGB as Mno stars. At variance with the first scenario, no star would transition between the two Mno families: those of the first Mno family would stay there until the end of their AGB life, those of the second Mno family would then transition to the Myes spectral type. The difference of periods between the two families would then be the simple result of the different star masses and the enhanced irregularity of the second Mno family would be the effect of stronger thermal pulses causing more active convection.

These two extreme scenarios form the basis on which more elaborate scenarios could be constructed. Both may allow some continuity between the two families: the first scenario with curves of stars in the process of transiting between the E- and TP-AGB; the second scenario as a result of the continuity of the star mass distributions, implying a corresponding continuity of the strength of the convective mixing following thermal pulses and eventually leading to third-dredge-up events.

\subsection{Presence of a hump on the ascending branch of some light curves} \label{sec7.2}
In Section 4.2, to each curve of oxygen-rich stars selected in sample A, we have assigned a profile type, ranking from \edit2{a} to \edit2{d}, the three latter displaying a hump on the ascending branch at increasing phases. We noted that the curves having oscillations of \edit2{a} and \edit2{b} profile types are more asymmetric, more dusty and more irregular; those having oscillations of \edit2{c} and \edit2{d} profile types are more symmetric, less dusty and more regular. In Section 4.3 we studied curves of sample B displaying a clear hump and we evaluated its magnitude, normalised to the oscillation amplitude, $y_{\rm hump}$. We found that its rms dispersion with respect to its mean is of the order of a tenth, and that the width of the hump is of the order of 14\% of the oscillation amplitude. On average (Figure \ref{fig10} right), we have shown that the parameters $\varphi_{\rm min}$, $C_0$ and $\Delta M_0$ are decreasing functions of $y_{\rm hump}$. We also established that the slowing-down caused by the hump is essentially compensated by speeding-up elsewhere on the ascending branch and does not cause an increase of the pulsation period. A weak, barely significant positive correlation between the strength of the hump and the amplitude of the descending branch of the preceding oscillation allowed for a possible contribution of a shock wave.

This suggests that when entering the TP-AGB, the curves develop a hump, first confused with the luminosity minimum and then progressively climbing the ascending branch when and if the star transitions to the Myes spectral type. However, while the hump reaches up to maximal light in the case of some carbon stars, such as S Cam, it does not always do so, far from it. Instead, each spectral type is seen to cover the whole corresponding $y_{\rm hump}$ range rather than clustering at its end. This is illustrated in Table \ref{tab4}, which lists, for each spectral type and each profile type separately, the number of curves selected in samples A and B. We recall that the assignment of type profiles is less reliable for sample A than for sample B, which, however, includes only half as many curves. But the assignments made in sample B can be trusted as reliable and their diversity was clearly illustrated in Figure \ref{fig15} in the case of carbon stars. It looks like if, when progressing along the TP-AGB from M to C spectral types, the star experiences something that causes an apparent slowing-down episode of the pulsation, accompanied by an episode of compensating speeding-up, when it is on the ascending branch of the light curve. The phase of the hump would start at minimal light and would increase together with the C/O ratio that measures the progression of the star on the TP-AGB. However, this phase would stop increasing at some point, the star continuing its evolution along the TP-AGB while the hump stays at a fixed phase.

Such humps are also observed on the light curves of other pulsating stars. In the case of RR Lyrae \citep[][and references therein]{Prudil2020} they track shock waves caused by a rapid compression of the star atmosphere: they are the result of a sudden stop of the infalling photosphere and immediate outward expansion. \citet{Kudashkina1994} have remarked that a similar mechanism could be at stake in the case of Mira variables. Using a simple model of the propagation of shock waves in the star atmosphere, heating the gas to temperatures sufficient for ionisation, they show that the duration of the resulting hump is of the order of a tenth of a period, in good agreement with observations. In the case of Cepheids \citep[][and references therein]{Bono2000, Marconi2024} the hump is on the ascending branch for periods larger than ~11 days and on the descending branch otherwise \citep[the so-called Hertzsprung progression][]{Hertzsprung1926}. State-of-the-art non-linear convective pulsation models, accounting for the movement of pressure waves in the star interior including reflections and resonances, provide descriptions of the Hertzsprung progression in agreement with observations \citep{Marconi2024}.

However, in the case of Mira light curves, and in spite of numerous and detailed descriptions that have been available for decades \citep{Schorr2016, Templeton2005, Marsakova2007, Lebzelter2011, MerchanBenitez2023}, we lack a reliable interpretation of what is causing them. The suggestion \citep[e.g.][]{Lebzelter2011} that they may simply result from a resonance between the fundamental and the first overtone modes is not satisfactory, the light curves of Mira stars having large oscillation amplitudes implying that the stars are in conditions allowing them to comfortably pulsate in the fundamental mode. \citet{MerchanBenitez2023} suggest instead that dust being formed inefficiently in the wake of an outwardly propagating shock wave would fall back toward the star instead of being accelerated outwards and would hit the outwardly moving shock wave of the next pulsation cycle \citep{Hinkle1982}, thereby causing humps or secondary maxima in the light curve. 

\begin{deluxetable*}{ccc ccc ccc ccc}
\tablenum{4}
\tablecaption{Number of light curves selected in samples A and B, listed according to their spectral type and profile type, separately. Five sample A curves have been excluded as having a too poorly defined profile type. \label{tab4}}
\tablehead{
\colhead{}&\multicolumn{5}{c}{Sample A}&\colhead{}&\multicolumn{5}{c}{Sample B}\\
\cline{2-6}
\cline{8-12}
\colhead{}&\colhead{\edit2{a}}&\colhead{\edit2{b}}&\colhead{\edit2{c}}&\colhead{\edit2{d}}&\colhead{\edit2{e}}&\colhead{}&\colhead{\edit2{a}}&\colhead{\edit2{b}}&\colhead{\edit2{c}}&\colhead{\edit2{d}}&\colhead{\edit2{e}}}
\startdata
Mno1&	6&	1&	0&	0&	0&	&6&	0&	0&	0&	0\\
Mno2&	3&	3&	0&	0&	0&	&1&	4&	0&	0&	0\\
Myes&	5&	2&	3&	3&	0&	&1&	1&	2&	4&	0\\
   S&	8&	3&	4&	7&	0&	&2&	1&	1&	3&	0\\
   C&	6&	0&	5&	3&	4&	&1&	0&	2&	2&	1\\
\enddata
\end{deluxetable*}

\subsection{Presence of correlations between regularity, colour index and symmetry of the curves} \label{sec7.3}
\begin{figure*}
  \centering
  \includegraphics[height=5cm,trim=0.cm 0cm 0cm 1cm,clip]{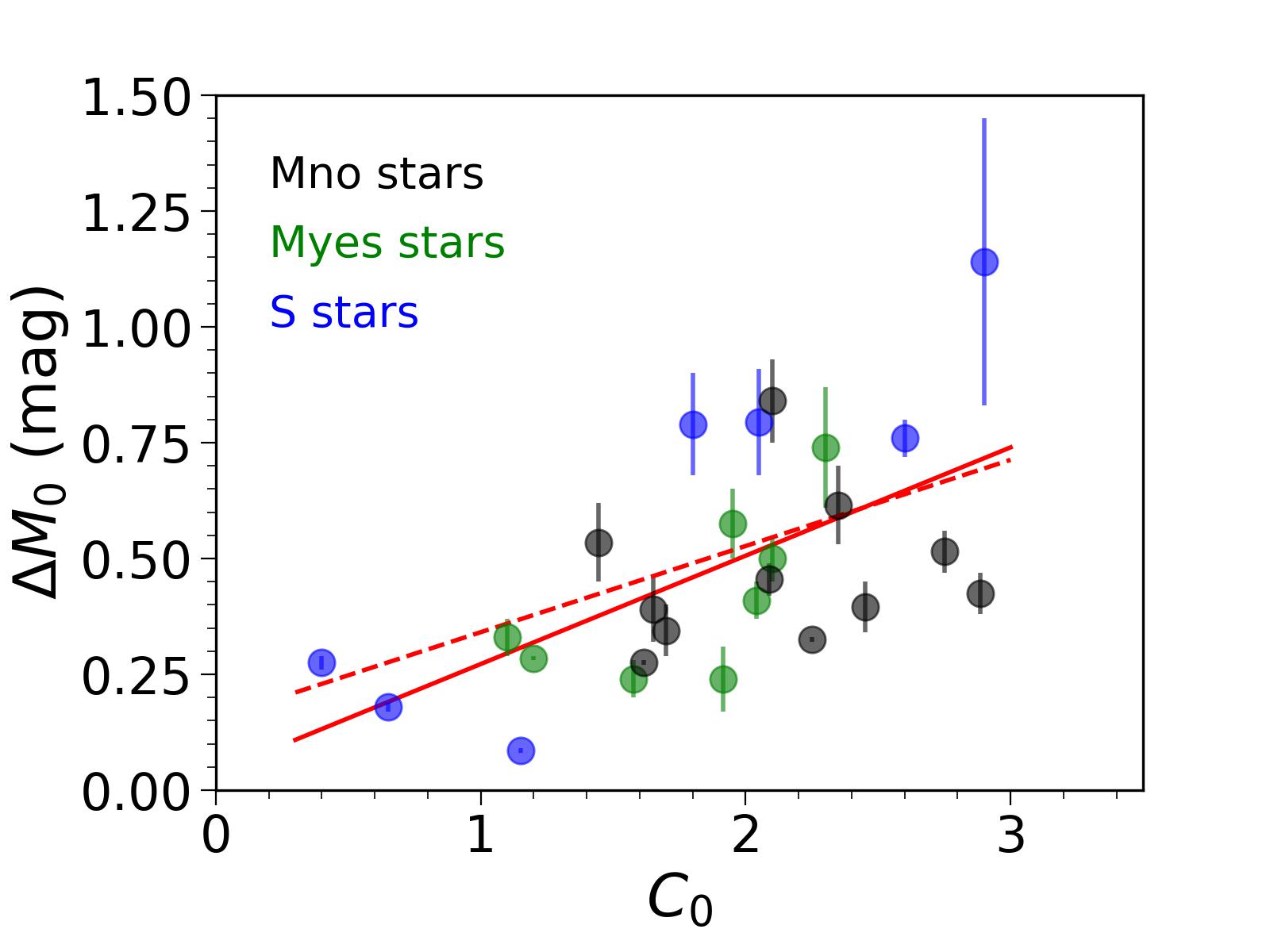}
  \includegraphics[height=5cm,trim=0.cm 0cm 0cm 1cm,clip]{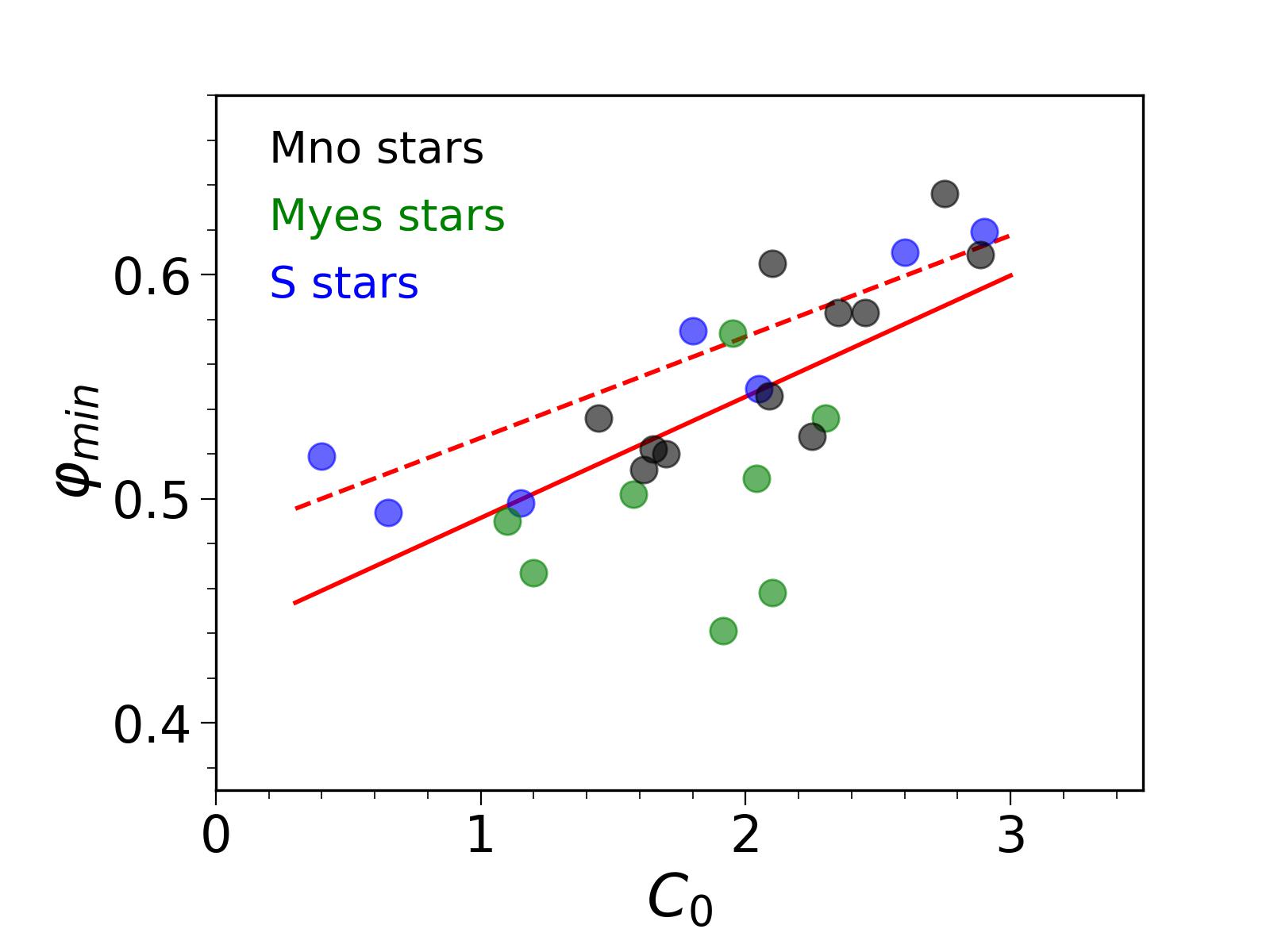}
  \caption{ Sample B curves. Dependence on colour index $C_0$ of the regularity parameters $\Delta M_{\rm max}$ and $\Delta'M_{\rm max}$ and of the phase at minimal light, $\varphi_{\rm min}$. A bar joins the values taken by $\Delta M_{\rm max}$ and $\Delta'M_{\rm max}$ at a same value of $C_0$. Different spectral-types are shown with different colours as shown in the inserts. The full red lines show linear fits to $\Delta M_0$=$\frac{1}{2}$($\Delta M_{\rm max}$+$\Delta'M_{\rm max}$) and to $\varphi_{\rm min}$ and the dashed red lines show the same linear fits obtained for sample A (Figure \ref{fig9}). }
 \label{fig22}
\end{figure*}

Figure \ref{fig9} has given evidence for significant correlations between the parameters describing the sample A light curves of oxygen-rich stars, spectral-types M to S. The more symmetric curves, exclusively made of \edit2{c} and \edit2{d} profile types, were observed to be more regular and have smaller colour indices than the more asymmetric curves, exclusively made of \edit2{a} and \edit2{b} profile types, which are irregular and have larger colour indices. Figure \ref{fig14} has shown that a similar trend is followed by the curves of carbon stars, which, however, for a same colour index, are more regular and more symmetric than the curves of oxygen-rich stars. Figure \ref{fig22} displays the dependence on colour index, $C_0$, of the regularity parameters, $\Delta M_{\rm max}$ and $\Delta'M_{\rm max}$ (left) and of the phase at minimal light, $\varphi_{\rm min}$ (right) for the curves of sample B. It qualitatively confirms the general trend found for sample A: the average slopes for sample A (Figure \ref{fig9}) were 0.19 mag of $\Delta M_0$ and 0.045 of $\varphi_{\rm min}$ per unit of magnitude of the mean colour index $C_0$. They are now, for sample B, 0.23 mag of $\Delta M_0$ and 0.054 of $\varphi_{\rm min}$.

While reliable and robust, the above result is only valid on average and the curve parameters shown in Figure \ref{fig22} display major dispersion with respect to the mean. This can be understood in part as an effect of the thermal pulses that occur at a typical rate of one every 10$^5$ years \citep{MillerBertolami2016, Marigo2013}. While very short, only a few years, these pulses are so violent, with temperatures up to a few 10$^8$ K, that they can be expected to disturb the interior and the atmosphere of the star for a significant fraction of the inter-pulse phase. In particular, this argument has been used by \citet{MerchanBenitez2023} to explain the occasional observation of period changes, the more drastic the closer to the thermal pulse. Some parameters, such as the C/O ratio, the amount of technetium in the spectrum, the $^{12}$C/$^{13}$C ratio, increase progressively after each third dredge-up episode and are not expected to display significant scatter. But others, such as temperature and turbulence, can be expected to be deeply affected, causing major temporary changes in the pulsation regime. However, such an interpretation seems unlikely to apply to the near totality of the curves illustrated in Figure \ref{fig22}. Such major scatter must be seen in the context of other puzzling features that Mira stars light curves display over the TP-AGB phase, such as major changes that affect their shape from one cycle to the next, as discussed in Section 6; these include, in particular, the value of $y_{\rm hump}$ and the intensity with which a hump may occur, but also, to a lesser extent, the maximal luminosity and the period. Another puzzling feature worth mentioning in this context is the absence of S- and C-types meandering period changes \citep{MerchanBenitez2023}. Also of relevance is the possibility for the curve of a given star to stop being a Mira variable, either momentarily or definitively. This may happen on the occasion of thermal pulses but may also happen in other occasions. In particular, we noted that many of the sample A curves of carbon stars are closer to the semi-regular type of pulsation than to the Mira type.

What makes it difficult to devise sensible interpretations of observations that must necessarily be made from outside the star, is that its appearance is the consequence of a long chain of causally connected events: it starts from nuclear burning around the inert core, continues with the properties of the convective zone and ends with the inner and outer atmospheres of the star. Ideally, one would like to understand nuclear burning and convection before having to deal with the complexity of the nascent wind, dust formation and radiation pressure. In this context, the latter, not to mention the impact of a possible companion, seem unfortunate complications that prevent a clear identification and description of the more basic mechanism at stake inside the star. In particular, quantities such as the mass loss rate \citep[][and references therein]{Uttenthaler2024} or the $^{12}$C/$^{13}$C isotopic ratio are known to be both episodic and anisotropic: their properties are very complex and their evaluation requires a detailed understanding of the radiative transfer taking place in the atmosphere, of the shock chemistry of relevance and of the dust formation process \citep[e.g.][]{Darriulat2024, Hoai2024}. As a result, very large uncertainties are attached to such quantities: apart from general trends, well documented in the published literature, they seem unlikely to help significantly with the understanding of the features addressed in the present article. Yet, while strongly affected by the properties of the star atmosphere, light curves are also apt at revealing features taking place inside the star, which makes a detailed study of their shapes particularly valuable.

\subsection{Uniformity of the descending branch} \label{sec7.4}
\begin{figure*}
  \centering
  \includegraphics[height=8cm,trim=0.cm 0cm 0cm 1cm,clip]{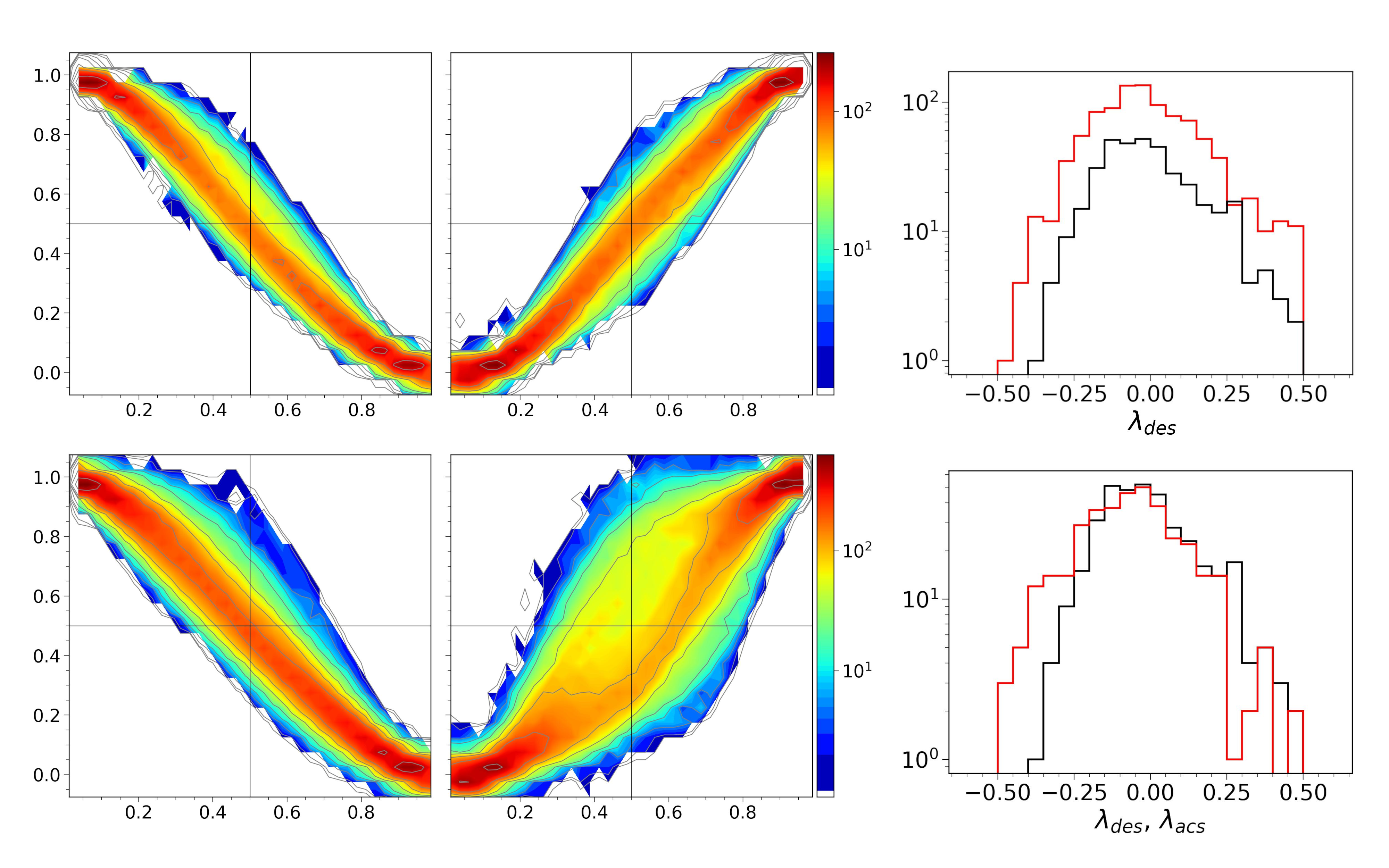}
  \caption{Left pair of panels: superimposed normalised profiles of the descending and ascending branches of sample B curves. The 8 curves of the first Mno family ($\sim$370 oscillations) are displayed on the upper row and the 24 other curves ($\sim$1000 oscillations) are displayed on the lower row. The colour scales are logarithmic. Right panels: distribution of the parameter $\lambda$ (see text) comparing the descending branches of the first Mno family (red) with the descending branches of the other curves (black, upper row) and with the ascending branches of the first Mno family (black, lower row).}
 \label{fig23}
\end{figure*}

In Section 3.3, we briefly commented that all curves of sample B display a hump-free and smooth normalised profile of the descending branch. This is important information, worth additional comments. Even if we know of a few, extremely rare cases of oscillations that seem to display a small hump on the descending branch, we can consider this as a robust and reliable result. The left pair of panels of Figure \ref{fig23} compares the superimposed normalised profiles of all oscillations of the ascending branches of the sample B curves with those of the descending branches. Curves of the first Mno family are shown separately from the other 24 curves. The figure offers a spectacular illustration of the difference between the ascending branches of these 24 curves and either their descending branches or both the ascending and descending branches of the first Mno family. In order to better quantify such a result, we fit each normalised profile of the ascending and descending branches observed in the oscillations of the 32 light curves of sample B with the form defined in Section 3.3; it uses a single parameter $\lambda$ that measures, for each branch separately, the deviation from mean magnitude ($y=\frac{1}{2}$) at mean phase ($x=\frac{1}{2}$). The distributions of the number of profiles as a function of $\lambda$ are displayed in the right panels of Figure \ref{fig23}. The (mean, rms) values of $\lambda$ are ($-$0.01, 0.16) for the descending branches of the first Mno family, ($-$0.07, 0.17) for its ascending branches and ($-$0.01, 0.17) for the descending branches of the 24 other curves. Of course, the similarity revealed by comparing these three distributions does not mean that all curves share a same branch profile. They span different fractions of the period, measured by the different values taken by $\varphi_{\rm min}$, and different fractions of the oscillation amplitude. But it means that no special event seems to take place when the luminosity of the star decreases, nor when it increases in the case of the first Mno family. This is at strong variance with the clear evidence for something specific happening, taking the form of an apparent slowing-down episode, when the luminosity increases, in the 24 other curves. It provides another illustration of the results presented in sub-sections 7.1 and 7.2.  

\subsection{Peculiarities} \label{sec7.5}
The results obtained in the present article deliberately ignored peculiarities that might have distracted us from trying to reveal general features. Yet several such peculiarities have been met and, while it is beyond the scope of the present article to address them, we know that science often progresses by detailed scrutiny of anomalous cases and we briefly list the main ones in the present sub-section.

X Oph is a star of spectral type Mno with a curve displaying an \edit2{a} profile type but its very small oscillation amplitude ($A$=1.7 mag compared with 5.1 mag) and its very broad minima (and correspondingly narrow maxima, $W_{\rm \frac{1}{2}}$=0.36 compared with 0.44) make it very different from the Mno curves which have been studied in the present article. The star has a bright K-type giant companion \citep{Fernie1959}.

R Cen and R Nor are also stars of spectral type Mno, with respective periods of 572 and 498 days, but they display a double-peaked oscillation profile. It has been recurrently argued in the literature that the periods are in fact half these values and that oscillations of higher and lower apparent luminosity alternate \citep{Templeton2004}. \citet{Hawkins2001} argue that R Cen is probably experiencing a helium-shell flash and that the most likely cause of double-peaking is a resonance between two modes, but \citet{GarciaHernandez2013} show instead that R Cen is a massive hot-bottom-burning AGB star.  One often reads in the published literature that ``R Cen used to have an unusual double-peaked light curve, but by 2001 this had reverted to an almost normal single-peaked curve'', but this statement is incorrect. From a study of lithium in the spectra of R Cen and R Nor, \citet{Uttenthaler2012} suggest that both are undergoing the hot-bottom-burning phase.

R Aql was found in Section 6 (Figure 20 right) to show evidence for a systematic variability of its period, which is indeed known to have declined from 320 days in 1915 to 264 days in 2010 \citep{ZhaoGeisler2012}. This is surprising for a star of spectral type M that displays no technetium in its spectrum, has a low $^{12}$C/$^{13}$C ratio of only 8, and has been identified in the present work as a clear candidate of the first Mno family. \citet{Wood1981} have shown that such a period change could very well be interpreted as the consequence of a recent thermal pulse, in which case the curve should belong to the second, rather than first, Mno family if the first scenario is retained: the case of R Aql seems therefore to favour the second scenario.

R Gem, a star of spectral type S with a curve displaying a clear a-type oscillation profile, stands out as being quite regular ($\Delta M_0$=0.27 mag, $R_{\rm \Delta M}$=1.3) together with a very large asymmetry ($\alpha$$\sim$$-$30\%), two properties that usually do not go together. It has a $K_{\rm MB}$ value of 0.8. As it has a period very close to a year (370 days) it suffers from being always observed at the same phase at a given time of the year, but this does not explain its peculiarity.

R Cyg, another star of spectral type S, has a light curve displaying spectacular episodes of alternating maxima of luminosity, that were discussed in Section 6 together with other extreme cases of irregularity including R Lep, R Cas and W Aql. Extreme values of the period include: below 200 days, T Her (165 days) and above 500 days, R Cen (572 days), BH Cru (529 days) and S Cas (613 days). R Lep is a carbon rich star which has been the target of many observations \citep[][and references therein]{Rau2017, Asaki2023}. W Aql is an S-type star, which has also been abundantly observed \citep{Danilovich2014, Brunner2018}; \citet{DeBeck2020}, from a study of the chemistry of the gaseous component of its CSE, have concluded that it appears considerably closer to a C-type star than to an S-type star. \citet{Mayer2013} have detected a companion at a separation of 160 au. BH Cru is known for the high variability of its period \citep{Loidl2001}. \citet{Zijlstra2004}, from a study of its evolution, argued that it is unlikely to be related to an ongoing thermal pulse. \citet{Walker2009} studied changes in the light curve and \citet{Uttenthaler2011} discussed the period changes and the transition from spectral type S to C. S Cas has been observed to have a high mass loss rate, of $\sim$1.8 10$^{-6}$ solar masses per year, by \citet{McDonald2018}. The pulsation period of R Hya has been observed to decline and \citet{Uttenthaler2010}, from a study of technetium and lithium in its spectrum, have argued that, while this decline may be related to a recent thermal pulse, such pulses would not be strong enough to generate a third dredge-up. The CSE has been a target for numerous observations \citep[][and references therein]{Homan2021, Joyce2024}.

\section{Summing up} \label{sec8}
In the present work, we have selected visual light curves of Mira variables that offer sufficient density and quality of observations for a reliable parameterisation of their shape. Depending on how strict the selection was, it produced a sample A of 71 curves or a sample B of 32 curves. While having corroborated most of the results of earlier studies, our work contributes new significant results: 1) the evidence for the presence of two distinct families of curves of spectral type M before the occurrence of third-dredge-up events, suggesting that they distinguish either between stars at the end of the E-AGB and stars on the TP-AGB or between low mass stars that will never transition to the Myes spectral type and higher mass stars that will. 2) the presence of a hump on the ascending branch of many curves, meaning an apparent slowing-down of the rate of luminosity increase on the ascending branch, compensated by an apparent speeding-up elsewhere on this same branch, and climbing it as the star evolves along the TP-AGB; it is seen as a broadening of the minimum when it starts deviating from it and as a broadening of the maximum if and when it reaches it. 3) The presence of correlations between the regularity, the colour index and the symmetry of the curves; while the evidence for their existence as a general average trend is a robust and reliable result, a very large scatter of the curve parameters with respect to such average significantly weakens their relevance. 4) The uniformity of the descending branch has been clearly demonstrated and shown to give evidence for the absence of specific event occurring on this phase of the pulsation.                
A sensible explanation of these results is still lacking. They suggest that something causes a change of shape of the ascending branch of the light curve when the star is on the TP-AGB phase, namely when it burns hydrogen around the helium shell in intervals separated by violent thermal pulses. It must be related to the details of the mechanism that governs the pulsation, being observed exclusively on the ascending branch of the light curve, and being seen to evolve along this branch in correlation with the time spent by the star on the TP-AGB. Understanding what is going on requires therefore understanding the nuclear processes at stake, the impact of convective mixing, the presence of shock waves and the morphology of the ionisation zone. For example, it could take the form of a shock wave that would repeat at each oscillation of the hydrogen-burning inter-pulses of the TP-AGB and would ionise part of the envelope as first speculated by \citet{Kudashkina1994}; or such a shock-wave could impact the mechanism of dust formation as suggested by \citet{Uttenthaler2024}.

The complexity and diversity of the light curves, the strong variability that they display from one cycle to the next, and the absorption caused by the dust-rich outer atmosphere, make it particularly difficult to provide reliable interpretations, in particular to make a clear distinction between events occurring inside the star and events occurring in its atmosphere. Our approach has been to reveal general trends at the price of ignoring apparent anomalies. One must not underestimate the speculative character of many of the considerations that we have been proposing. As we are unable to state reliable interpretations with sufficient confidence, our aim has more modestly been to trigger reactions and inspire new studies that would either confirm or deny our findings. Yet, the rich set of results that we have been able to obtain confirms our feeling that the outstanding set of high-quality observations of visual light curves of Mira variables has not yet been exploited as thoroughly as it deserves.

\section*{acknowledgments}
  
To the extent that the analyses presented in the present article may have contributed some progress, the credit belongs to the innumerable observers around the world and to the AAVSO, without whom none of the results could have been obtained. We are accordingly deeply indebted to them: to the many observers for the high quality of their observations and to the AAVSO for the outstanding handling and reduction of the data that makes their use particularly easy and efficient.
We are deeply grateful to Dr Stefan Uttenthaler, who published, together with colleagues, several seminal articles that largely inspired our work, for having patiently listened to and answered our many questions. We thank him deeply for having shared with us his views and understanding of many of the issues at stake in the present work and for a careful and critical reading of the manuscript. 
Financial support from the World Laboratory, the Rencontres du Vietnam and the Vietnam National Space Center is gratefully acknowledged. This research was funded by Vingroup Innovation Foundation (VINIF) under project code VINIF.2023.DA.057.

\bibliography{lightcurves}{}
\bibliographystyle{aasjournal}

\appendix
\renewcommand\thefigure{A\arabic{figure}}
\setcounter{figure}{0}

\begin{figure*}[h]
  \centering
  \includegraphics[width=1\linewidth,trim=0.cm 0cm 0cm 0cm,clip]{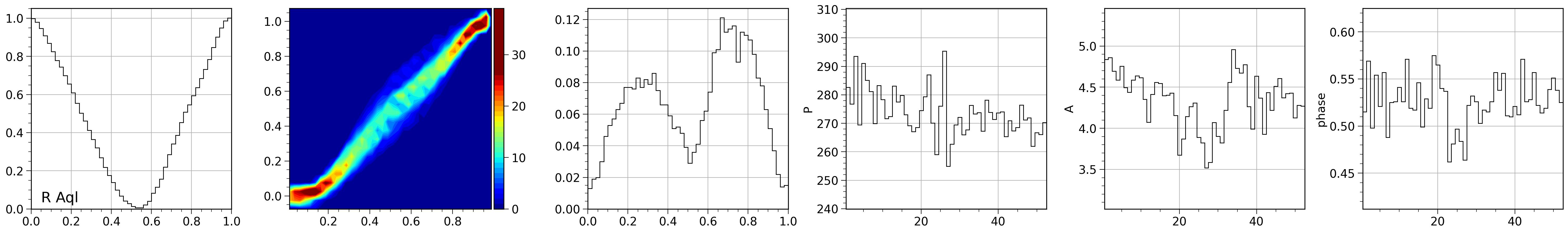}
  \includegraphics[width=1\linewidth,trim=0.cm 0cm 0cm 0cm,clip]{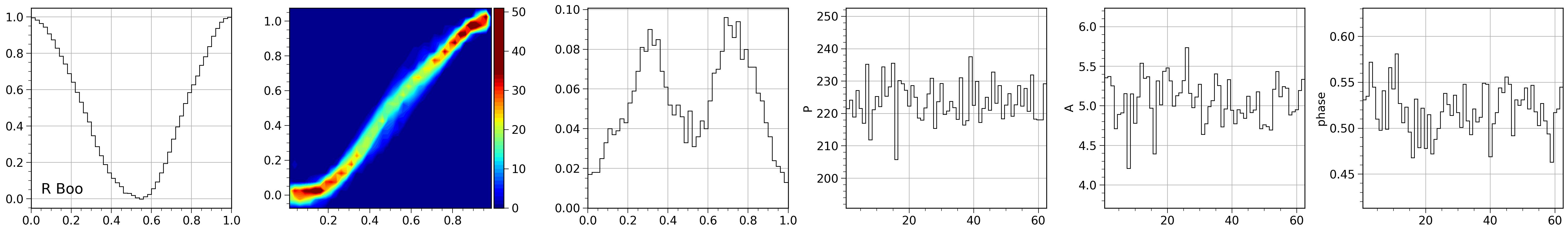}
  \includegraphics[width=1\linewidth,trim=0.cm 0cm 0cm 0cm,clip]{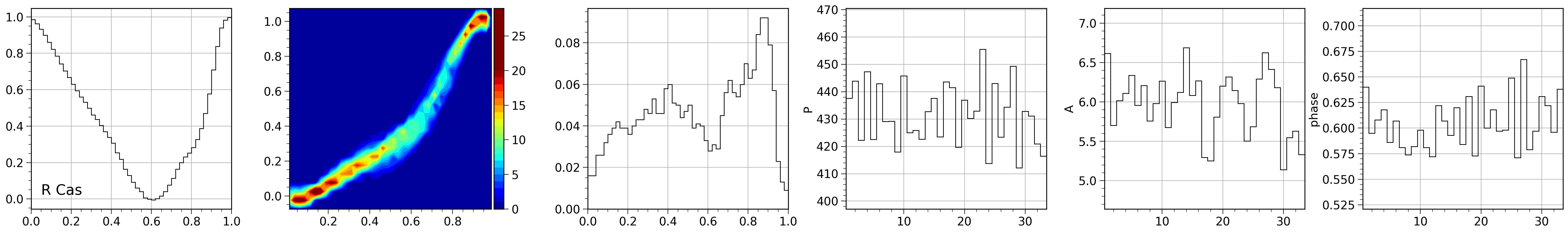}
  \includegraphics[width=1\linewidth,trim=0.cm 0cm 0cm 0cm,clip]{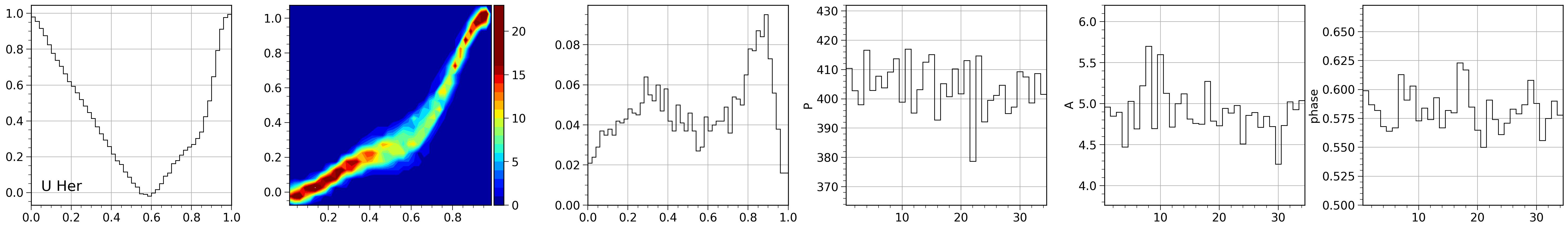}
  \includegraphics[width=1\linewidth,trim=0.cm 0cm 0cm 0cm,clip]{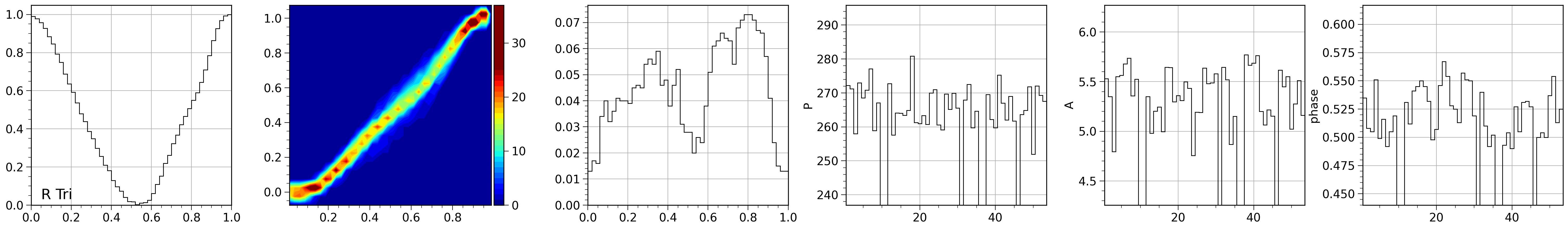}
  \includegraphics[width=1\linewidth,trim=0.cm 0cm 0cm 0cm,clip]{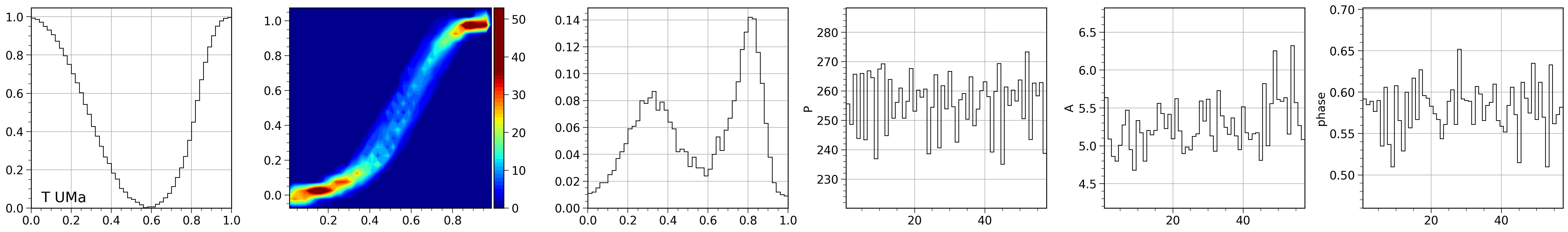}

  \caption{32 stars of sample B. From left to right: the mean normalised profile, the superimposed normalised profiles of the ascending branch, the rms deviation of each normalised pulse profile with respect to the mean normalised oscillation profile, the dependence on oscillation number of the period $P$ (days), the amplitude $A$ (mag) and the phase $\varphi_{\rm min}$ as listed in Table 1 of the main text. The curves are ranked in alphabetic order for each spectral type separately, ending with the 8 additional M-type curves introduced in Section 4.4.}
 \label{figA1}
\end{figure*}
\setcounter{figure}{0}
\begin{figure*}[h]
  \centering
  \includegraphics[width=1\linewidth,trim=0.cm 0cm 0cm 0cm,clip]{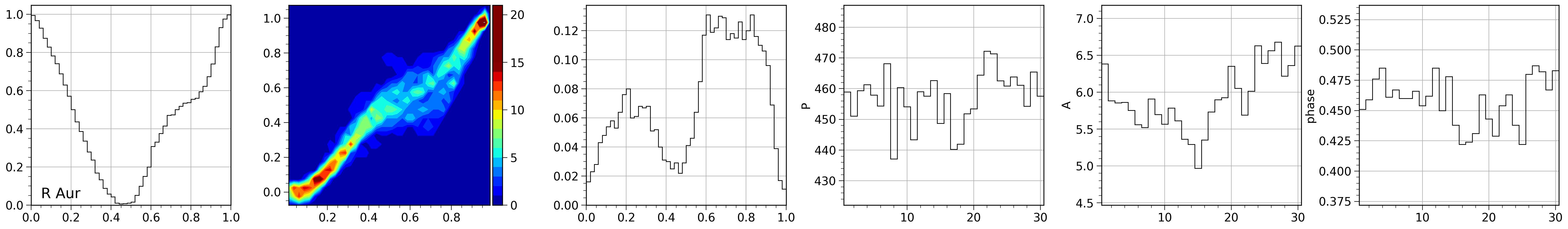}
  \includegraphics[width=1\linewidth,trim=0.cm 0cm 0cm 0cm,clip]{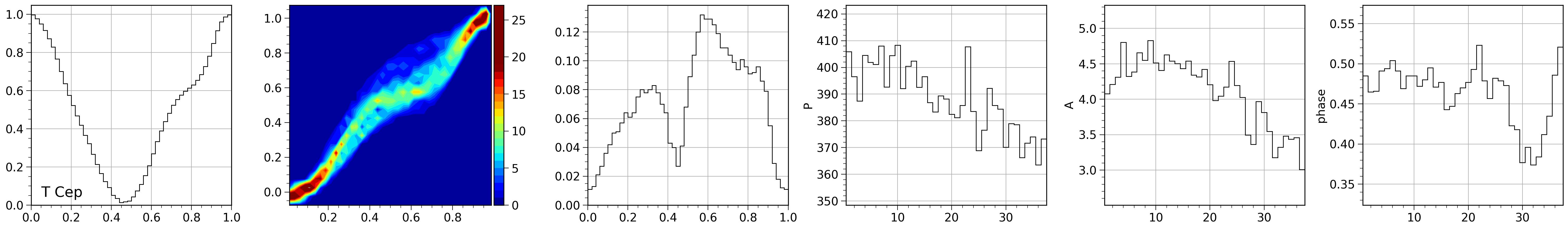}
  \includegraphics[width=1\linewidth,trim=0.cm 0cm 0cm 0cm,clip]{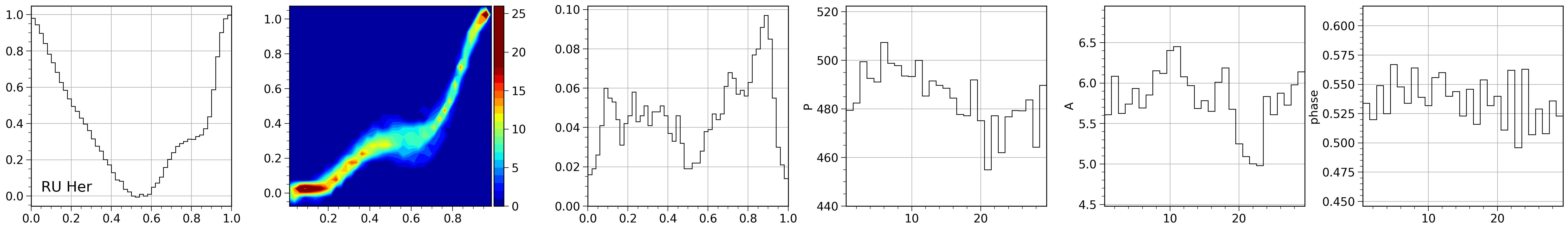}
  \includegraphics[width=1\linewidth,trim=0.cm 0cm 0cm 0cm,clip]{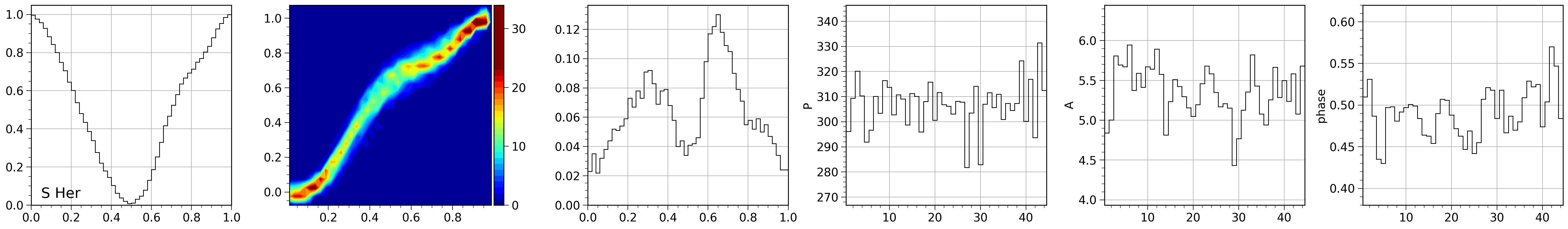}
  \includegraphics[width=1\linewidth,trim=0.cm 0cm 0cm 0cm,clip]{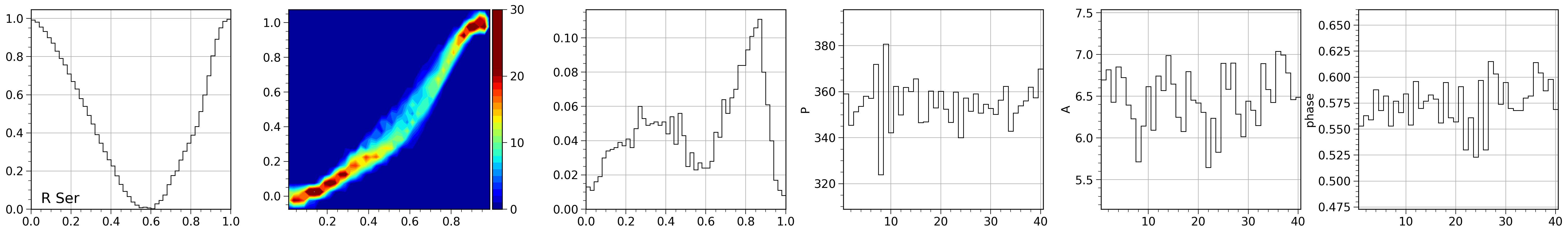}
  \includegraphics[width=1\linewidth,trim=0.cm 0cm 0cm 0cm,clip]{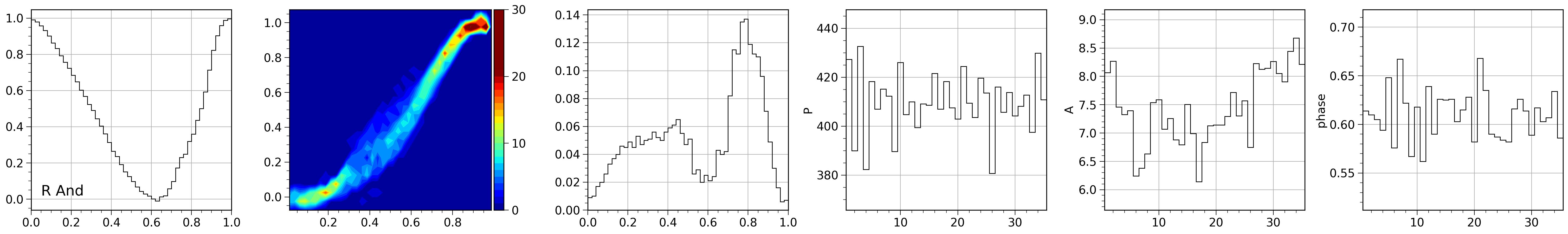}
  \includegraphics[width=1\linewidth,trim=0.cm 0cm 0cm 0cm,clip]{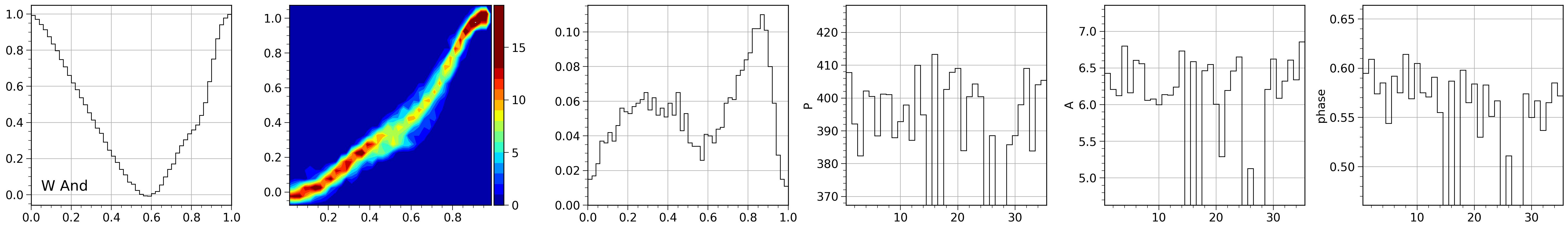}
  \includegraphics[width=1\linewidth,trim=0.cm 0cm 0cm 0cm,clip]{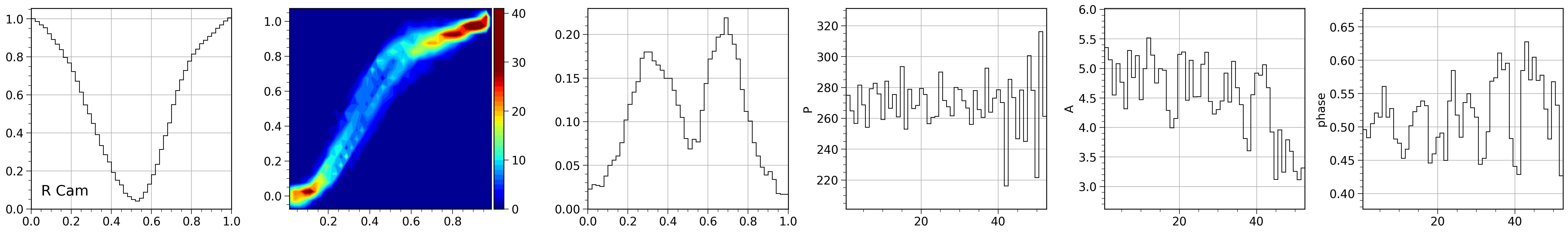}
 
   \caption{\edit1{(continued)}}
\end{figure*}
\setcounter{figure}{0}
\begin{figure*}[h]
  \centering
  \includegraphics[width=1\linewidth,trim=0.cm 0cm 0cm 0cm,clip]{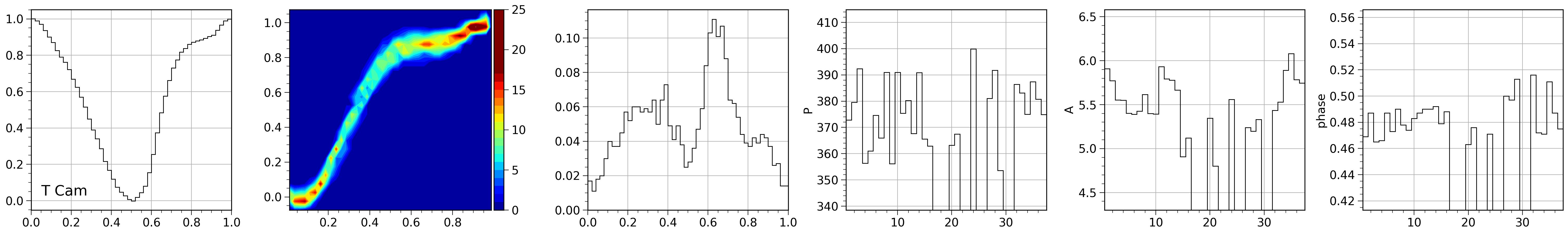}  
  \includegraphics[width=1\linewidth,trim=0.cm 0cm 0cm 0cm,clip]{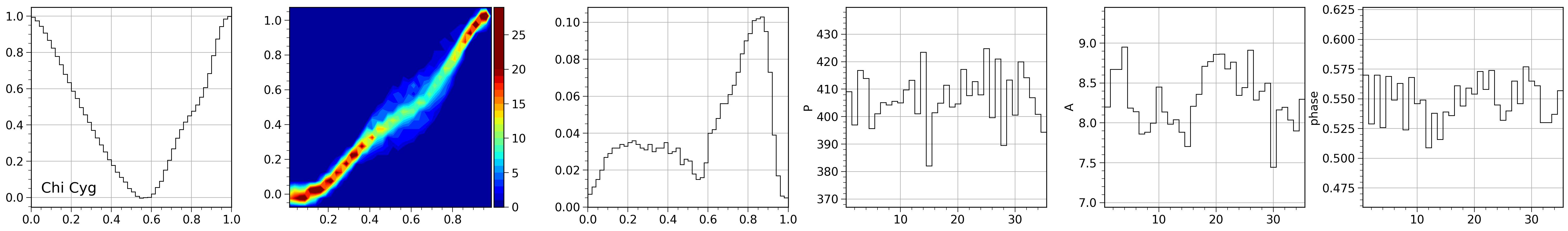}
  \includegraphics[width=1\linewidth,trim=0.cm 0cm 0cm 0cm,clip]{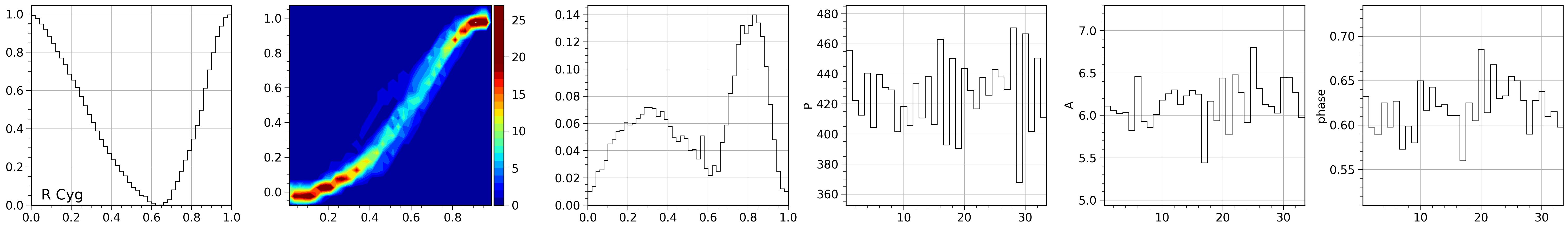}
  \includegraphics[width=1\linewidth,trim=0.cm 0cm 0cm 0cm,clip]{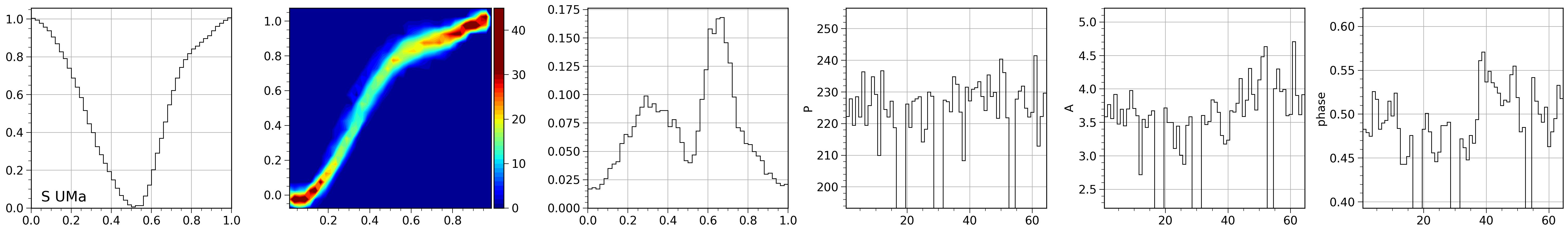}
  \includegraphics[width=1\linewidth,trim=0.cm 0cm 0cm 0cm,clip]{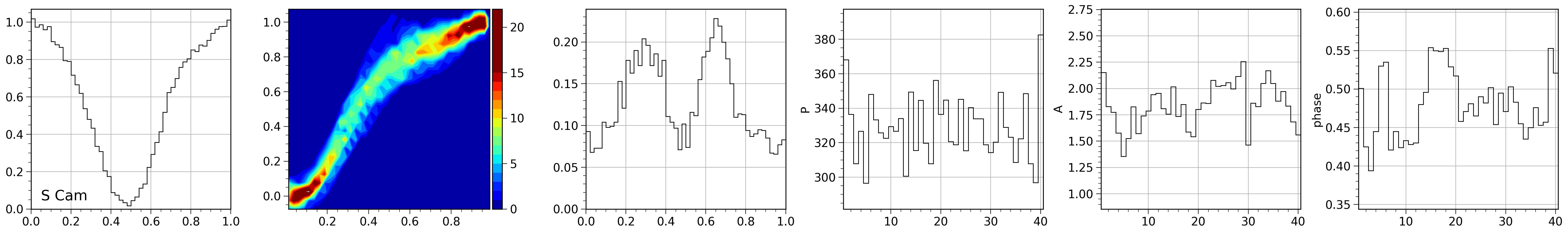}
  \includegraphics[width=1\linewidth,trim=0.cm 0cm 0cm 0cm,clip]{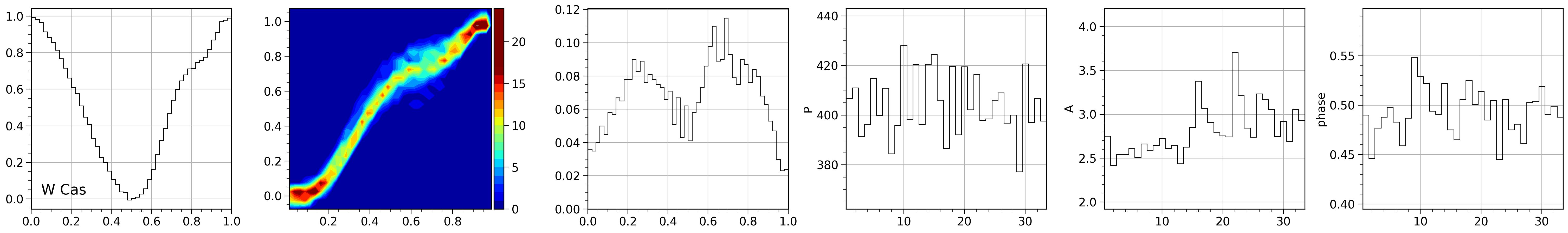}
  \includegraphics[width=1\linewidth,trim=0.cm 0cm 0cm 0cm,clip]{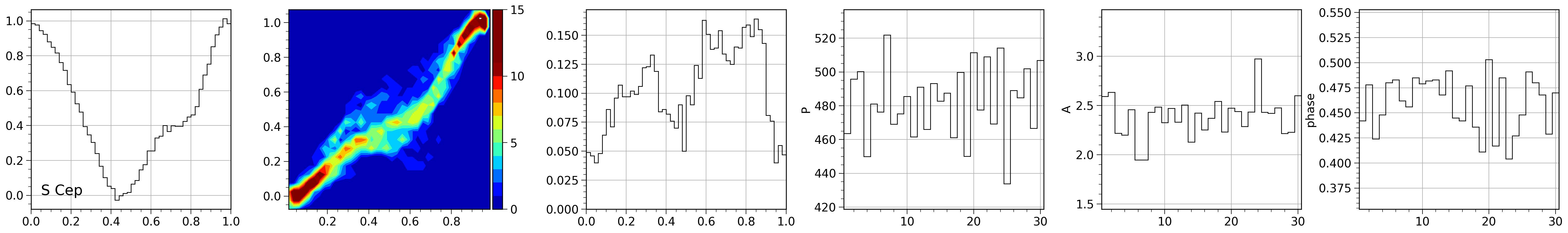}
  \includegraphics[width=1\linewidth,trim=0.cm 0cm 0cm 0cm,clip]{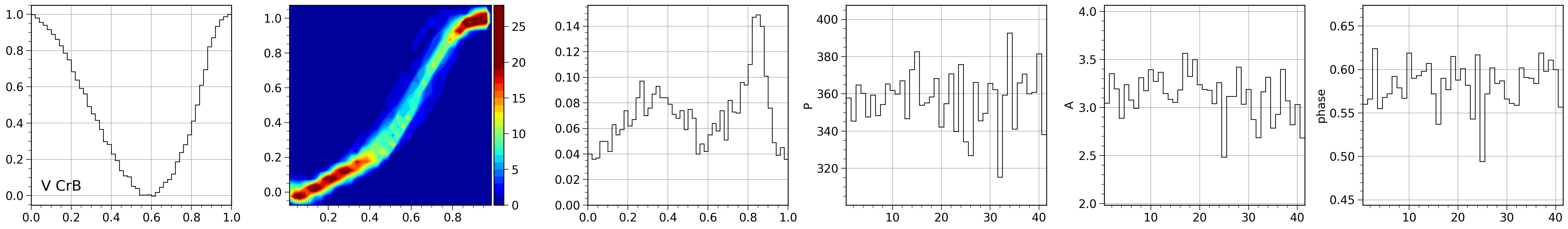}
  
  \caption{ \edit1{(continued)}}
\end{figure*}
\setcounter{figure}{0}
\begin{figure*}[h]
  \centering
  \includegraphics[width=1\linewidth,trim=0.cm 0cm 0cm 0cm,clip]{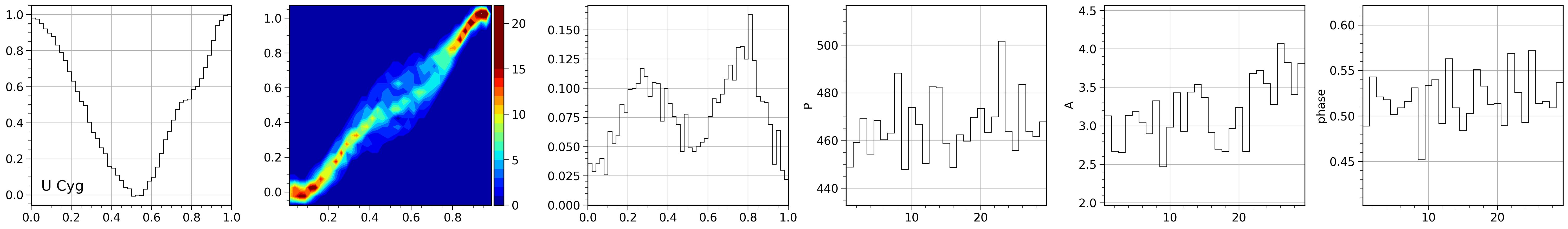}
  \includegraphics[width=1\linewidth,trim=0.cm 0cm 0cm 0cm,clip]{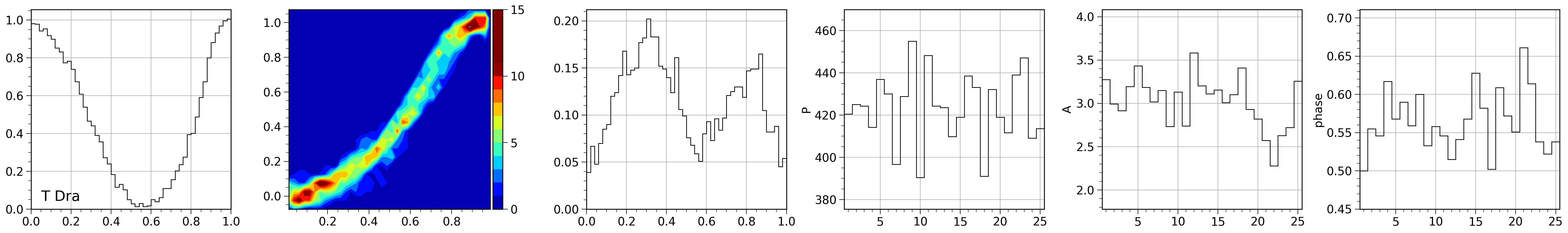}
  \includegraphics[width=1\linewidth,trim=0.cm 0cm 0cm 0cm,clip]{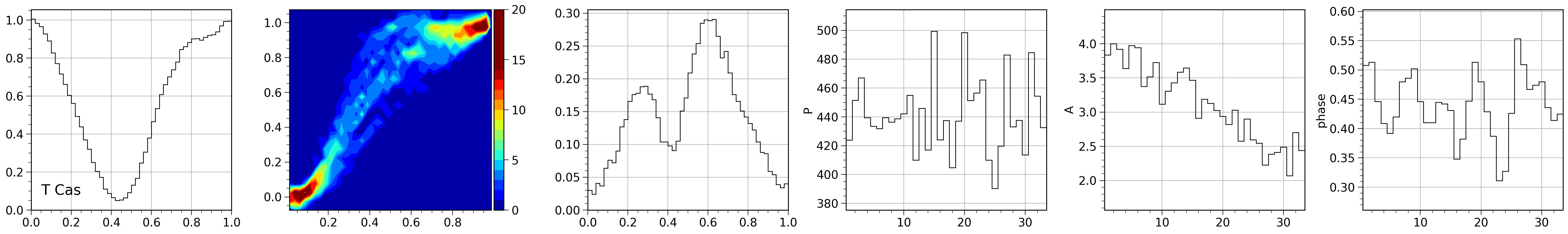}
  \includegraphics[width=1\linewidth,trim=0.cm 0cm 0cm 0cm,clip]{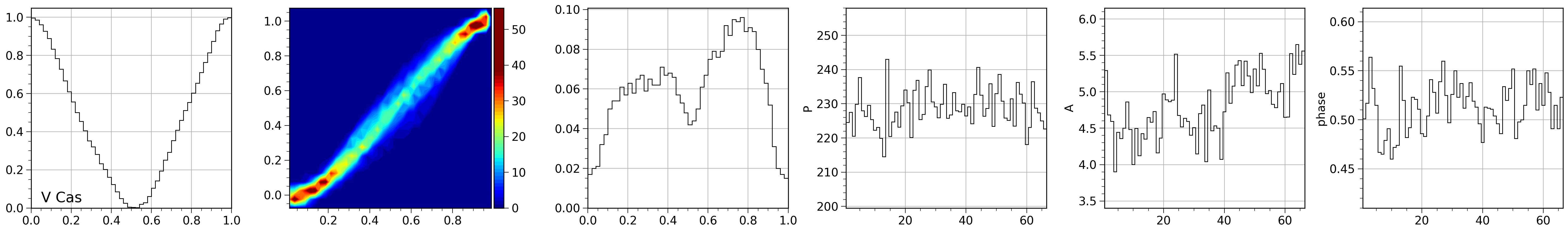}
  \includegraphics[width=1\linewidth,trim=0.cm 0cm 0cm 0cm,clip]{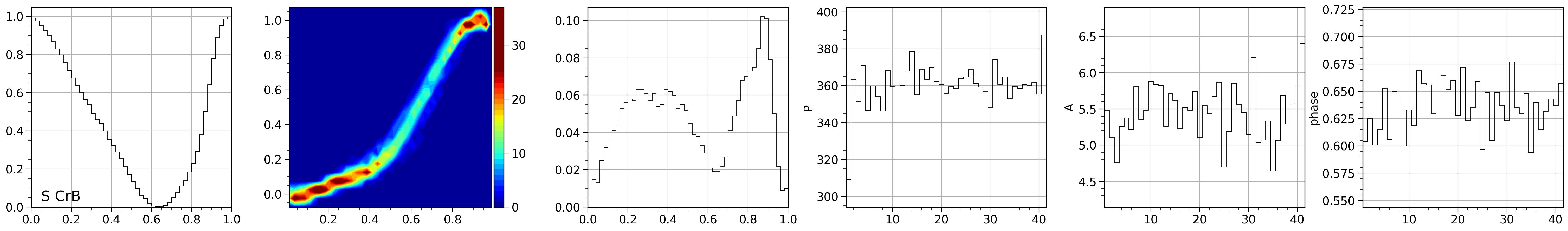}
  \includegraphics[width=1\linewidth,trim=0.cm 0cm 0cm 0cm,clip]{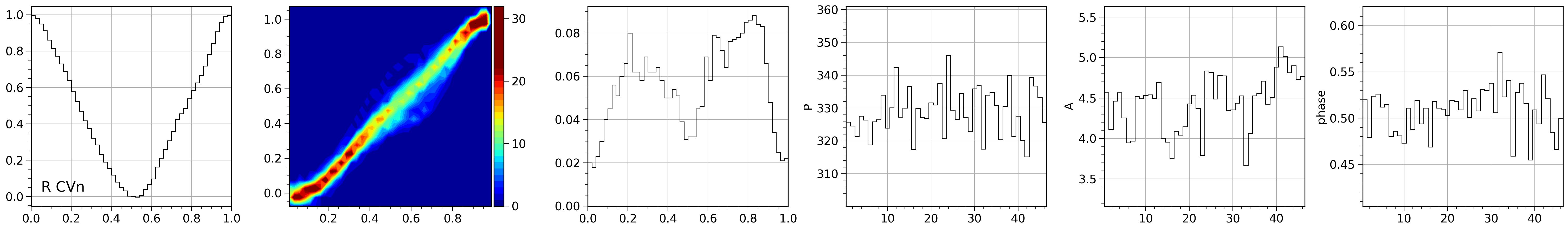}
  \includegraphics[width=1\linewidth,trim=0.cm 0cm 0cm 0cm,clip]{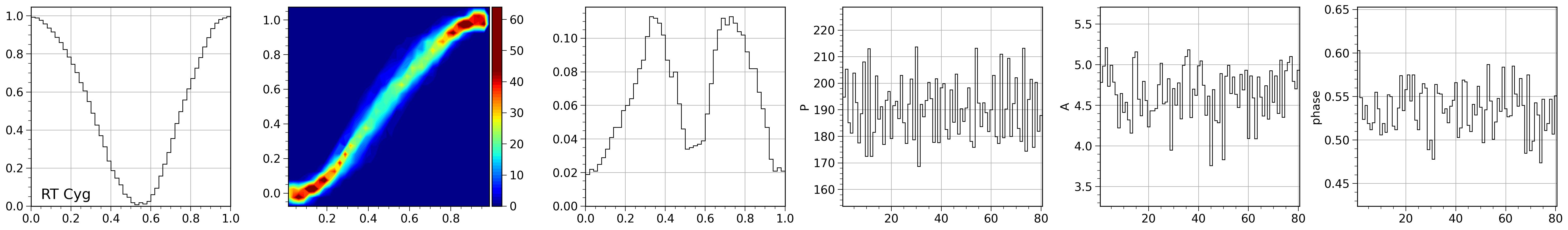}
  \includegraphics[width=1\linewidth,trim=0.cm 0cm 0cm 0cm,clip]{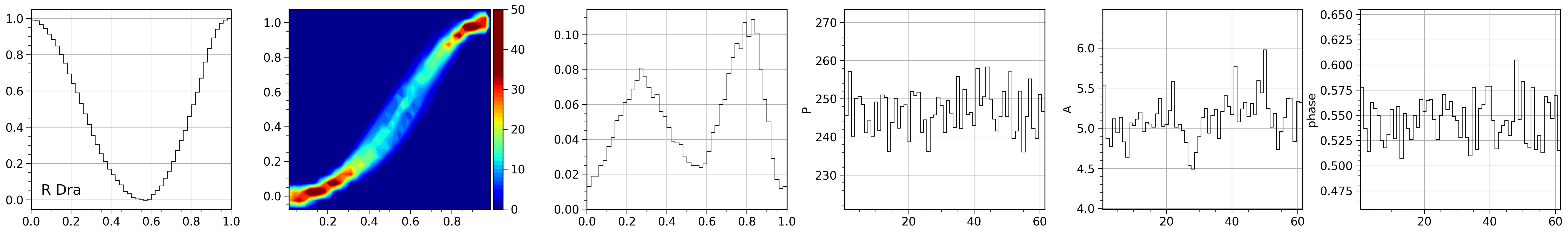}
  \caption{\edit1{(continued)}}
\end{figure*}

\setcounter{figure}{0}
\begin{figure*}[h]
  \centering
  \includegraphics[width=1\linewidth,trim=0.cm 0cm 0cm 0cm,clip]{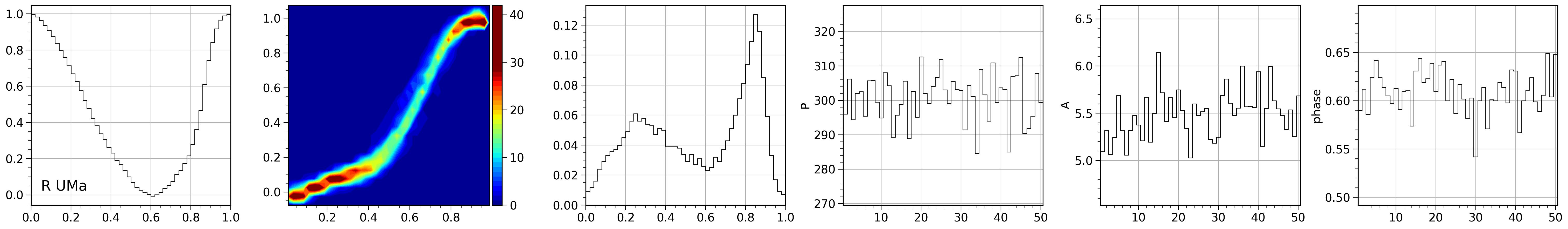}
  \includegraphics[width=1\linewidth,trim=0.cm 0cm 0cm 0cm,clip]{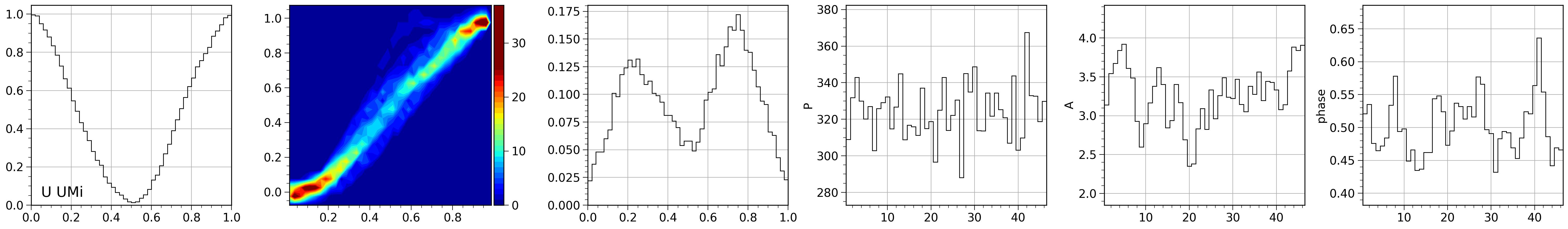}
  \caption{\edit1{(continued)}}
\end{figure*}

\startlongtable 
\begin{deluxetable*}{lcccccccccc}
\tablenum{A1}
\tablecaption{Parameters of the 71 curves of sample A. Columns 2 ($P/dP/P$), 3 (K-[22]/[3.4]-[22]), 6 (spectral type) and 7 ($^{12}$C/$^{13}$C ratio) are copied from Merchán Benítez et al. (2023). Columns 4 (asymmetry parameter $\alpha$, in \%), 5 (irregularity parameters $\Delta M_{\rm max}$ and $\Delta ^\prime M_{\rm max}$, in units of 0.01 mag), 8 (oscillation amplitude $A$, in mag), 9 (Merchán Benítez index $K_{\rm MB}$), 10 (full width at half maximum of the oscillation profile in units of period, $W_{\rm 1/2}$) and 11 (profile type) are evaluated approximately as described in Sections 2 and 3. The curves are ranked by alphabetic order in each of the four spectral-type families, Mno, Myes, S and C. \label{tab5}}
\tablehead{  \colhead{Name} & \colhead{$P/$} & \colhead{$K-$[22]/} & \colhead{$\alpha$} & \colhead{$\Delta M_{\rm max}$/} & \colhead{Type} & \colhead{$^{12}$C/$^{13}$C} & \colhead{$A$} & \colhead{$K_{\rm MB}$} & \colhead{$W_{\rm 1/2}$} & \colhead{Profile}\\
\colhead{} & \colhead{$dP/P$} & \colhead{[3.4]$-$[22]} & \colhead{} & \colhead{$\Delta ^\prime M_{\rm max}$} & 
\colhead{} & \colhead{} & \colhead{} & \colhead{} & \colhead{} & \colhead{}}
\startdata
R Aql	&	281/17	&	2.3/2.2	&	$-$7	&	36/38	&	Mno	&	8	&	5.0	&	$-$0.6	&	45	&	\edit2{a} 	\\
R Boo	&	224/2	&	1.6/1.8	&	$-$10	&	30/38	&	Mno	&	-	&	5.5	&	$-$0.5	&	45	&	\edit2{a} 	\\
R Cas	&	431/2	&	2.4/1.8	&	$-$25	&	75/99	&	Mno	&	12	&	6.5	&	1.0	&	45	&	\edit2{b} 	\\
X CrB	&	241/3	&	1.6/1.8	&	$-$7	&	33/38	&	Mno	&	-	&	4.5	&	$-$0.3	&	44	&	\edit2{a} 	\\
R Del	&	286/2	&	1.8/1.5	&	$-$14	&	35/41	&	Mno	&	-	&	4.5	&	$-$0.4	&	39	&	\edit2{a} 	\\
W Dra	&	279/15	&	3.0/2.4	&	$-$18	&	49/51	&	Mno	&	-	&	5.0	&	$-$1.3	&	51	&	\edit2{a} 	\\
T Her	&	165/2	&	1.3/1.6	&	$-$13	&	40/50	&	Mno	&	-	&	5.0	&	$-$0.9	&	49	&	\edit2{a} 	\\
U Her	&	406/2	&	2.4/2.5	&	$-$20	&	36/41	&	Mno	&	19	&	4.5	&	0.7	&	45	&	\edit2{b} 	\\
S Lac	&	240/3	&	2.1/1.9	&	$-$7	&	29/39	&	Mno	&	-	&	4.5	&	$-$0.8	&	44	&	\edit2{a} 	\\
R Leo	&	312/3	&	1.3/1.0	&	$-$13	&	37/47	&	Mno	&	10	&	5.0	&	0.8	&	38	&	\edit2{b} 	\\
R LMi	&	373/2	&	3.0/2.4	&	$-$19	&	56/66	&	Mno	&	12	&	5.0	&	$-$0.3	&	41	&	\edit2{b} 	\\
R Tri	&	267/1	&	1.8/1.5	&	$-$11	&	32/39	&	Mno	&	-	&	5.7	&	$-$0.2	&	44	&	\edit2{a} 	\\
T UMa	&	256/2	&	2.4/2.3	&	$-$20	&	50/69	&	Mno	&	-	&	5.5	&	$-$1.0	&	47	&	\edit2{a} 	\\
U Ari	&	372/2	&	2.1/2.1	&	$-$26	&	57/61	&	Myes	&	-	&	6.2	&	0.6	&	44	&	\edit2{a} 	\\
R Aur	&	457/4	&	2.4/1.8	&	5	&	42/53	&	Myes	&	33	&	6.0	&	1.3	&	62	&	\edit2{d}	\\
T Cep	&	390/7	&	1.2/1.2	&	5	&	28/29	&	Myes	&	33	&	4.5	&	1.7	&	59	&	\edit2{c} 	\\
S CMi	&	333/2	&	2.0/1.7	&	$-$1	&	31/27	&	Myes	&	18	&	5.2	&	0.3	&	59	&	\edit2{d}	\\
V CMi	&	366/2	&	2.5/2.0	&	$-$17	&	76/101	&	Myes	&	-	&	6.0	&	0.2	&	35	&	\edit2{a} 	\\
V Gem	&	275/5	&	1.5/1.5	&	$-$14	&	48/53	&	Myes	&	-	&	6.7	&	0.2	&	45	&	\edit2{a} 	\\
RU Her	&	486/5	&	2.2/2.4	&	$-$11	&	60/78	&	Myes	&	25	&	6.5	&	1.8	&	34	&	\edit2{b} 	\\
S Her	&	306/5	&	1.0/1.2	&	$-$2	&	32/39	&	Myes	&	-	&	5.5	&	1.0	&	58	&	\edit2{d}	\\
R Hor	&	405/3	&	2.4/1.8	&	$-$21	&	46/53	&	Myes	&	-	&	7.0	&	0.7	&	43	&	\edit2{a} 	\\
R Hya	&	388/15	&	0.8/0.9	&	7	&	32/26	&	Myes	&	26	&	3.5	&	2.1	&	55	&	\edit2{c} 	\\
S Ori	&	424/9	&	1.7/1.9	&	3	&	41/23	&	Myes	&	45	&	5.0	&	1.6	&	48	&	\edit2{c} 	\\
Z Peg	&	328/4	&	1.6/1.7	&	0	&	28/33	&	Myes	&	-	&	5.0	&	0.6	&	45	&	-	\\
R Ser	&	356/2	&	1.9/2.0	&	$-$20	&	49/67	&	Myes	&	14	&	7.0	&	0.6	&	42	&	\edit2{a} 	\\
S Vir	&	378/3	&	2.0/1.9	&	$-$12	&	35/48	&	Myes	&	-	&	6.2	&	0.8	&	41	&	\edit2{b} 	\\
R And	&	410/3	&	2.9/2.3	&	$-$24	&	77/71	&	Syes	&	40	&	7.0	&	0.2	&	46	&	\edit2{a} 	\\
RR And	&	328/2	&	1.3/1.2	&	$-$6	&	47/52	&	S?	&	-	&	6.0	&	0.9	&	53	&	\edit2{a/-}	\\
W And	&	396/2	&	1.8/1.8	&	$-$20	&	73/91	&	Syes	&	-	&	6.5	&	1.2	&	38	&	\edit2{b} 	\\
X And	&	346/2	&	2.2/2.1	&	$-$26	&	39/47	&	S?	&	-	&	5.5	&	0.2	&	42	&	\edit2{a} 	\\
W Aql	&	487/3	&	3.0/3.4	&	$-$25	&	102/58	&	Syes	&	26	&	6.0	&	1.0	&	44	&	\edit2{a} 	\\
R Cam	&	270/4	&	0.2/0.6	&	$-$3	&	29/23	&	S?	&	-	&	5.0	&	1.4	&	57	&	\edit2{d}	\\
T Cam	&	374/4	&	0.8/0.5	&	$-$3	&	18/14	&	Syes	&	31	&	5.5	&	1.9	&	63	&	\edit2{d}	\\
S Cas	&	613/3	&	4.1/3.1	&	$-$20	&	65/51	&	S?	&	32	&	6.0	&	1.3	&	34	&	\edit2{b} 	\\
U Cas	&	277/2	&	1.3/1.0	&	$-$11	&	31/37	&	S?	&	-	&	6.5	&	0.4	&	48	&	\edit2{a} 	\\
WY Cas	&	479/1	&	3.6/2.8	&	$-$23	&	58/54	&	Syes	&	-	&	6.0	&	0.3	&	38	&	\edit2{b} 	\\
SZ Cep	&	329/2	&	2.0/1.5	&	$-$8	&	50/34	&	S?	&	-	&	5.5	&	0.2	&	49	&	\edit2{d}	\\
V Cnc	&	272/2	&	1.8/1.0	&	$-$17	&	33/36	&	Syes	&	-	&	5.0	&	$-$0.2	&	57	&	\edit2{a} 	\\
Chi Cyg	&	408/2	&	2.4/1.7	&	$-$17	&	71/96	&	Syes	&	36	&	9	&	0.7	&	49	&	\edit2{c} 	\\
FF Cyg	&	325/2	&	0.9/1.0	&	$-$9	&	21/19	&	S?	&	-	&	4.5	&	1.3	&	62	&	\edit2{d}	\\
R Cyg	&	427/2	&	2.9/2.9	&	$-$29	&	92/144	&	Syes	&	29	&	5.8	&	0.4	&	49	&	\edit2{a} 	\\
S Cyg	&	323/3	&	0.7/0.9	&	$-$4	&	49/67	&	S?	&	-	&	5-6	&	1.5	&	49	&	\edit2{c} 	\\
Z Del	&	305/2	&	1.6/1.2	&	$-$5	&	42/54	&	Syes	&	-	&	5.5	&	0.4	&	44	&	\edit2{c} 	\\
R Gem	&	370/2	&	1.9/2.0	&	$-$31	&	24/31	&	Syes	&	22	&	7.0	&	0.8	&	43	&	\edit2{a} 	\\
T Gem	&	289/4	&	0.1/0.8	&	$-$4	&	27/22	&	Syes	&	-	&	5.5	&	1.7	&	56	&	\edit2{d}	\\
R Lyn	&	378/2	&	1.6/1.1	&	$-$14	&	44/44	&	Syes	&	-	&	6.0	&	1.2	&	53	&	\edit2{d}	\\
T Sgr	&	393/4	&	1.7/1.2	&	$-$8	&	33/32	&	Syes	&	-	&	4.0	&	1.2	&	43	&	\edit2{c} 	\\
S UMa	&	226/4	&	1.2/1.1	&	$-$1	&	14/14	&	Syes	&	-	&	3.5	&	$-$0.1	&	58	&	\edit2{d}	\\
AZ Aur	&	415/3	&	3.0/2.0	&	$-$10	&	68/51	&	Cyes	&	-	&	$\sim$3	&	0.2	&	50	&	-	\\
UV Aur	&	395/1	&	3.0/2.5	&	$-$18	&	42/38	&	Cyes	&	-	&	2.2	&	0.0	&	48	&	-	\\
V Aur	&	353/3	&	1.7/1.5	&	4	&	21/23	&	C?	&	-	&	2.5	&	0.8	&	48	&	\edit2{a/-}	\\
S Cam	&	330/3	&	0.9/1.2	&	2	&	16/14	&	C?	&	14	&	2.0	&	1.4	&	65	&	\edit2{d}	\\
W Cas	&	408/2	&	0.7/1.2	&	$-$5	&	22/19	&	Cyes	&	25	&	2.5	&	2.4	&	60	&	\edit2{d/c}	\\
X Cas	&	424/4	&	1.6/1.3	&	10	&	29/22	&	C?	&	19	&	2	&	1.7	&	56	&	\edit2{e/c}	\\
AX Cep	&	398/2	&	2.7/2.6	&	$-$10	&	79/51	&	C?	&	-	&	2.5	&	0.3	&	53	&	\edit2{e/a}	\\
S Cep	&	484/3	&	2.7/1.7	&	6	&	47/50	&	C?	&	224	&	2.5	&	1.2	&	47	&	\edit2{c} 	\\
R CMi	&	340/2	&	1.1/1.4	&	$-$12	&	46/34	&	Cyes	&	11	&	2.7	&	1.3	&	53	&	\edit2{e/a}	\\
V CrB	&	358/4	&	2.3/1.9	&	$-$20	&	50/41	&	Cyes	&	-	&	3	&	0.3	&	45	&	\edit2{a} 	\\
BH Cru	&	529/10	&	1.6/1.4	&	$-$3	&	25/22	&	SCyes	&	8	&	3	&	2.8	&	65	&	\edit2{c/e}	\\
U Cyg	&	463/5	&	1.9/2.2	&	$-$4	&	48/43	&	Cyes	&	14	&	3.0	&	1.8	&	47	&	\edit2{c} 	\\
V Cyg	&	421/2	&	3.2/2.3	&	$-$14	&	61/47	&	C?	&	20	&	3.5	&	0.1	&	45	&	\edit2{a/e}	\\
WX Cyg	&	411/2	&	1.5/1.5	&	$-$7	&	20/21	&	Cyes	&	5	&	4.7	&	1.6	&	60	&	\edit2{e/d}	\\
T Dra	&	422/4	&	3.4/2.6	&	$-$9	&	54/44	&	Cyes	&	24	&	3	&	$-$0.1	&	50	&	\edit2{e/a}	\\
R Lep	&	427/5	&	3.0/2.0	&	$-$2	&	84/47	&	C?	&	34	&	$\sim$2	&	0.3	&	50	&	\edit2{c/e}	\\
T Lyn	&	409/6	&	2.1/1.8	&	$-$10	&	53/37	&	C?	&	-	&	3	&	1.0	&	51	&	\edit2{a/-}	\\
U Lyr	&	452/4	&	2.4/1.8	&	4	&	45/46	&	C?	&	23	&	$\sim$2	&	1.2	&	60	&	\edit2{c/d}	\\
V Oph	&	297/4	&	1.6/1.4	&	$-$1	&	28/29	&	Cyes	&	11	&	2.5	&	0.3	&	63	&	\edit2{d/e/a}	\\
RZ Peg	&	439/2	&	2.2/2.3	&	$-$16	&	36/40	&	Cyes	&	9	&	3.5	&	1.3	&	51	&	\edit2{a/d}	\\
Y Per	&	249/4	&	1.6/1.1	&	10	&	23/21	&	C?	&	-	&	$\leq$1	&	$-$0.2	&	70	&	\edit2{-/e}	\\
BD Vul	&	430/3	&	1.8/1.6	&	$-$1	&	46/52	&	C?	&	-	&	2.5	&	1.6	&	51	&	\edit2{-/e}	\\
\enddata
\end{deluxetable*}



\end{document}